# Early Architecture Concepts for the Habitable Worlds Observatory – System Design, Modeling, and Analysis

Alice (Kuo-Chia) Liu,[a,*] Marie Levine,[b] Charley Noecker,[b] Jon Lawrence,[a] Joshua Abel,[a] Michael Akkerman,[c] Eric Anstadt,[d] Ruslan Belikov,[e] Pin Chen,[b] Kenneth Dziak,[c] Jordan Effron,[a] Lee Feinberg,[a] Alan Gostin,[a] James Govern,[c] Cameron Haag,[b] Joseph Howard,[a] Brian Kern,[b] Gary Kuan,[b] Milan Mandic,[b] Carson McDonald,[f] Connor Mulrenin,[a] Bijan Nemati,[g] Jon Papa,[a] Fang Shi,[b] Samuel Sirlin,[b] Breann Sitarski,[a] Cory Smiley,[c] J. Scott Smith,[a] Philip Stahl,[h] Christopher Stark,[a] Gregory Walsh,[a] John Ziemer[b]

[a]NASA Goddard Space Flight Center, 8800 Greenbelt Rd., Greenbelt, USA

[b]Jet Propulsion Laboratory, California Institute of Technology, 4800 Oak Grove Dr., La Cañada Flintridge, USA

[c]Aerodyne Industries, LLC, 8910 Astronaut Blvd., Suite 208, Cape Canaveral, USA

[d]Quartus Engineering, 2300 Dulles Station Blvd., Suite 650, Herndon, USA

[e]NASA Ames Research Center, Moffett Field, USA

[f]Vertex Aerospace, LLC, P.O. Box 192, Grasonville, USA

[g]Tellus1 Scientific, LLC, 8401 Whitesburg Dr. SE, Unit 4662, Huntsville, AL, USA

[h]NASA Marshall Space Flight Center, Martin Rd. SW, USA

The acknowledgements section, at the end, recognizes additional individuals and organizations whose work contributed to this paper's content.

**Abstract**. The Habitable Worlds Observatory (HWO), NASA's next flagship science mission, follows in the tradition of the Nancy Grace Roman Space Telescope (Roman) and other preceding great observatories. HWO will directly image and characterize Earth-like exoplanets and their atmospheres, with the capability to detect biosignatures and potentially answer the question "are we alone?" HWO will also serve as a powerful general astrophysics observatory, enabling breakthroughs in galaxy evolution, stellar astrophysics, and dark matter studies. Currently in pre-formulation, the project has established Exploratory Analytic Cases (EACs), a series of architectural concept designs used to assess the mission's demanding science objectives while exploring challenging engineering parameters. This paper describes the first three EACs, starting with observing strategies and error budget formulation and then progressing to design formulations, trade studies, and lessons learned; this paper also discusses the integrated modeling pipeline, a key multidisciplinary system-level analysis capability, and analysis findings as applied to the first EAC. These activities set the stage for the follow-on EACs 4 and 5, which will further explore the trade space and prepare for the baseline design that will support the Mission Concept Review (MCR).

*Alice Liu, E-mail: alice.liu@nasa.gov

## 1 Introduction

Direct imaging of exoplanets, which requires blocking host starlight ten billion times brighter than the Earth-sized planets orbiting them, remains a technologically challenging science goal. Achieving the necessary contrast and stability entails carefully designed and manufactured coronagraph masks, an ultrastable telescope with mirror wavefront stability at the picometer level, and advanced low-noise detectors - these are just some of the technologies required for mission success. Fortunately, previous observatories have positioned HWO technology development as evolutionary rather than revolutionary, greatly reducing mission risk. Examples include segmented mirror and wavefront sensing and control heritage from the James Webb Space Telescope (JWST), coronagraph demonstration from Roman, and design for serviceability from the Hubble Space Telescope (HST). These precursor achievements define a clear path forward for realizing HWO mission objectives.[1]

As HWO entered the pre-formulation phase in 2024, the Technology Maturation Project Office (TMPO) tasked the Systems Engineering Team (SET), consisting of members from NASA Ames, Goddard, JPL, and Marshall, to translate science goals into implementable designs, enable observatory performance evaluation, and identify technology gaps. This work, in parallel with industry teams' efforts to advance HWO technologies and mission designs, focuses on developing and evaluating architecture options, referred to as Exploratory Analytic Cases (EAC), from a holistic mission perspective and determining technology requirements. These EACs represent notional system architectures and are the SET's main tool, through design, trades, and analyses, for achieving its objectives; the first three EACs (covered in detail in this paper) accommodate a variety of subsystem and interface implementation approaches to maximize design exploration.

Architecture design, in this context, must consider the interconnectedness among error budgets, observation strategies (e.g. Concept of Operations, or ConOps), and post-processing techniques, all things that require close collaboration between science and engineering, more so than most previous missions. For example, the ConOps directly influences system design and drives observatory capabilities, while post-processing offers promising options to relax stability requirements; consequently, because ConOps and post-processing are often closely linked, we refer to them as COPP. For performance assessment, the SET began by formulating the coronagraph Flux Ratio Noise (FRN) budget to establish top-down allocations for various contributors to raw contrast, throughput, and contrast stability; for this, the SET adopted a ConOps similar to Roman's Angular Differential Imaging (ADI) as a point of departure.

This paper provides a summary of the different ConOps under consideration and the initial FRN budget, in Sec. 1.1 and Sec. 1.2, respectively. Details of both topics will be addressed in future dedicated conference papers. The paper then steps through the design evolution of each of the first three EACs and shares lessons learned in Sec. 0. Due to time constraints, only the EAC1 design was developed for system analysis; however, key analysis results and findings are often common for all three EACs. Sec. 3 and Sec. 4 describe the HWO system modeling and analysis (i.e. integrated modeling) approaches, respectively; Sec. 4 in particular highlights initial results, including critical trade studies that inform technology requirements. Ultimately, the culmination of SET work will inform selection of the next round of EACs, eventually narrowing to a single baseline design that demonstrates complete mission feasibility at the Mission Concept Review (MCR).

## 1.1 Observation Strategies and Post Processing

Given the challenges with exoplanet detection, the plans for observatory observation must be coordinated with the observatory design features to provide the best possible advantage. Using modulation and synchronous detection to isolate weak signal represents one such optimization; such strategies have system engineering and programmatic benefits, including:

- Defining a maximum time scale for stability requirements,
- Defining clear metrics for performance requirements of all subsystems,
- Enabling clear partitioning of interface requirements among subsystem vendors, and
- Developing clear testing criteria at all subsystem levels during flight integration and test

The Roman coronagraph and telescope were designed for two methods of modulation using the observatory attitude, captured in an Observing Scenario known as OS-11.

- Reference Differential Imaging (RDI) means alternating between a science star and a reference star, and subtracting the two sets of images
- Angular Differential Imaging (ADI) means alternating between two roll angles around the line of sight to the science star, and subtracting images at the two roll orientations.

The strategy for RDI and ADI is to hold the Roman observatory extremely stable during these images, so that the speckles of residual starlight can be subtracted to reveal the light of one or more exoplanets. OS-11 embodies a series of observatory attitude maneuvers representing observations of one "science" star and one "reference" star at a specific epoch. We developed an error budget approach for RDI and ADI that mirrors the arithmetic of image subtraction as post-processing, and allows budgeting of telescope and coronagraph errors to keep track of the most serious problems.

We will not use the RDI technique for HWO, because its inherent biases make it impractical. The major biases are the sensitivities to the color (spectral distribution) and the angular size of the observed star; these sensitivities make it nearly impossible to compare a science star to a reference star accurately enough to allow Earth-size planet detection and characterization.

The first observing scenario (OS-1) considered for HWO EAC assessment is based on OS-11, as shown in Fig. 1. The "roll" angle is rotation around X, the telescope symmetry axis. "Pitch" is rotation around Y, turning the line of sight toward or away from the Sun. "Yaw" is a rotation around the Sun-normal-line—the one rotation that has no thermal effect. The science observation starts on the reference star before slewing to the target star. The observatory rolls between two orientations to support Angular Differential Imaging (ADI). This sequence alternates between the reference star, target star, and back to the reference star a total of 4 times over 48 hours. Although post-processing will not use RDI, the ConOps still requires observing a brighter reference star to establish and maintain the dark hole.

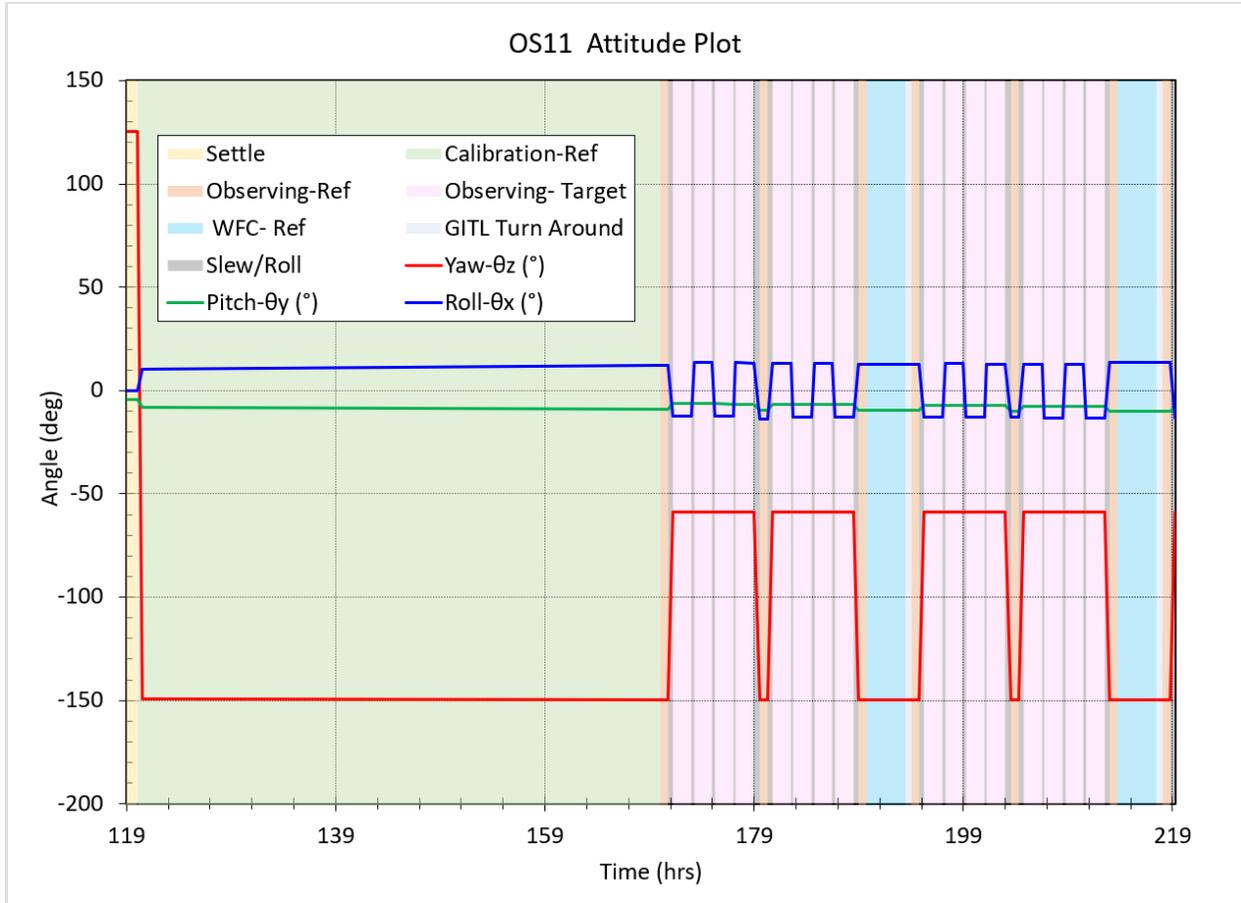

**Fig. 1** EAC1 Observing Scenario 1 Attitude Plot, based on Roman OS-11.

The science and SE teams are exploring other COPP strategies; among them are:

1. Dark Zone Maintenance,[2] a procedure for continually updating the deformable mirror settings to counteract any thermo-mechanical drifts in the telescope and coronagraph

2. Coherence Differential Imaging[3], exploiting the optical coherence of residual starlight and incoherence of the exoplanet light to help isolate the exoplanet signal from variability in starlight speckles

3. Polarization Difference Imaging[4], using the polarization of exoplanet light to help distinguish it from residual starlight

4. Medium Resolution Imaging Spectroscopy[5], using spectroscopy at resolution ~ 1000-2000 to detect exoplanets by the presence of molecular signatures, and characterize them by more precise quantification of those signatures

The SET is in the process of evaluating each COPP option to assess observatory accommodations requirements, including measurement sequences, instrument specifications, and performance drivers. In some cases, the SET developed the OS to run through IM and use the stability results to exercise various post-processing algorithms. The goal is to identify algorithms that can significantly relax stability requirements and/or reduce system design complexity.

*1.2 Error Budget Structure and Sub-Allocations to Integrated Modeling*

The error budget approach for HWO follows the methodology developed for the Roman Coronagraph Instrument (CGI);[6] the foundation for this approach is to compute the noise in a photometric measurement of the exoplanet. For direct detection and imaging, this would be photometry of the planet light within the observation band of the coronagraph, while for spectroscopy it would be for a spectral element, taking also into account the appropriate number of detector pixels involved in that measurement.

Considering first, the simpler case of direct detection, a signal is an enhancement of light against the dark hole background which is quantified via some processing technique. The simplest would be aperture photometry, where the counts within a "core" region centered on the planet PSF peak are taken as the signal $S$. More sophisticated techniques, such as matched filter methods, offer more accuracy and lower background noise, but for error budgeting purposes aperture photometry offers simplicity and clarity in assessing the noise.

The signal $S$ is proportional to the photometric quantity of interest, the flux ratio $\xi$:

$$S = \xi F_\lambda \Delta\lambda A \tau \eta t. \tag{1}$$

In the above expression, $F_\lambda$, is the spectral flux from the star, $\Delta\lambda$ is the spectral band width, $A$ the collecting area, $\tau$ the throughput of the instrument into the PSF core, $\eta$ the detector quantum efficiency, and $t$ the integration time. This simple linear correspondence lends itself well to error budgeting. We can write the flux ratio as $\xi = \kappa \cdot S$ where $\kappa$ contains all the observational and instrument factors. The errors are obtained by first differentiating this equation, then interpreting the differentials as standard deviations, and replacing addition operations with root-sum-square (RSS, symbolized by $\oplus$), so that we obtain $\delta\xi = \delta\kappa \oplus \delta S$. On the left is the flux ratio noise (FRN). On the right, the first term, $\delta\kappa$, corresponds to calibration errors, knowledge errors in the star flux, the throughput, and the detector efficiency. The second term, $\delta S$, is the noise in the extraction of the signal $S$. There are two major components to $\delta S$: random noise, including photon noise and detector noise, and speckle noise $\delta S = \delta S_{ran} \oplus \delta S_{spc}$. Thus, the FRN error budget's top level can be summarized by the equation

$$\delta\xi \;=\; \delta\kappa \oplus \delta S_{ran} \oplus \delta S_{spc}\,. \qquad (2)$$

The three terms on the right are each the top-level roll-up of one branch of the error budget. The FRN branches are called calibration, random noise, and speckle stability.

Noise estimation for the three branches is increasingly more challenging going from left to right: calibration factors are mostly multiplicative and the error propagation for these is simple. Random noise is a bit involved but well understood and straightforward. Speckle noise, on the other hand, is the result of the nonlinear processes that apply to the alteration, propagation, and perturbation of the field as it passes through the coronagraph and experiences change over time as the telescope and instrument undergo disturbances.

Figure 2 shows the top level FRN error budget for imaging and exo-Earth at ~ 60 mas separation, in a 20% wavelength band centered at 600 nm. The spectroscopy case is similar in structure, but the optical band is limited to a spectral element. The modifications needed for a spectroscopy FRN error budget are provided in Sec. 4.6 of Ref. 6.

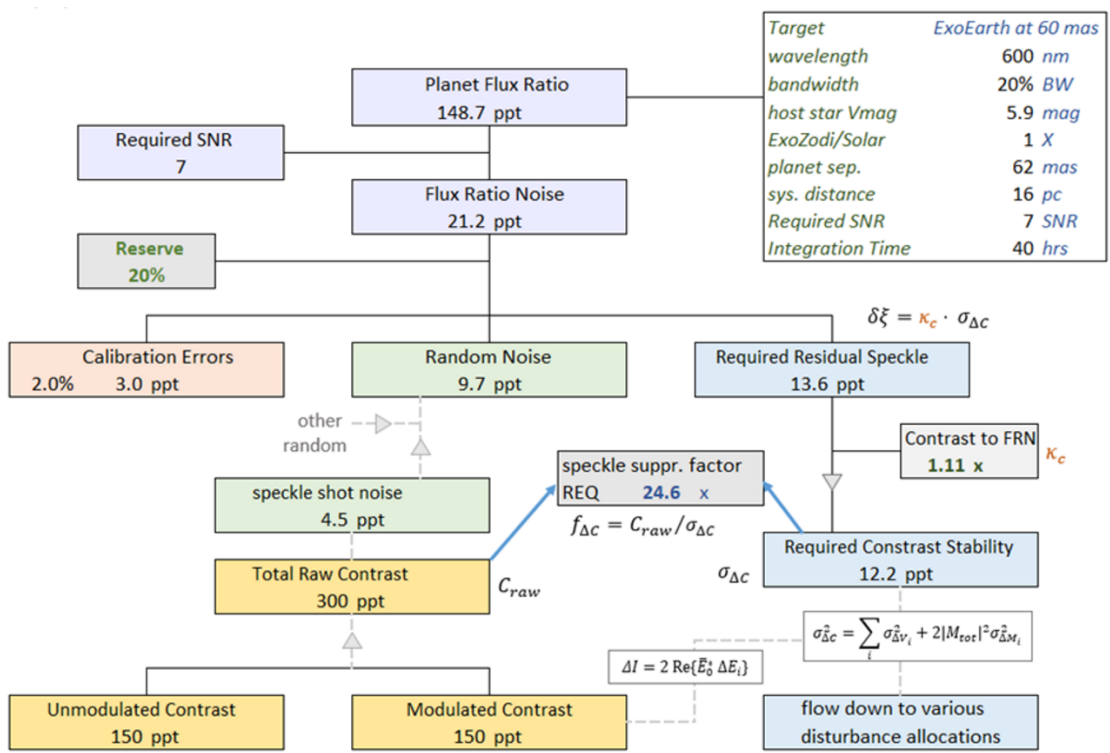

**Fig. 2** Top Level Error Budget for, Units in parts per trillion (ppt).

Speckle noise estimation is one of the most important uses of the integrated model. The integrated model, augmented with high- and low-order wavefront sensing models and algorithms, such as Electric Field Conjugation (EFC) for digging a dark hole[7], can predict first the generation of the dark hole speckles, including their chromatic properties, and then their temporal evolution as the observation progresses. The mean speckle level is measured by contrast, and its time evolution is called contrast stability. The third error budget branch flows into contrast stability.

Two assumptions are implied in the word *stability* in the context of speckles:

- They are lumpy, and can potentially confuse the signal
- They are changing, and that change affects how accurately the signal is measured

For the second point to be true, some sort of differential imaging is being assumed. HWO currently only assumes angular differential imaging (ADI) as described in Sec. 1.1 to decouple the speckles from the planet: the speckles would stay fixed to the spacecraft and land on the same part of the focal plane, while the planet would appear at two locations, corresponding to the two roll angles.

Other concepts of operation are currently being discussed, such as speckle smoothing, dark zone maintenance, and various methods for real-time speckle estimation. Perhaps when these mature, some or all of the contrast stability branch would be relaxed. Until that maturity is attained, speckle stability remains the most important branch of the error budget, and the most resource-consuming branch to estimate.

To estimate speckle instability noise the FRN model computes the ensemble variance of the speckles expected under the signal after differential imaging. In this model, four statistics are used to describe the residual starlight field and its evolution, and from these it computes both the total contrast and its instability pertaining to FRN. The four parameters of interest are computed for each disturbance source $i$, at the location of the planet, within a planet PSF core sized region:

- The (complex) mean field, $M_i$
- The field temporal variance, $V_i$
- The change in the mean field, $\Delta M_i$, across the differential imaging time
- The change in the field temporal variance, $\Delta V_i$, across the differential imaging time

In the context of an ADI roll scenario, the differential imaging time is from the mid-point of the integration at $+22°$ to the mid-point of the integration at $-22°$, a few hours. The differential image speckle noise is computed from an ensemble variance given by

$$\sigma_{\Delta I}^2 = \sum_{i=0}^{n} \sigma_{\Delta V_i}^2 + 2|M_{tot}|^2 \sigma_{\Delta M_i}^2, \tag{3}$$

where $\sigma^2_{\Delta I}$ is the differential image ensemble variance for the region of interest.

As described in Ref. 6, the calculation also involves computing the sensitivity of the complex field to each disturbance source. Then, the system is perturbed according to the disturbance in question. For example, for ACS error, the telescope pointing is changed slightly. This disturbance we can label $\Delta x_i$. Then, the resulting mean field change is recorded. The sensitivity is then given by

$$S_i = \frac{1}{I_{pk}} \cdot \left| \frac{\Delta E_i}{\Delta x_i} \right|^2 , \tag{4}$$

where $I_{pk}$ is a normalization factor and corresponds to the peak field intensity when the source (the star) is off-axis and centered at the location of the planet. The integrated model provides the disturbance statistics $\Delta x_i$ for a given observing scenario. A static optical model of the system, such as one based on PROPER,[8] provides the sensitivity.

Table 1 lists the current sources of contrast stability; note that the term "residual" refers to quantities after improvement by any real-time wavefront sensing and control. Table 2 below summarizes the errors allocated to IM predicted performance.

**Table 1** Contrast Stability Sources.

| |
|---|
| Residual full-aperture Zernike disturbances (non-jitter, i.e. drift) |
| Changes in tip-tilt / segment / WFE jitter |
| Residual segment disturbances |
| LOWFSC (low order wavefront sensing and control system) sensing noise |
| LOWFSC actuation errors |
| LOWFSC camera stability |
| Pointing repeatability |
| Deformable mirror (DM) thermal instability |
| Deformable mirror creep (if using certain deformable mirror architectures) |
| Pupil shear |
| OTE beamwalk |
| Coronagraph-internal beamwalk, on the DM |

**Table 2** Errors Allocated to IM Predicted Performance.

| Performance Metric Name | Performance Goals (preliminary) | Units |
|---|---|---|
| Mean | 2.0 | pm |
| Variance | 4.0 | pm$^2$ |
| Delta Mean | 0.2 | pm |
| Delta Variance | 3.2 | pm$^2$ |
| Pointing Stability: ACS | 4.0 | mas RMS |
| Pointing Stability: ACS + FSM | 0.1 | mas RMS |
| Line of Sight (LOS) Jitter | 0.1 | mas RMS |
| WFE Jitter | 1.0 | pm RMS |

## 2 EAC 1-3 Design Formulation

### 2.1 Overall Design Approach

The SET conceived the EACs as "strawman" conceptual designs to enable trade space exploration and develop integrated modeling methods for performance evaluation. The EAC concepts are not requirements-driven formulations but rather generated to address the most demanding science objectives of the mission, such as yield of terrestrial exoplanet discoveries, contrast ratio, and contrast ratio stability, while exploring the most challenging engineering parameters of the trade space, such as telescope aperture and size, instrument volume, wavebands, and detector temperatures. It is generally agreed that contrast stability will be the most demanding aspect of the mission, which is driven by mechanical, thermal and pointing stability aided by closed-loop control; therefore, the EAC concepts were formulated primarily to support evaluation of these aspects. The following specific objectives for this first round of EACs provide additional context for design formulation:

- Perform observatory "sizing" to estimate basic resource needs such as mass and mass properties; power; and volume (stowed and deployed)
- Gain information on the relations between design and performance, particularly for contrast stability
- Perform design reconnaissance to uncover technical problems, particularly those involving the complex opto-mechanical control system interfaces
- Practice integrated modeling to uncover necessary process improvements, including those related to file transfers, data interfaces, and organizational interfaces

The TMPO provided additional, specific architecture configuration guidance that influenced the EAC designs:

- A common feature to all the EACs is a deployed baffle/barrel around the telescope to protect the optics from micrometeoroid impacts, informed by JWST lessons learned. The first set of EACs omits a large flat sunshade (JWST heritage and LUVOIR[9] design) with the assumption that a properly designed baffle could provide both meteoroid protection and thermal conditions conducive to the required optical stability. This assumption will be tested through the EAC1 IM cycle.

- The effort should consider current and planned future launcher possibilities, including the NASA Space Launch System (SLS), SpaceX Starship, and Blue Origin New Glenn 7x2 and 9x4. The design must be compatible with at least two launchers to maintain flexibility. The highly successful HST clearly demonstrated the benefits of serviceability, not considered practical for JWST in L2 orbit. Given rapid advances in this area for other NASA, defense, and commercial missions and the obvious advantages for mission longevity, initial EAC studies will consider options for the replacement and installation of science instruments as well as refueling and avionics replacement. However, servicing will not be allowed to drive the EAC concepts; instead, potential problem areas for servicing concepts will be identified for future work.

- The primary mirror and secondary mirror will be Ultra-Low Expansion (ULE) Glass or a similar material, with a composite backplane structure operating at room temperature to significantly simplify telescope integration and testing, similar to ATLAST.[10]

- Fine Steering Mirrors (FSM) will be part of the science instruments, rather than the telescope assembly; this architecture will therefore not penalize the throughput of UV instruments that do not need the pointing accuracy afforded by an FSM.

- The Observatory will accommodate a Field of Regard (FOR): ±45° pitch and ±22.5° roll. Refence Sec. 2.5.2 and Fig. 29 for additional FOR detail.

- The design effort for the first three EACs was primarily schedule-driven, with only 3-4 months allocated to mature each design from its initial definition. The primary goal was not to complete a design that closes all requirements but rather to understand design challenges, identify design issues, and find technology gaps.

Figure 3 illustrates initial concepts, configuration parameters, and an optical view for the first three EACs.

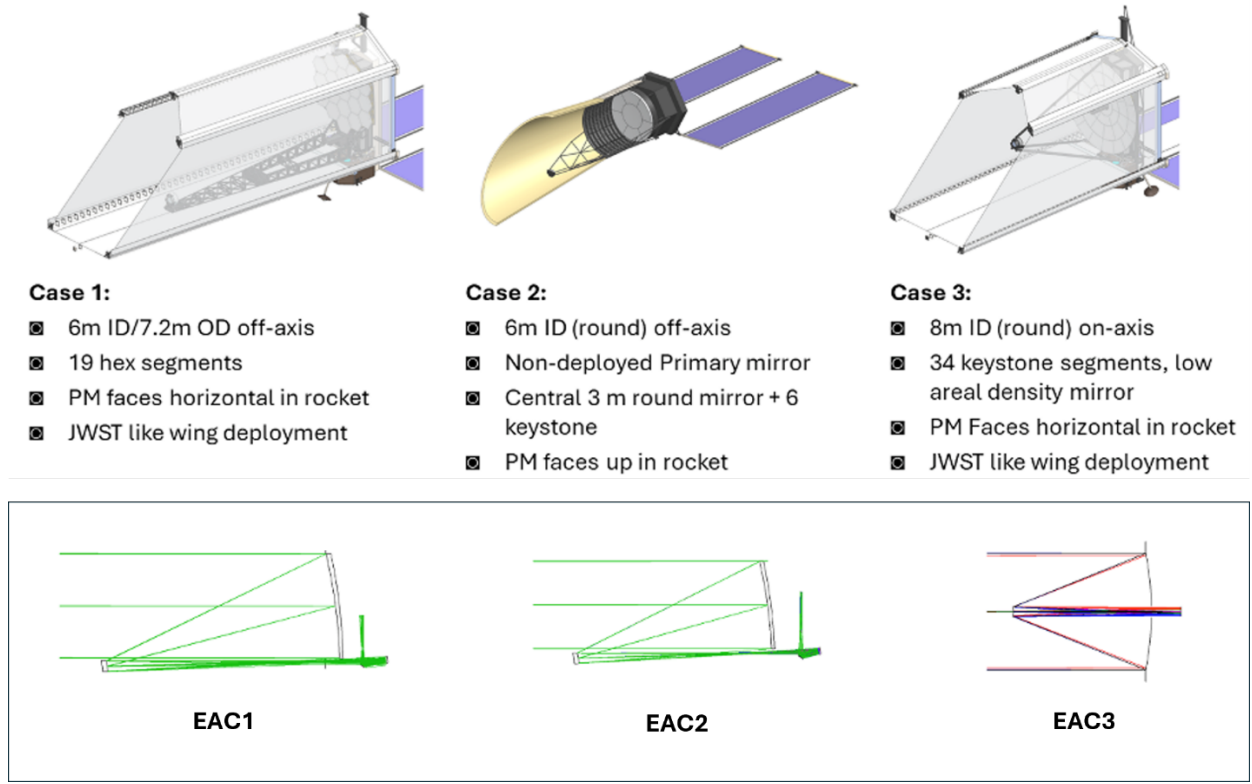

**Case 1:**
- ⊠ 6m ID/7.2m OD off-axis
- ⊠ 19 hex segments
- ⊠ PM faces horizontal in rocket
- ⊠ JWST like wing deployment

**Case 2:**
- ⊠ 6m ID (round) off-axis
- ⊠ Non-deployed Primary mirror
- ⊠ Central 3 m round mirror + 6 keystone
- ⊠ PM faces up in rocket

**Case 3:**
- ⊠ 8m ID (round) on-axis
- ⊠ 34 keystone segments, low areal density mirror
- ⊠ PM Faces horizontal in rocket
- ⊠ JWST like wing deployment

EAC1          EAC2          EAC3

**Fig. 3** HWO Exploratory Analytic Cases; Optical Layouts all to same Scale.

The telescope serves as the starting point and foundation for each EAC design iteration. These three telescopes were highly constrained by four driving optical specifications: 1) aperture or entrance pupil diameter (EPD), 2) whether obstruction was allowed, 3) the maximum angle of incidence (AOI) allowable on the front optics, and 4) the field of view needed to accommodate the science instruments. The EPD is driven by science needs, and it fundamentally sets the scale for the entire observatory. The choice of allowing obstruction effectively doubles the length between primary (PM or M1) and secondary (SM or M2) mirrors, and the maximum allowable Angle of Incidence (AOI) of light on the telescope optics also factors into this distance by establishing a minimum value. Finally, the fields of view drive the size of the aft optics of the telescope, as well as the package volumes of the instruments themselves. The five separated fields of view (FOV) for each of the science instruments (SIs) are listed in Table 3. At the top-level, EAC1 features an off-axis telescope, where the secondary mirror does not obscure the aperture, with a 6-meter inscribed diameter and a 7.2-meter circumscribed diameter. The aperture uses hexagonal segments with foldable wings for stowage inside the launch vehicle fairing. EAC2 has a 6-meter round aperture using keystone segments, no foldable wings, and is also off-axis. EAC3 is on-axis with an 8-meter outer diameter using keystone segments. Ultimately, a minimum of four powered mirrors were prescribed for each telescope concept to meet the above constraints, which continues the increasing trend for NASA's flagship mission telescopes: from HST (2 mirrors), to JWST and Roman (3 mirrors), and now to HWO (4 mirrors).

**Table 3** Science Instrument Field of View.

| Instrument | FOV Area | Units |
|---|---|---|
| Coronagraph Instrument (CI) | 1.4 | arcsec^2 |
| Ultraviolet Instrument (UVI) | 4 | arcmin^2 |
| High-Resolution Imager (HRI) | 6 | arcmin^2 |
| eXtra Instrument (XI) | 1 | arcmin^2 |
| Guider Instrument (GI) | 9 | arcmin^2 |

Figure 4 demonstrates the wide range of trade options considered by the first three EACs, spanning science, mission system, observatory, payload, and spacecraft, with some trades continuing to future EAC iterations.

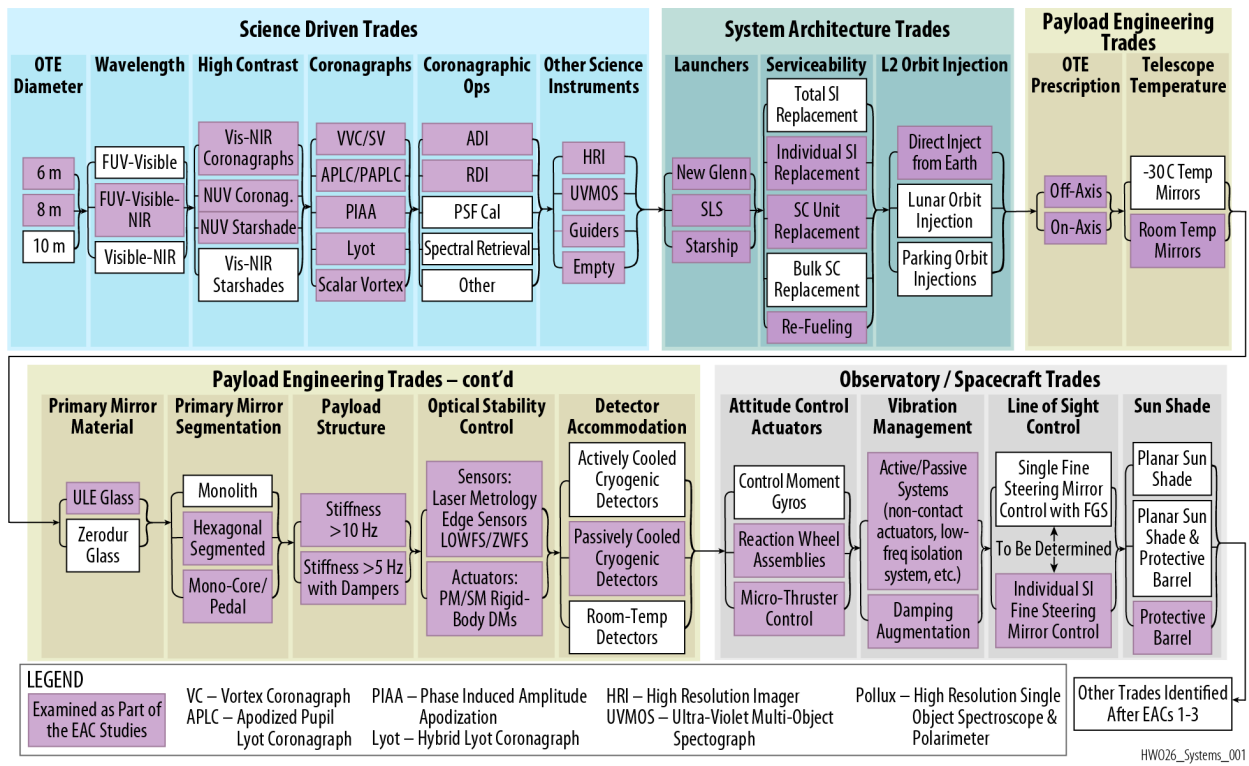

**Fig. 4** Trade Options Considered by the First Three EACs.

## *2.2 EAC1 Design Formulation*

EAC1 is a 6 m inscribed and 7.2 m circumscribed diameter class UV-Optical-Near Infrared observatory that operates at the Sun-Earth 2nd Lagrange (L2) Point. Figure 5 provides an overview of the EAC1 observatory architecture and major components; while design implementation may differ, components shown are common across all EACs. Figure 6, following, shows the observatory coordinate systems, as shown in the EAC1 stowed configuration; note V3 is aligned with boresight, the orientation of the M1 vertex coordinate system may differ between EACs.

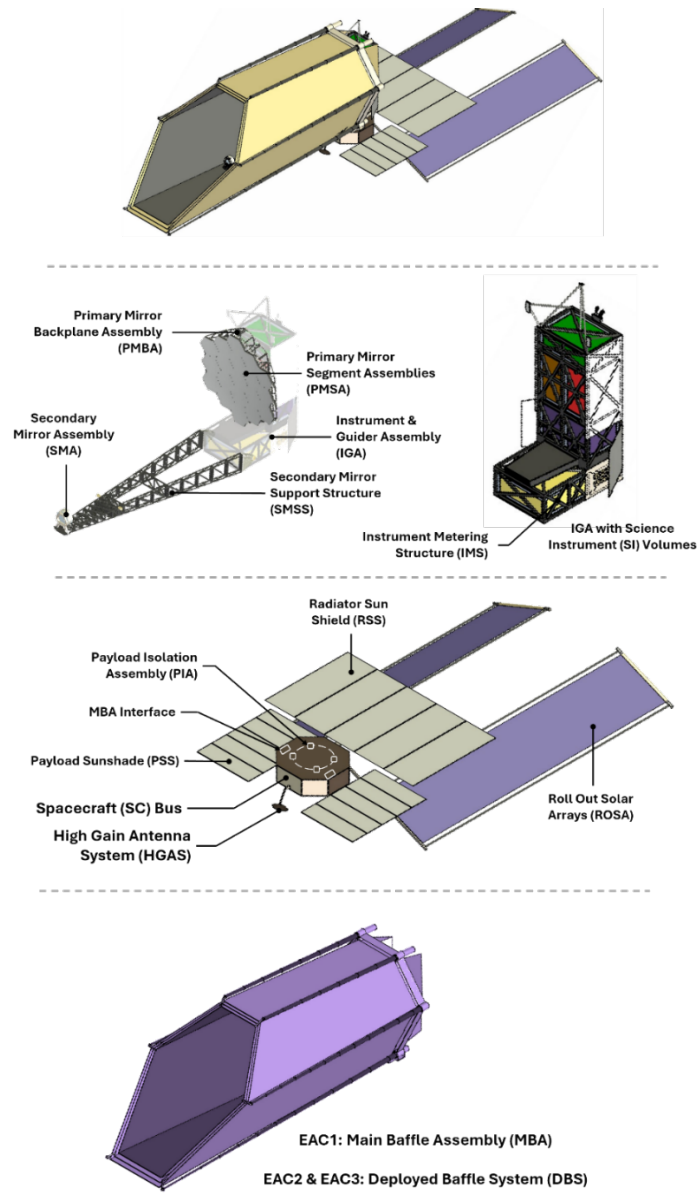

**Fig. 5** HWO Physical Configuration Overview (EAC1 shown).

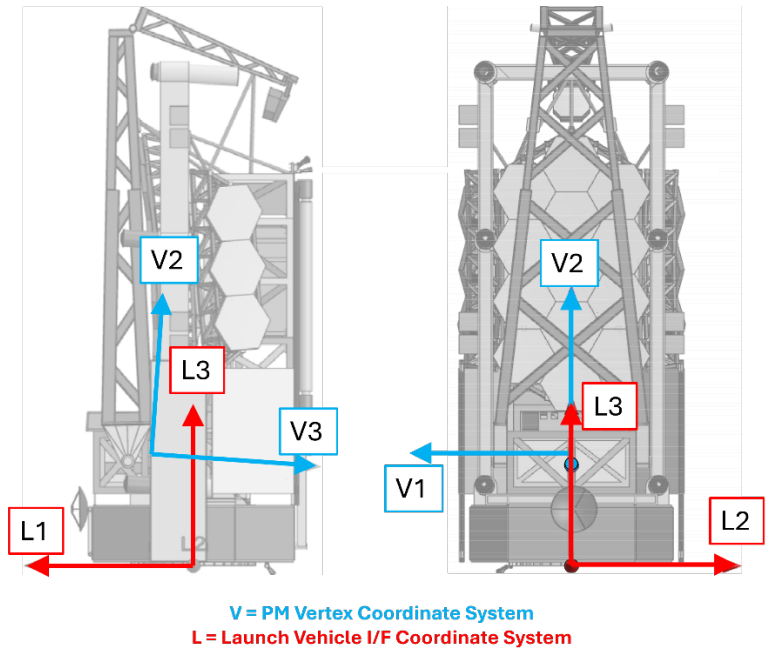

**Fig. 6** Observatory Coordinate System, EAC1.

### 2.2.1 EAC1 Optical Configuration

The EAC1 Optical Telescope Element (OTE) design was driven by the relatively wide Field-of-View (FOV) necessary to accommodate the four instruments plus two guiders, and the desire to have an unobstructed entrance pupil for the CI while maintaining a maximum angle of incidence on the primary and secondary mirrors.

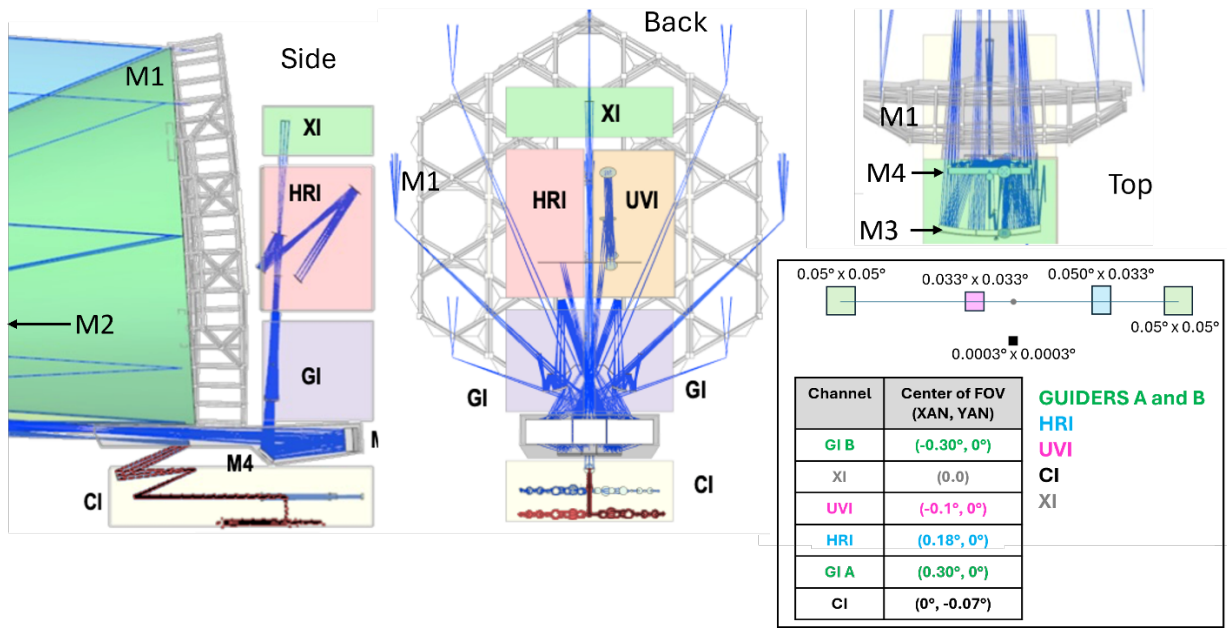

**Fig. 7** EAC1 Optical Design and FOV Allocations for the Science Instruments.

Figure 7 illustrates the light paths through the OTE and the Science Instruments (SI). The OTE is a four-mirror prescription with the Primary Mirror (PM) designated M1, the Secondary Mirror (SM) M2, and M3 and M4 are relay optics in the aft section. The telescope feeds light to the CI directly from the SM, in the plane of symmetry of the parent PM conic. Light is fed to the other SIs via M3 and M4 relays. These relay mirrors are freeform optics defined as standard Zernike aspheres to provide the best flat-field performance to UltraViolet Instrument (UVI) Micro Shutter Array (MSA).

The PM aperture is illustrated in Fig. 8, which shows the projected geometry in the entrance pupil plane. The PM consists of 19 hexagonal ULE glass segments with an inscribed diameter of 6 m and an outer circumscribed diameter of 7.2 m; note, ULE provides several manufacturing advantages and benefits from mature, proven design heritage, see Ref. 1 for additional information. The center of the OTE entrance pupil is 4.8 m from the axis of symmetry of its parent conic.

Each hex segment has a 1.7 m point-to-point diameter, well within the 1.8 m maximum boule size for ULE manufacturing, yielding a 33.6 m$^2$ overall light gathering area. The segment-to-segment gaps are sized at 6 mm to facilitate segment edge sensors.

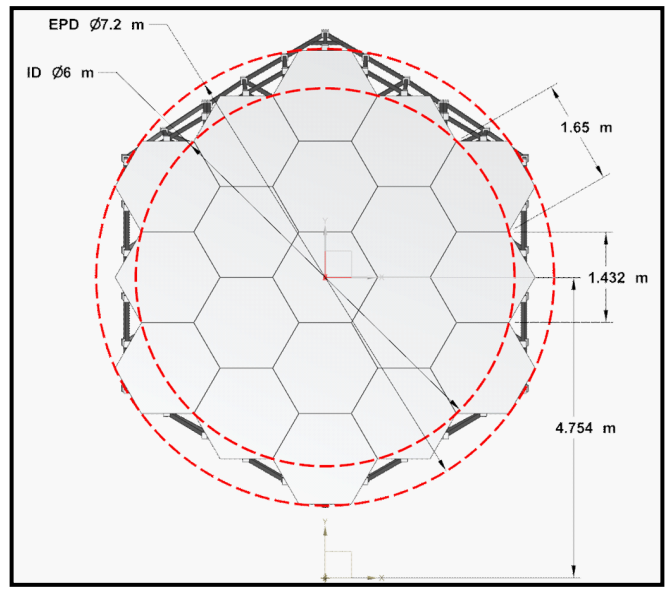

**Fig. 8** EAC1 Primary Mirror Geometry.

### 2.2.2 EAC1 Science Instruments

Science instrument architecture focuses on significantly maturing the CI design while leveraging LUVIOR design heritage to implement and package designs for the astrophysical instruments.

This sections provides top level summaries for each instrument, with detailed descriptions for the CI provided in Ref. 11.

The optical design of the CI consists of four channels. These channels receive collimated light from its own three-mirror collimator that picks off light from the Cassegrain focus of the telescope after M2. Two of the CI channels are separate polarization channels for the visible wavebands and two are separate polarization channels for the Near InfraRed (NIR) channels. Each channel incorporates its own Deformable Mirrors (DM) and Wavefront Sensor. The two DMs are spaced 1.8 m apart and are each a 96x96 actuator array with 1 mm pitch. Figure 9 illustrates the initial optical layout of the CI concept.

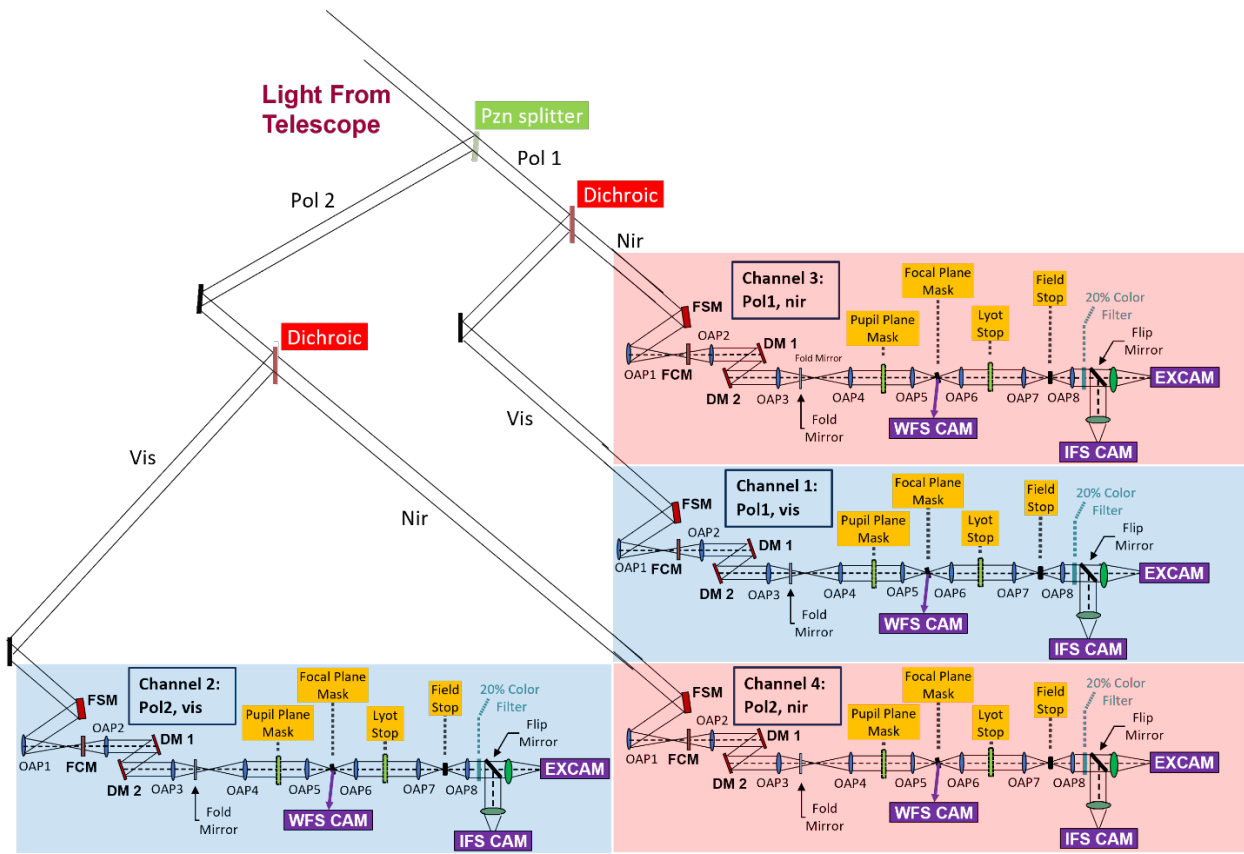

**Fig. 9** The Initial Four Channel CI Concept.

Figure 10 illustrates the Ultra Violet Instrument (UVI) optical layout. The UVI has the following three optical channels: 1) Far UV (FUV) Multi-Object Spectrometer (MOS), 2) Near UV (NUV) MOS, and 3) an FUV Imager. Channels 1 and 2 share a common Micro Shutter Array (MSA).

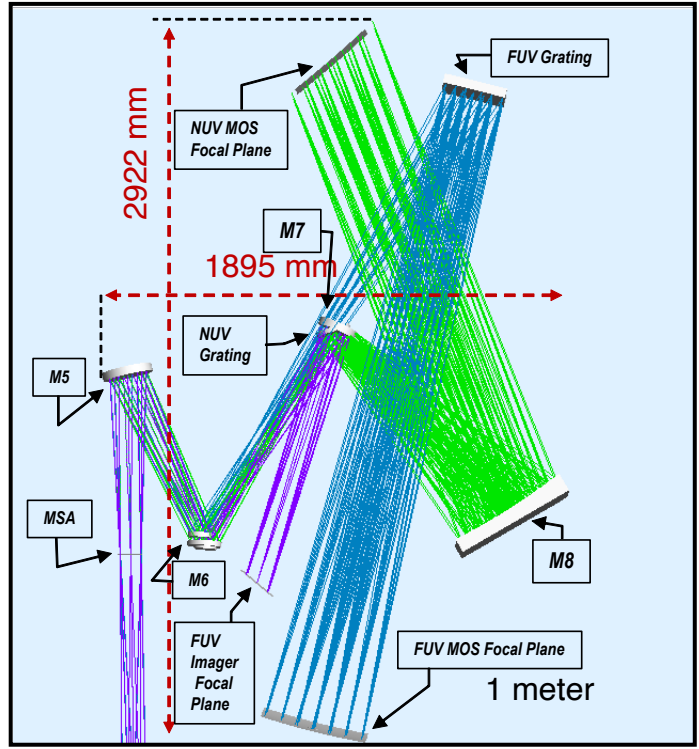

**Fig. 10** Optical Layout of the EAC1 UVI.

The High Resolution Imager (HRI) provides UV / Visible and NIR imagery over a 3 X 2 arcmin FOV; Fig. 11 illustrates its optical layout. The light coming from the OTE interface is collimated by a four-mirror collimator that forms an accessible pupil image. At this pupil image, in both the UVIS or NIR channel, there are filter wheels. After passing through the filter wheels the light is focused by a four-mirror imager.

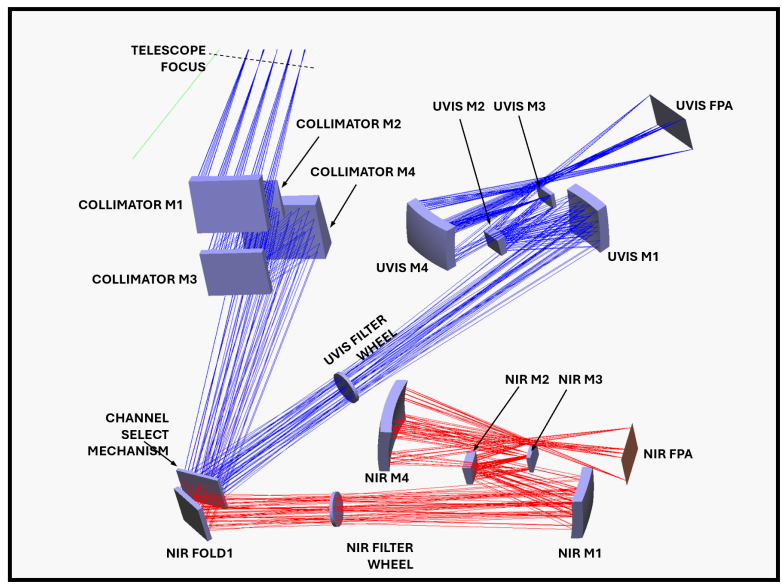

**Fig. 11** Optical Layout of the EAC1 HRI.

Finally, the two Guiding Instruments (GI-A and GI-B), provide the fine pointing signals to the observatory pointing control system. These three-mirror imagers operate in the waveband from 400 nm to 1000 nm over a 3 x 3 arcmin FOV. Figure 12 illustrates their optical layout.

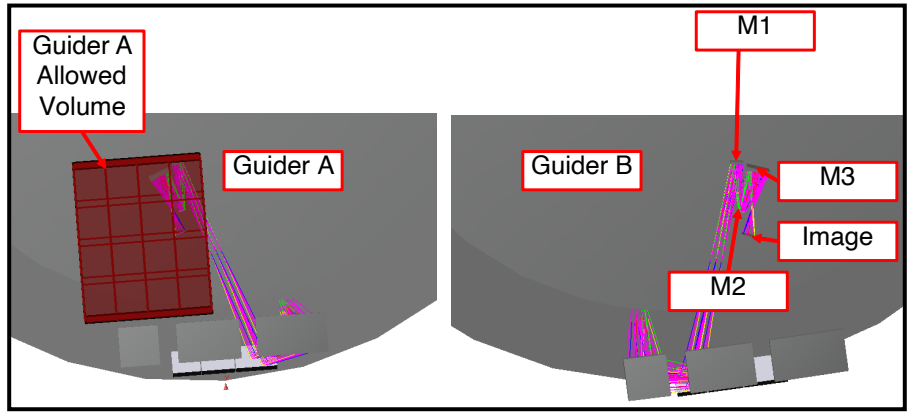

**Fig. 12** Optical Layout of the EAC1 GIs.

### 2.2.3 EAC1 Mechanical Configuration

Figure 13 and Fig. 14 illustrate the EAC1 observatory deployed and stowed configurations, respectively.

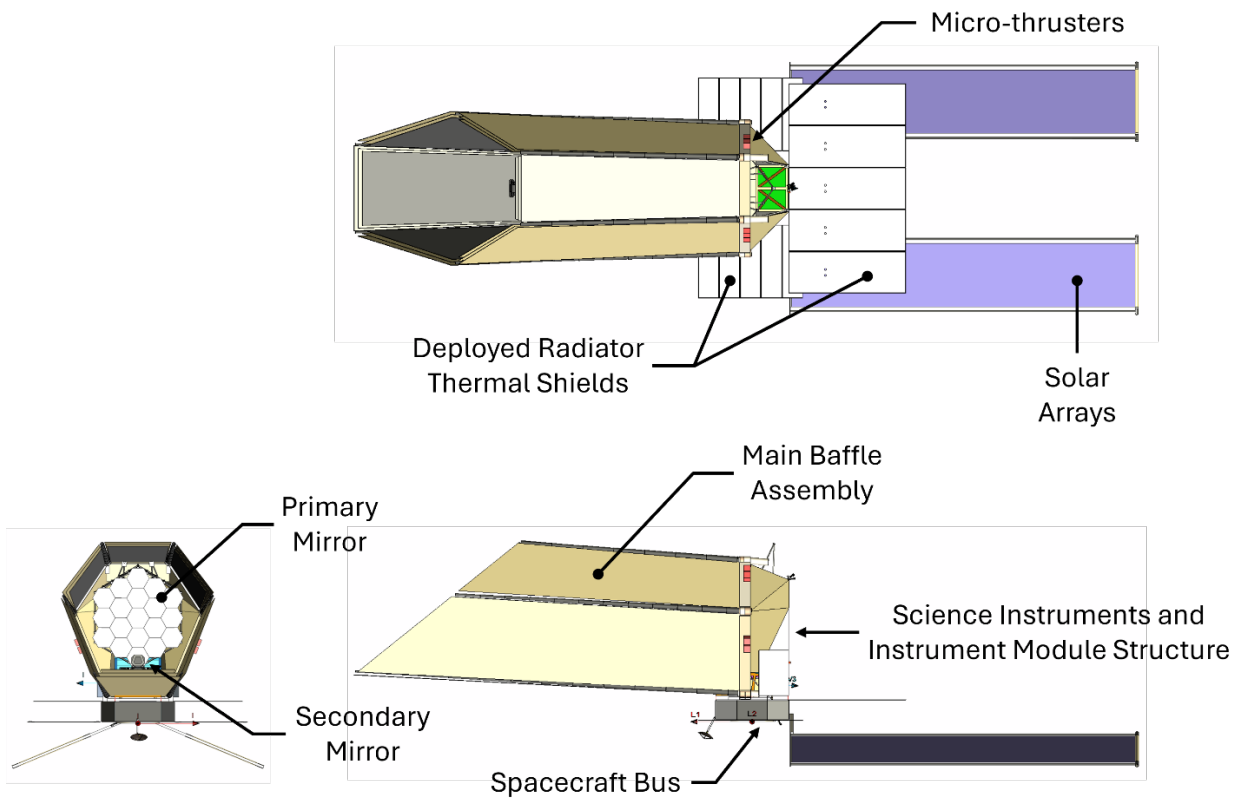

**Fig. 13** EAC1 Deployed Configuration.

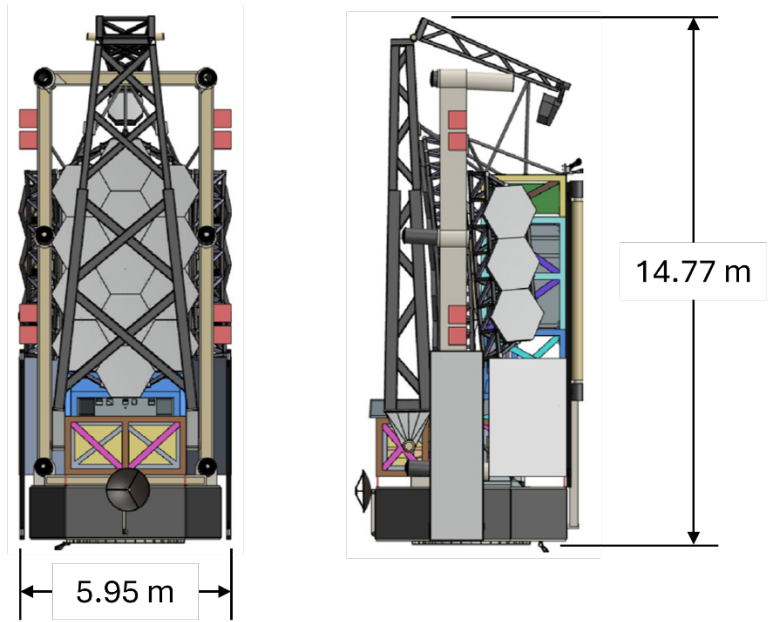

**Fig. 14** EAC1 Stowed Configuration.

Two primary challenges drive EAC1 mechanical design: 1) the need to fit within the New Glenn Launch Vehicle (7 m diameter) and 2) the need to provide sufficient design detail to enable adequate discipline level modeling to support the first round Integrated Modeling.

To facilitate on-orbit replacement, the design mounts the SIs to the IGA structure with a series of rails and latches in a semi-kinematic approach, similar to HST; accessibility to the SIs is from the -L1 direction of the observatory. Although it was initially desired to have the individual radiators attached to their SIs and inserted with them as an integrated unit, structural stiffness considerations for launch drove the need for an individual Instrument and Guider Assembly (IGA) Aft Door for each instrument with the corresponding radiators mounted to it. Note, the observatory has deployable radiator shields and canted solar arrays to reduce thermal backloads on the SI radiators; these radiators as well as the "strong-back" support and stowage compartments for the Baffle are structurally mounted to the spacecraft bus.

The major observatory deployments, as shown in Fig. 15, are carried out in five major steps: 1) Solar array and High Gain Antenna (HGA), 2) Radiator and payload shields, 3) Secondary Mirror Support Structure is deployed by rotations about two hinge lines, 4) the six containment boxes of the Main Baffle Assembly (MBA), followed by the two PM wings about their hinge lines, and finally 5) MBA membranes are pulled out by six telescoping booms driven by Storable Tubular Extendible Member (STEM) mechanisms, in a synchronous manner.

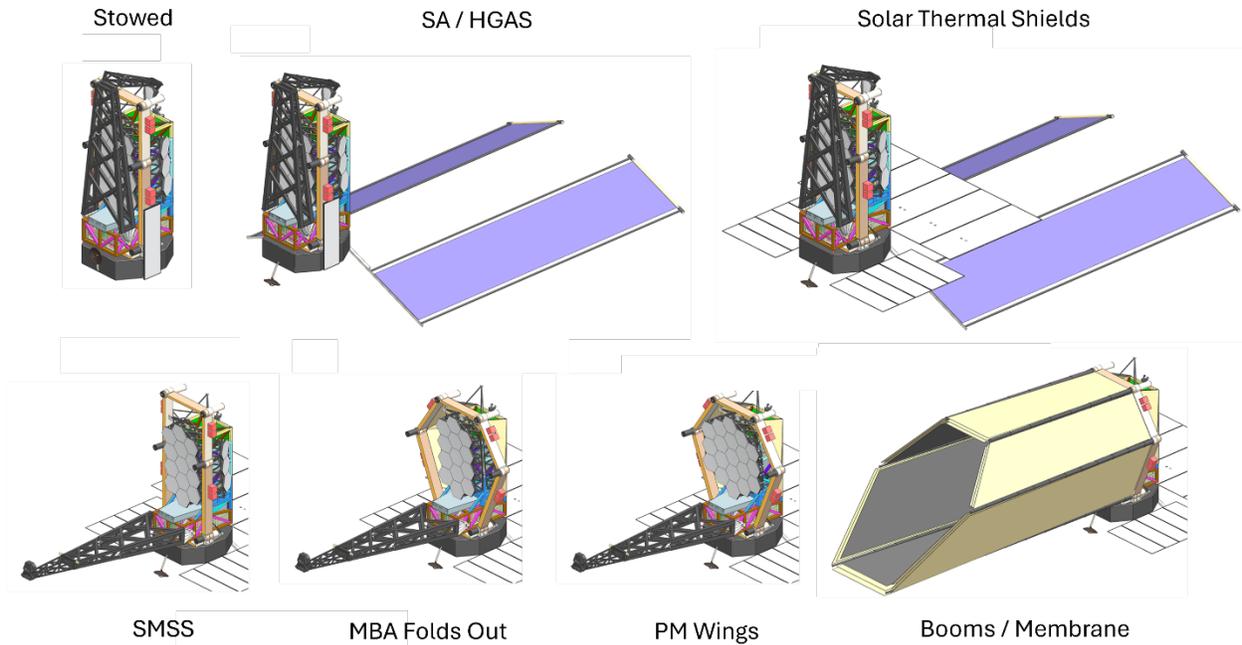

**Stowed**　　　**SA / HGAS**　　　**Solar Thermal Shields**

**SMSS**　　**MBA Folds Out**　　**PM Wings**　　**Booms / Membrane**

**Fig. 15** EAC1 Deployment Sequence.

### 2.2.4  EAC1 Structural Configuration

Structural modeling and analysis supported EAC1 mechanical design with early deployed frequency estimations, with emphasis on structure sizing. To guide this exercise, the team developed a set of deployed natural frequency goals, influenced by past experience, specifically from JWST given similar hexagonally segmented PM and overall architecture (reference Table 4). Regardless of whether final EAC1 designs meet these goals, they enabled the team to contrast different design concepts and focus optimization efforts. Stowed structural modeling and analysis, however, was not as extensive, with the exception of expected first launch condition modes comparison against launch vehicle requirements and a simple check of stress levels in the Spacecraft Bus.

For reference, Fig. 16 below shows the EAC1 deployed Finite Element Model (FEM). The model includes isolation systems between the payload and spacecraft as well as at the reaction wheels to attenuate disturbance sources by providing a "low pass filter" effect. With attenuation beginning at frequencies beyond 1.4 times the isolation frequency, it becomes advantageous to design the telescope side to be as stiff as possible; this concept guided much of the structural analysis focus for the design effort.

The 15-meter long SMSS was recognized immediately as likely the lowest natural frequency subsystem on the telescope. The truss type design was chosen due to the expected uniform coefficient of thermal expansion (CTE) in truss members along with an aggressively stiff truss member stiffness achieved through ply layup design of the composite beam elements; Fig. 17 below shows this truss structure. Optimization runs were performed early to achieve a first cantilevered mode at 14 Hz. Later work was done to optimize the SMSS in the actual operational configuration (deployed in the telescope model), achieving a final frequency prediction of 10.7 Hz with an SMSS mass just over 2,000 kg.

**Table 4** EAC1 Natural Frequency Goals.

| Component | Frequency Goal, Low (Hz) | Frequency Goal, High (Hz) | Lowest Mode (Hz) | Boundary Conditions / Comments |
|---|---|---|---|---|
| SMSS | 10 | 20 | 14 | Fixed Base (Observatory level at 11 Hz) |
| Mirror Segment Fixed Base, Tip-Tilt | 100 | 120 | 72 | Fixed Base |
| Mirror Segment on Backplane, Tip-Tilt | 75 | 90 | 50 | Observatory Level |
| Mirror Segment, First Flexible Mode | 200 | 400 | 364 | Free-Free Mirror |
| PMBA First Flexible Mode | 50 | 100 | 40 | This is Tuned Wing Flap Mode, Observatory Level |
| Telescope Side First Mode (excluding SMSS) | 30 | 40 | 20 | Observatory Level |
| SI First Mode on Latches, Mounted to IMS | 35 | | | HWO Free-Free |
|     Coronagraph Instrument | | | 16 | Flex Mode on Axial Latches |
|     AOS - Aft | 35 | | 70 | Fixed Bipods at IGA I/F |
|     AOS – Forward | 35 | | 71 | Fixed Bipods at IGA I/F |
|     Guiders (2) | 35 | | 39 | Flex Mode on Axial Latches |
|     HRI | 35 | | 32 | Flex Mode on Axial Latches |
|     UVI | 35 | | 29 | Flex Mode on Axial Latches |
|     XI | 35 | | 47 | Rigid XI on Radial Latches |
| Isolation Systems | | | | |
|     Spacecraft Isolator (Modes 1-6) | 0.5 | 1.0 | 0.51 | With lumped SC and OTE mass |
|     Reaction Wheel Isolator (Modes 1-6) | 2.5 | 6.5 | 3.1 | Fixed base with RW and housing mass included |
| Spacecraft | | | | |
|     MBA Primary Structure First Mode | 0.02 | 0.04 | 0.046 | Cantilevered on the SC |
|     Solar Arrays | 0.05 | 0.10 | 0.063 | Cantilevered on the SC |
|     Deployed Sunshade (Large) | 0.10 | 0.15 | 0.116 | Cantilevered on the SC |
|     Deployed Sunshade (Small) | 0.15 | 0.45 | 0.166 | Cantilevered on the SC |

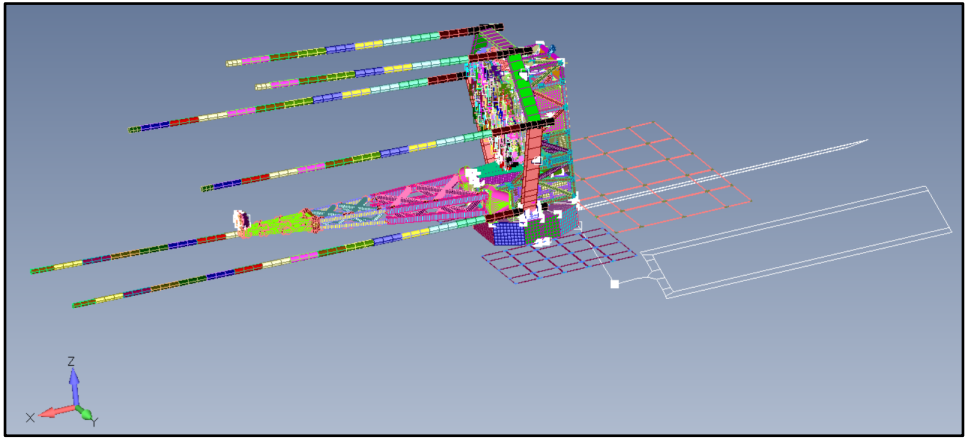

**Fig. 16** Deployed FEM, EAC1.

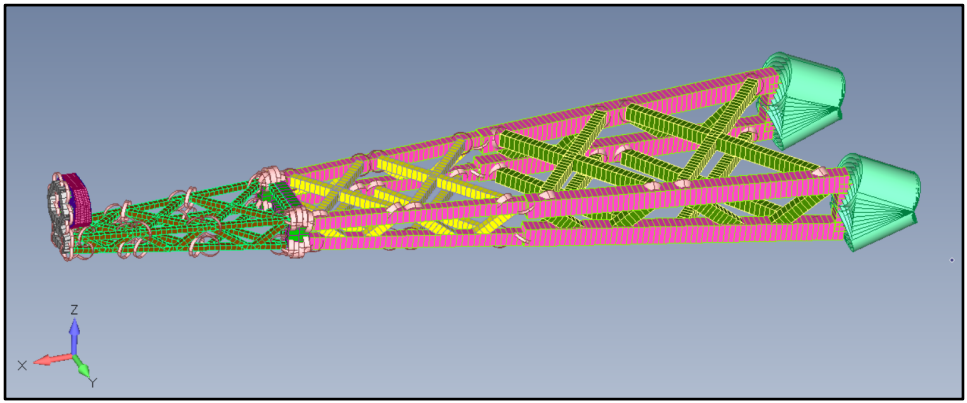

**Fig. 17** SMSS Truss Design FEM, EAC1.

On the spacecraft side, the Main Baffle Assembly (MBA) posed the most challenging design constraint due to its size, cantilevered interface to the spacecraft, and the expectation that its first natural frequency would be the lowest EAC1 resonance. Figure 18 shows the MBA FEM. Much of the model drew on JWST experience, though the membrane is not mechanically tensioned as on JWST. Note, the model does not include a geometric representation of the membrane, instead includes additional mass spread on the Deployable Boom Assembly (DBA). The MBA launch (stowed) configuration is launch locked to the IGA and has a removable support strut at the cantilevered spacecraft interface. On-orbit, these connections break and the entire MBA is cantilevered at the spacecraft top deck. The combination of a long cantilever design, a small footprint on the spacecraft deck, and a large and conservatively modeled total mass of over 4,000 kg yielded the lowest observatory resonance at 0.046 Hz. Later designs, with more favorable spacecraft interfaces, achieved first mode frequencies 10x higher.

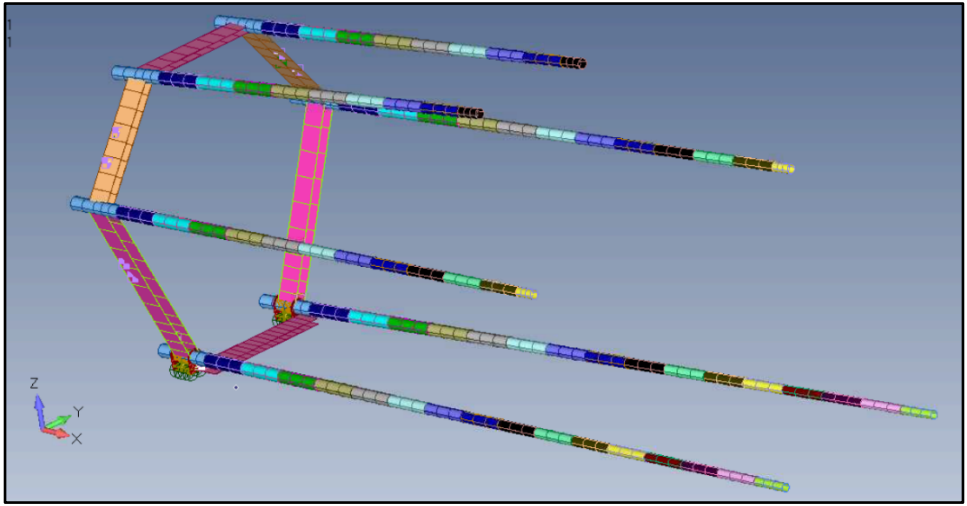

**Fig. 18** Deployed MBA FEM, EAC1.

Analysis work for the PMA proceeded in three stages: 1) a component level analysis of the structure holding the mirror segments, the PMBA, with specific focus on the hinge joint about which the PM "wings" fold, 2) analysis addressing the mirror segments themselves, the PMSAs, and 3) component level models integrated with the IGA to form a complete PMA model. The deployed/latched wings represent the lowest PMA frequency. Based on JWST experience, this was set at 40 Hz (achievable in practice) to provide separation from lower EAC1 frequencies. The PMSAs on the backplane have a 50 Hz first mode, also separated from lower telescope modes.

The final EAC1 system level deployed model provided a satisfactory starting point for subsequent thermal distortion and jitter analysis work as well as generated several lessons learned applicable to future EAC design and modeling efforts. Of note, initial designs and models did not always consider sufficient interface structure details, notably at the SMSS to IGA and PMBA to IGA connections, a shortcoming addressed in early design efforts for EAC2 and EAC3.

### 2.2.5 *EAC1 Thermal Configuration*

EAC1 Thermal design architecture followed a passive cooling philosophy centered on effective heat rejection through radiators, while minimizing thermal disturbances to the telescope; efforts concentrated on three main areas: baffle system design to provide a stable sink for the room temperature PM, analysis to determine radiator sink temperatures and estimated areas for each SI radiator, and evaluation of other observatory areas that affect thermal stability.

The environmental thermal load remains consistent across all of the EACs, with the observatory orbiting at Lagrange point 2 (L2), with solar heating hot biased (1,392 W/m²), and a field of regard (FOR) based on ±45° pitch and ±22.5° roll (reference Sec. 4.3.1.1 for additional FOR detail, in this context).

Several challenges, identified during early thermal architecture consideration, drove individual EAC thermal designs:

1) Utilization of a deployable baffle to exclusively provide telescope thermal stability and Micro-meteoroid (MMOD) protection

2) Stringent ~5 mK baffle cavity thermal stability, based on a previous study

3) Large FOR, which results in highly variable solar impingement levels and locations

4) Stringent instrument thermal operational requirements, difficult to achieve in a fully passive system especially where potential radiator surfaces have either direct solar impingement or Infra-Red (IR) backloading

Given its importance, thermal design efforts focused first on the baffle surrounding the telescope, especially considering unique constraints levied by mechanical design: MMOD protection required at least three separated layers of material and the stowed launch configuration limited options for the overall deployed shape. The final MBA thermal design included a five-layer baffle, with the layers slightly diverging axially to improve thermal performance (largest separation between the layers is at the end of the MBA). As expected, more layers and larger divergence angles produced stabler telescope environments; however, mechanical packaging and deployment challenges set an upper bound, the difference between the thermal five-layer design and the mechanical three-layer design remains a known lien on the EAC1 design.

The following lays out the iterative approach used for SI radiator sizing and evaluation:

1) Collaboration with SET to determine an initial set of instrument operational temperature needs and corresponding thermal dissipations (heat load rejection involved a series of parasitic head load calculations on SI components below room temperature; IGA assumed at room temperature to match the telescope)

2) Evaluation with the mechanical team to assess available radiator areas

3) Thermal modeling and analysis to determine baseline sink temperatures and assess design features needed to achieve sink temperatures for adequate heat rejection capability throughout the FOR

Preliminary results made clear that direct solar impingement and IR backloading from the solar arrays drive unacceptably high sink temperatures. Several follow-on studies assessed thermal shield configurations to reduce the sink temperatures; Fig. 19 below shows the final, baseline configuration with the shield elements noted.

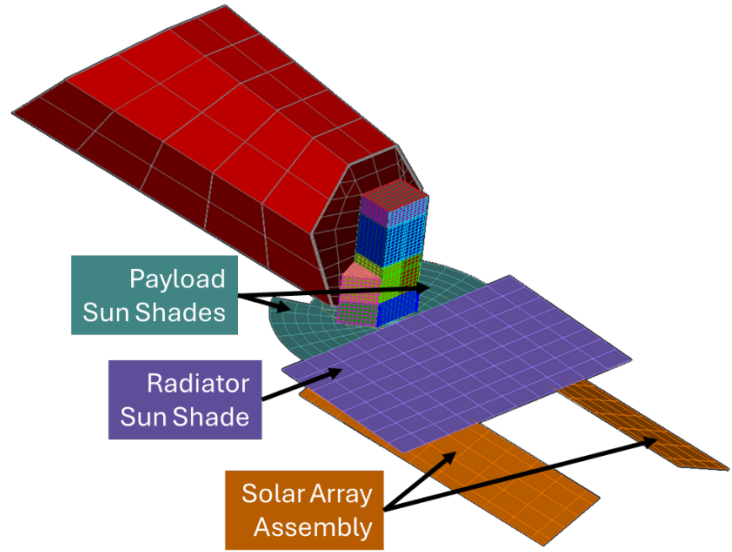

**Fig. 19** EAC1 Thermal Model Architecture.

While the UVI, HRI, and GIs achieved necessary radiator sizes within allocated areas, the CI radiator design did not close. Despite raising the original, desired detector temperature from 50 K to 65 K (CI NIR channel uses HgCdTe detectors), sink temperatures remained stubbornly high and insufficient to reject heat and parasitics through passive means. This remains a known lien on the EAC1 design and yielded a key lesson learned for EAC2 and EAC3 design efforts – the EAC1 CI location at the bottom of the instrument module, close to the shields and solar arrays, moderated clean, direct views to space. An alternative to relocation of the CI to the top of the instrument stack is to introduce an active cryocooler at the CI. Ongoing analysis is being performed to determine if the pump disturbance and corresponding jitter are acceptable.

Exiting the EAC1 system architecture effort, the team proposed several items for consideration heading into EAC2 and EAC3, including refinement of parasitic calculations and assumed coatings, optimization of solar shielding of instrument radiators, and more thorough evaluation of instrument temperature needs. Further, the payload aft closeout required further study as it represents a large area of deployable structure with MLI that impacts the telescope radiative sink temperatures within the baffle.

## 2.3 EAC2 Design Formulation

### 2.3.1 EAC2 Optical Telescope Configuration

The entrance pupil diameter (EPD) and instrument locations were the primary differences influencing the EAC2 optical design as it departed from the EAC1 design. Otherwise, this telescope is essentially a scaled version of EAC1. Figure 20 illustrates the light paths through the EAC2 OTE and the SIs.

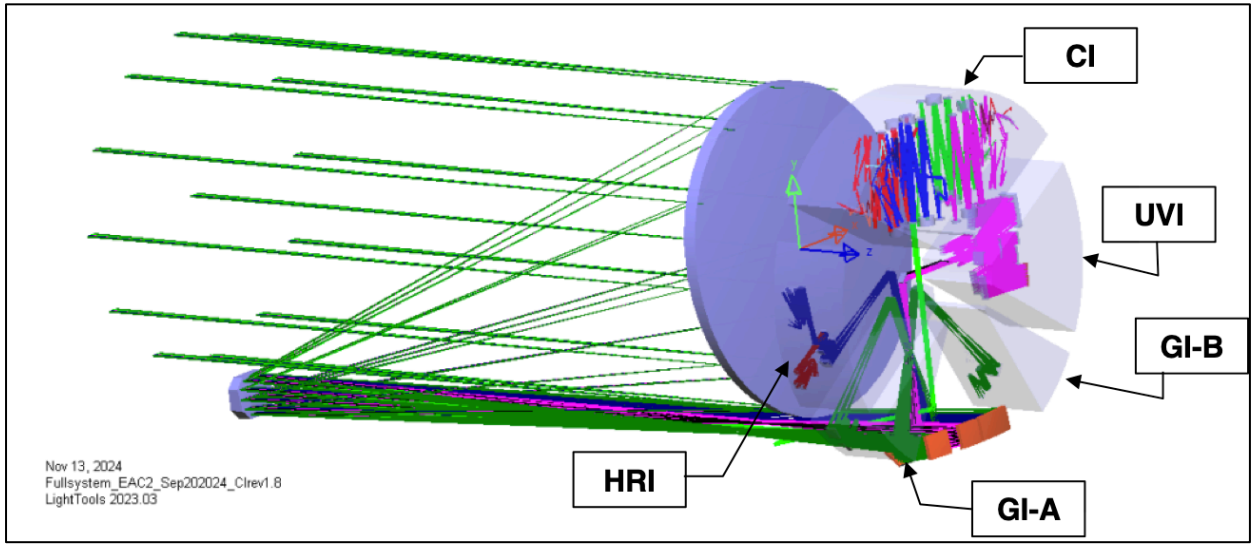

**Fig. 20** FOV Allocations for the EAC Instruments.

Figure 21 illustrates the PM geometry and shows the projected geometry in the entrance pupil plane. The 6 m diameter PM consists of six "keystone" segments surrounding a central circular 3 m diameter mirror. The segments are separated by 4 mm physical gaps. The center of the OTE entrance pupil is 4.272 m from the axis of symmetry of its parent conic.

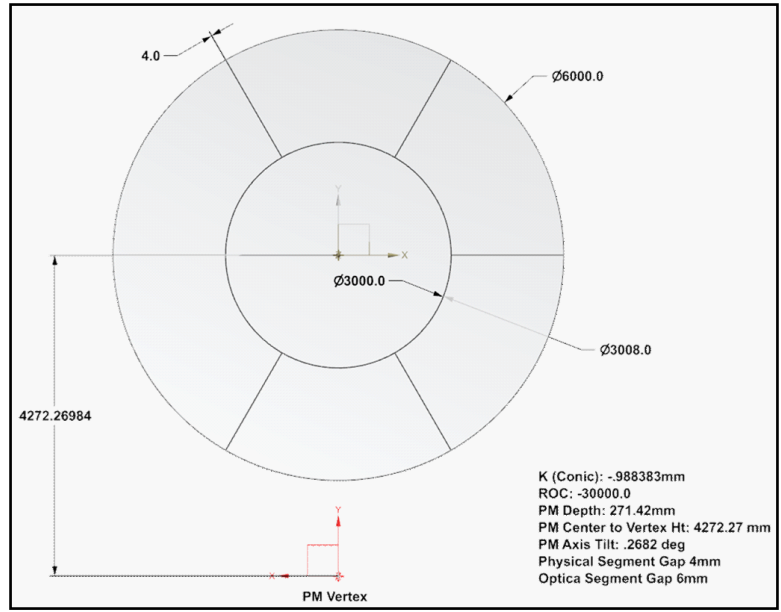

**Fig. 21** EAC2 Primary Mirror Geometry.

The EAC2 keystone segments have a maximum linear dimension of 3 m along the outer secant line. This exceeds the maximum boule size for ULE manufacture, therefore these segments would need to be formed from reflowed ULE facesheets with multiple cores, which would force a reconsideration of the ULE Coefficient of Thermal Expansion requirements.

### 2.3.2 EAC2 Science Instruments

Similar to the EAC1 CI, the optical design of the EAC2 CI consists of four channels. Two are separate polarization channels for the Visible wavebands and two are separate polarization channels for the NIR channels. Each channel incorporates its own Deformable Mirror (DM) pair, Wavefront Sensor, and Integral Field Spectrometer (IFS). The optical layout of the CI concept, illustrated in Fig. 22, shows the packaging all four CI channels into its allocated volume. Figure 23 shows the optical layout of one typical CI channel.

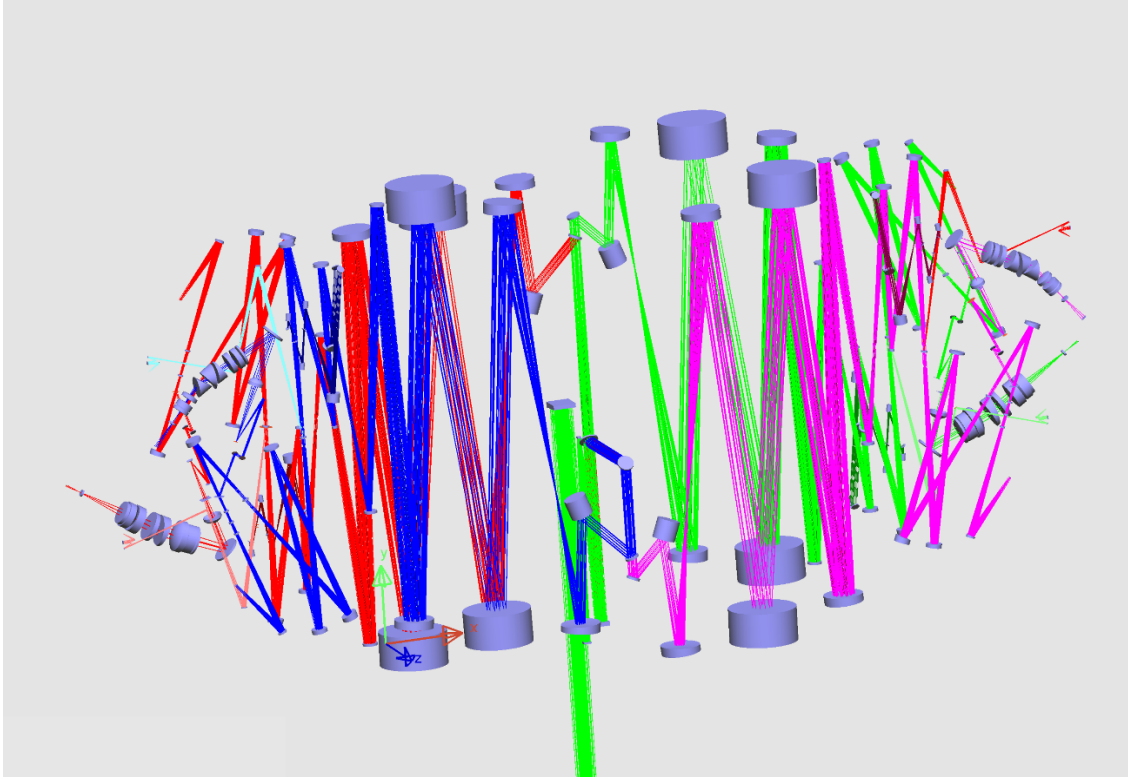

**Fig. 22** Packaging of the Four Channels for the EAC2 CI.

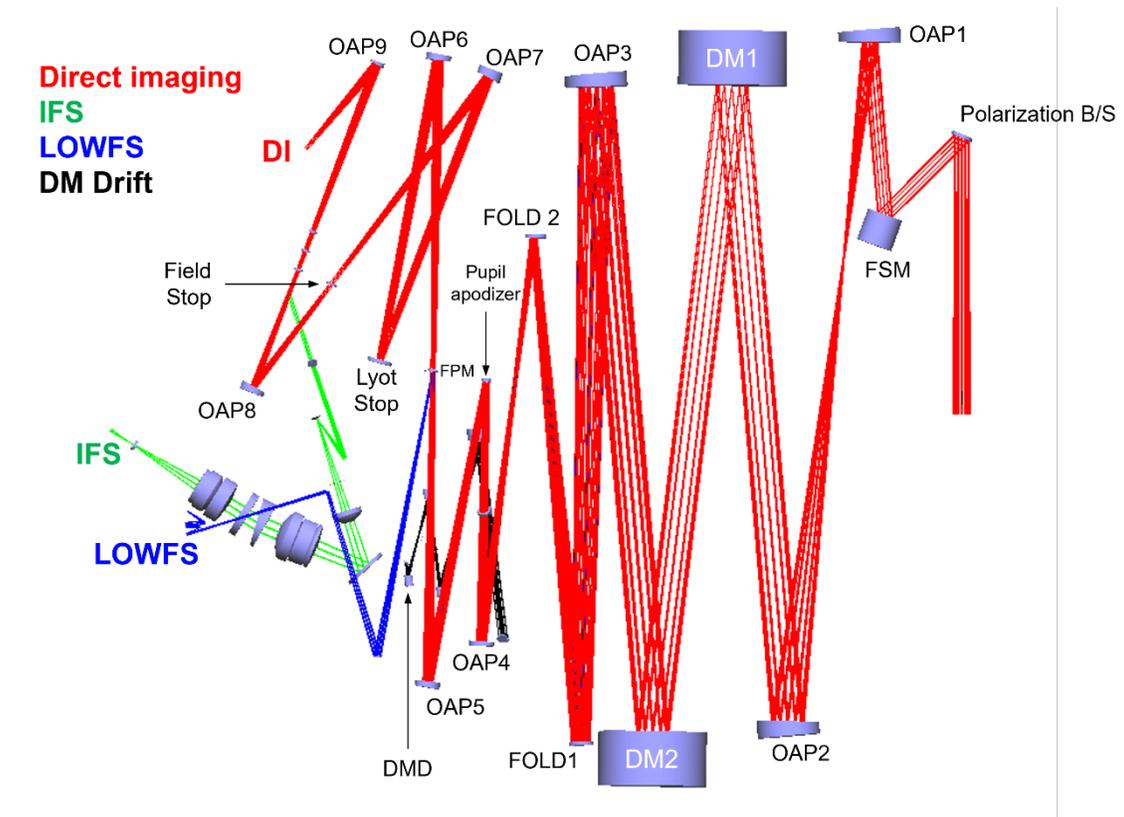

**Fig. 23** Optical Layout of a Typical CI Channel.

Figure 24 illustrates the UVI optical layout. The UVI has the following four optical channels: 1) FUV Imager, 2) FUV Multi Object Spectrometer (MOS) with short wavelength grating modes, 3) FUV MOS with medium wavelength grating modes, and 4) the NUV MOS with longer wavelength grating modes. All four channels pass through the same MSA and then into their individual optics. Due to severe packaging constraints, one of the FUV MOS grating modes was not included but will be considered for future EACs by adding a fifth channel and manipulating the location of other channels.

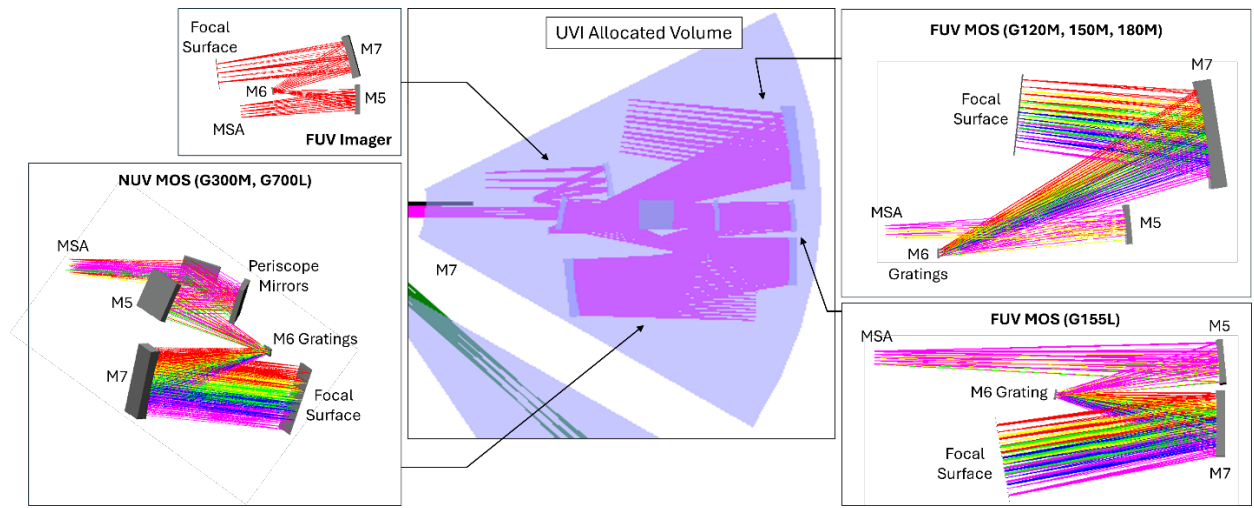

**Fig. 24** Optical Layout of the EAC2 UVI.

The HRI provides UV / Visible and NIR imagery over a 3 x 2 arcmin FOV. Figure 25 illustrates its optical layout.

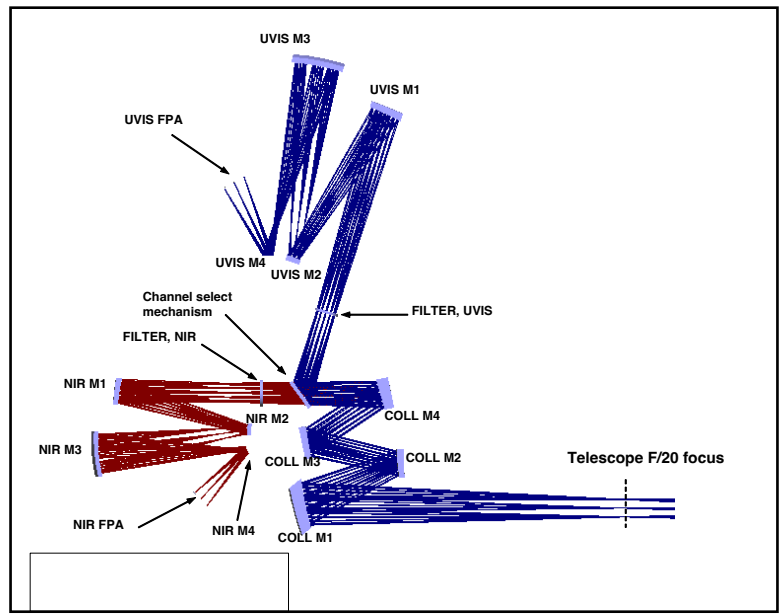

**Fig. 25** Optical Layout of the EAC2 HRI.

Finally, the two Guiding Instruments (GI-A and GI-B), provide the fine pointing signals to the observatory control system. These three-mirror imagers operate in the waveband from 400 nm to 1000 nm over a 3 x 3 arcmin FOV. Figure 26 illustrates their optical layout.

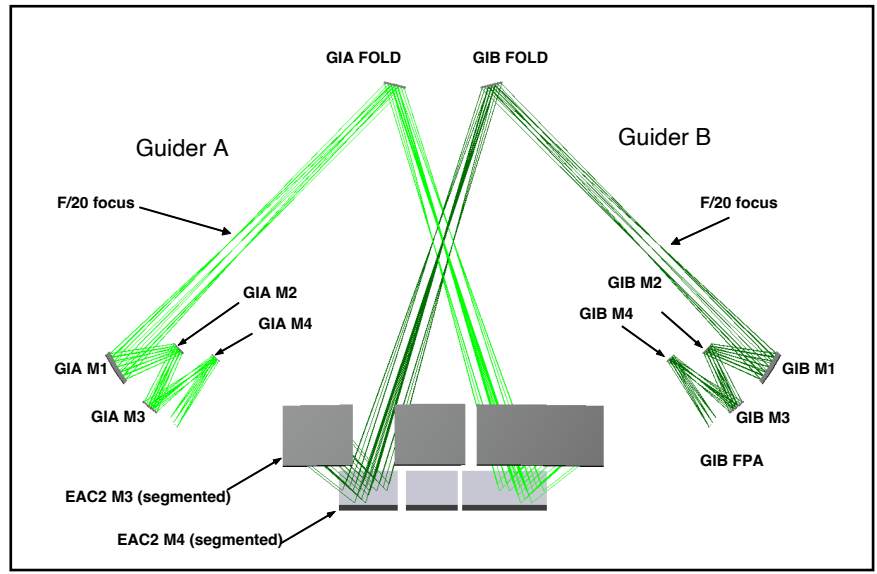

**Fig. 26** Optical Layout of the EAC2 GIs.

### 2.3.3 EAC2 Mechanical Configuration

The EAC2 design effort leveraged lessons learned from EAC1, in particular the positioning of SIs to better accommodate the cryogenic needs of some detectors, specifically the CI. Figure 27 shows the overall EAC2 configuration evolution, in part driven by the lessons learned.

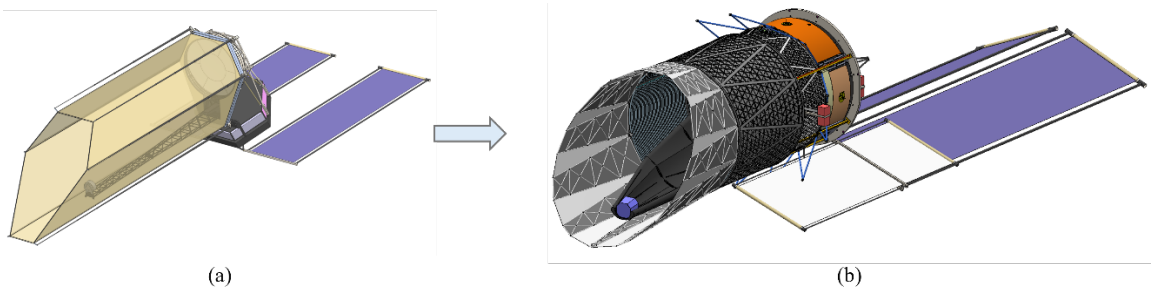

**Fig. 27** Evolution of EAC2 Concept: (a) Initial Design and (b) Final Design

This evolution included the following major changes from the initial configuration:

- Significant changes to the deployable baffle from an EAC1 like concept to a less complex sliding cylinder "camping cup" concept

- A modified SMSS to occupy the maximum available volume to optimize its deployed stiffness; stiffness of the EAC1 SMSS fell well below the goal, the EAC2 effort opted to determine expected upper limits for a relatively stiff SMSS design
- A Spacecraft Bus "in-line" with the telescope, this simplifies load paths and volumes for the SIs
- Deployable Sunshades added to shield SI radiators over the roll ranges and minimize back-loading from the solar arrays

Figure 28 and Fig. 29 illustrate the EAC2 deployed and stowed configurations, respectively.

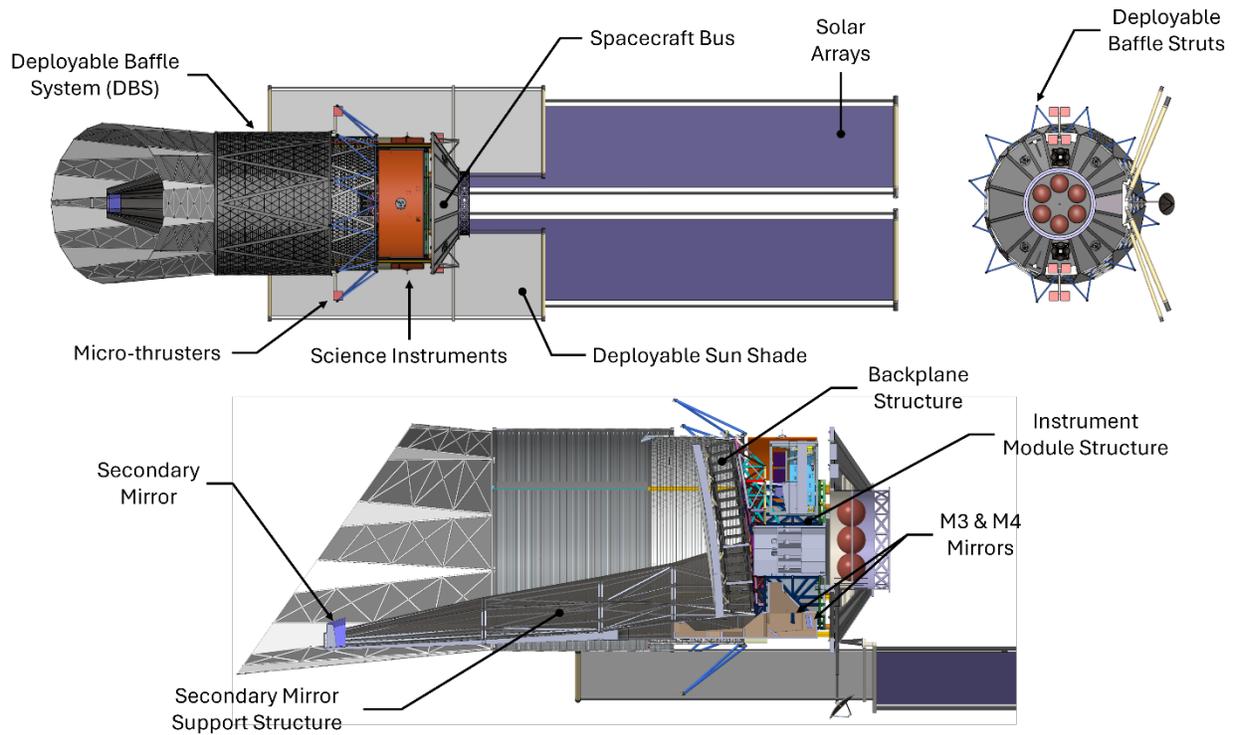

**Fig. 28** EAC2 Deployed Configuration.

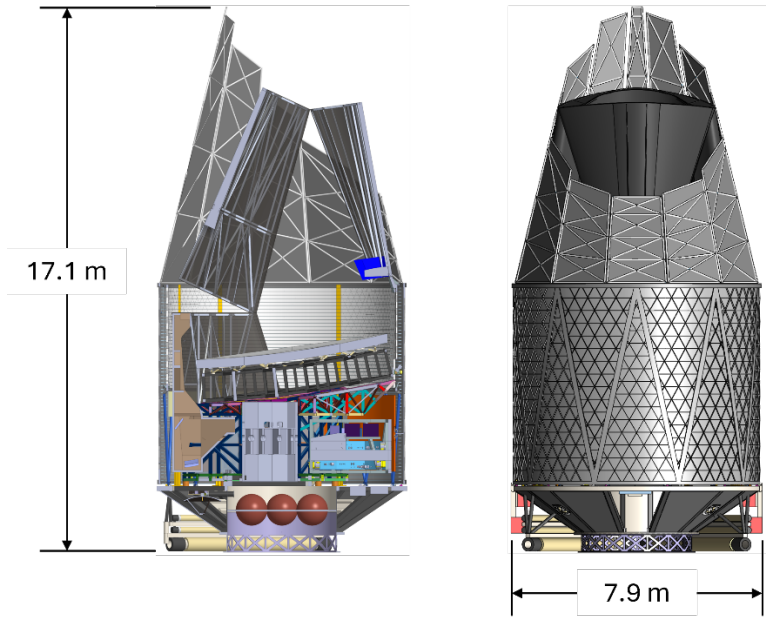

**Fig. 29** EAC2 Stowed Configuration.

The complement of SIs are functionally the same as EAC1 but with modified layouts to fit within the cylindrical volume allocations behind the telescope; the SI locations were designed to allow those with the coldest detector requirements, such as the CI, to have the best radiator view factors to cold space. To allow on-orbit replacement of the SIs, the Sunshade and Solar Arrays are actuatable to allow access; each SI and its corresponding radiator are extracted radially.

Two initial assumptions, both of which allow for a simplified baffle design, simplify the deployment sequence, as shown in Fig. 30: compatibility with only the wider launcher fairings, notably the Starship and SLS fairings, and a 6 m diameter OTE. Instead of multiple, complex deployments steps, the EAC2 baffle deployment consists of a "spreadable" forward petal deployment followed by the forward slide of a rigid cylindrical baffle segment over a fixed cylindrical section in a "camping cup" style design. The overall deployment sequence includes: 1) Solar array and the HGA, 2) Unrolling of the Deployable Sunshades to shield SI radiators, 3) Forward deployable petals are spread out, 4) SMSS deployment, first by a rotation about a root hinge at its interface with the IGA and then by a rotation about a mid-hinge assembly, and finally 5) DBS outer cylinder slides forward; as part of this final deployment, supporting struts, that support the fixed cylinder against launch loads are swung forward out of the view of the SI radiators.

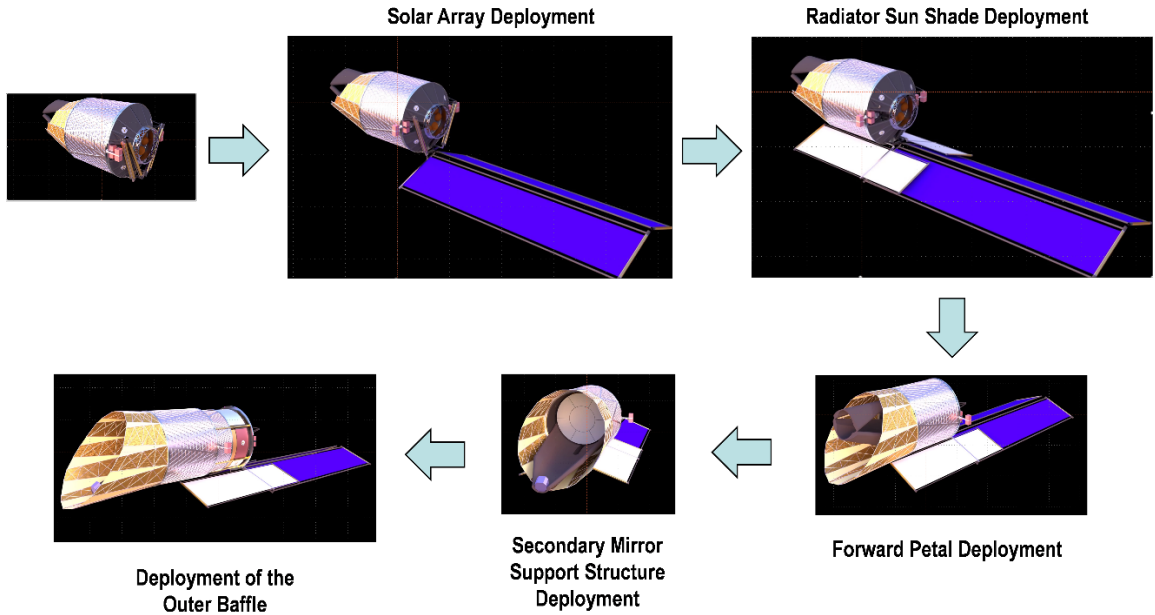

**Fig. 30** EAC2 Deployment Sequence.

### 2.3.4 EAC2 Structural Configuration

EAC2 structural modeling and analysis efforts were more limited in scope compared to EAC1, with no final, system level FEM developed for system level analysis (Fig. 31 depicts the deployed system model as-is). The structural team focused instead on key component models and analysis to support the mechanical design effort, with attention on new concepts. Similar to EAC1, design goals were established for critical components and subsystems, as shown in Table 5; though, note predictions are available for a only a smaller set of components.

**Table 5** EAC2 Deployed Frequency Goals.

| Component | Frequency Goal, Low (Hz) | Frequency Goal, High (Hz) | Lowest Mode (Hz) | Boundary Conditions / Comments |
|---|---|---|---|---|
| SMSS | 10 | 20 | 17.5 | Free-Free with telescope CONM2 |
| Mirror Segment, First Flexible Mode | 200 | 300 | 201 / 271 | Free-Free Mirror |
| PMBA First Flexible Mode | 50 | 100 | 40-58 | Free-Free |
| Isolation Systems | | | | |
|     Spacecraft Isolator (Modes 1-6) | 0.5 | 1.0 | 0.51 | With lumped SC and OTE masses |
|     Reaction Wheel Isolator (Modes 1-6) | 2.5 | 6.5 | 3.1 | Fixed base with RW and housing mass included |
| Spacecraft | | | | |
|     DBS Primary Structure First Mode | | | 0.6 | Fixed base, composite version |
|     Solar Arrays | 0.05 | 0.10 | 0.063 | Cantilevered on the SC |

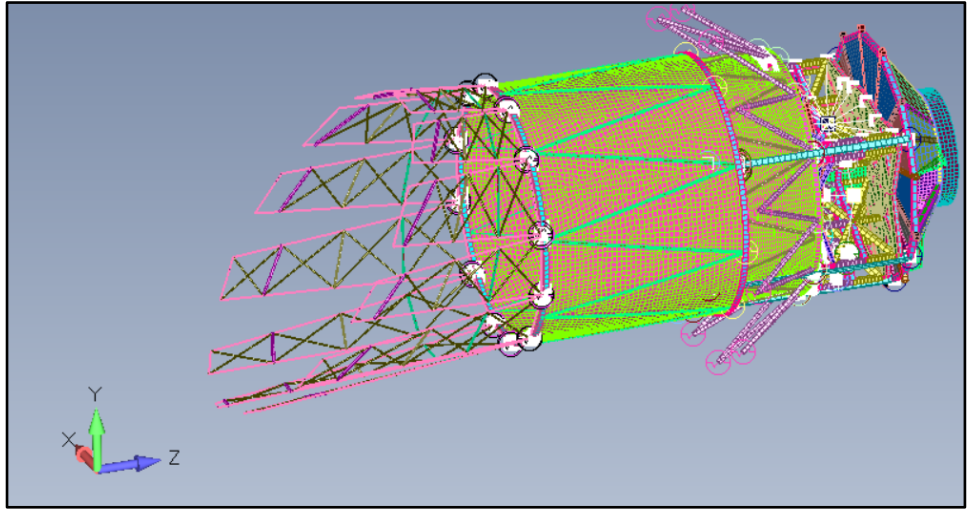

**Fig. 31** Deployed EAC2 FEM.

The EAC2 DBS connects to the Spacecraft Bus through a barrel-like structure, which results in an 0.6 Hz first natural frequency, more than an order of magnitude higher than the EAC1 DBS. The second DBS mode, a much larger modal mass resonance, is 2 Hz. The solar array bending modes now represent the lowest observatory natural frequency at 0.06 Hz.

Other notable advances included designs for the SMSS, Instrument Module Structure (IMS), PMBA, and PMSA. Like EAC1, the SMSS model began as a truss structure; however, the SMSS had to fit into a volume with less depth (Y direction) compared to EAC1, though with more space available circumferentially. After an initial 9.5 Hz result, the truss concept changed to a panelized approach, which improved the first mode to 17.5 Hz after some optimization effort. Figure 32 shows the EAC2 SMSS model before and after this optimization.

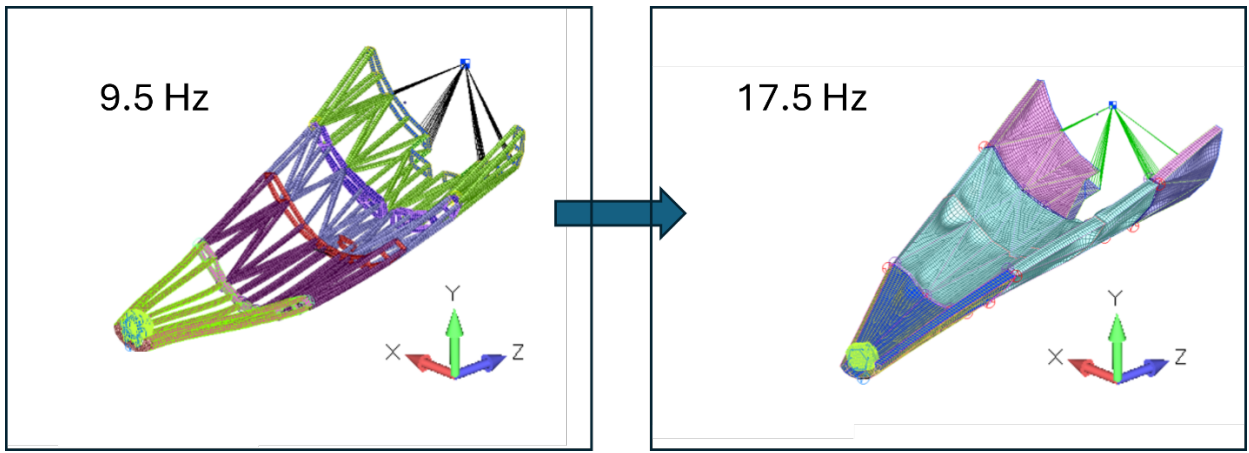

**Fig. 32** EAC2 SMSS Design Evolution.

The IMS and PMBA each needed additional shear stiffness to account for early mechanical design choices. Specifically, the mechanical design implemented large, open bays in the IMS structure to accommodate SI servicing. Similarly, the PMBA structure mechanical design removed diagonal elements in the backplane, with the goal of eventually contrasting the thermal distortion performance with EAC1. For the IMS, shear plates were added to available surfaces, especially around the inner ring, raising the first mode from 49 Hz to 93 Hz (before instrument mass). For the PMBA, the ends of struts were not terminated at a traditional bonded joint fitting, rather, stiffening gussets were introduced to the base of the struts to add a small bending/shear stiffness. Figure 33 and Fig. 34 below show the resulting frequency progression for the IMS and PMBA, respectively.

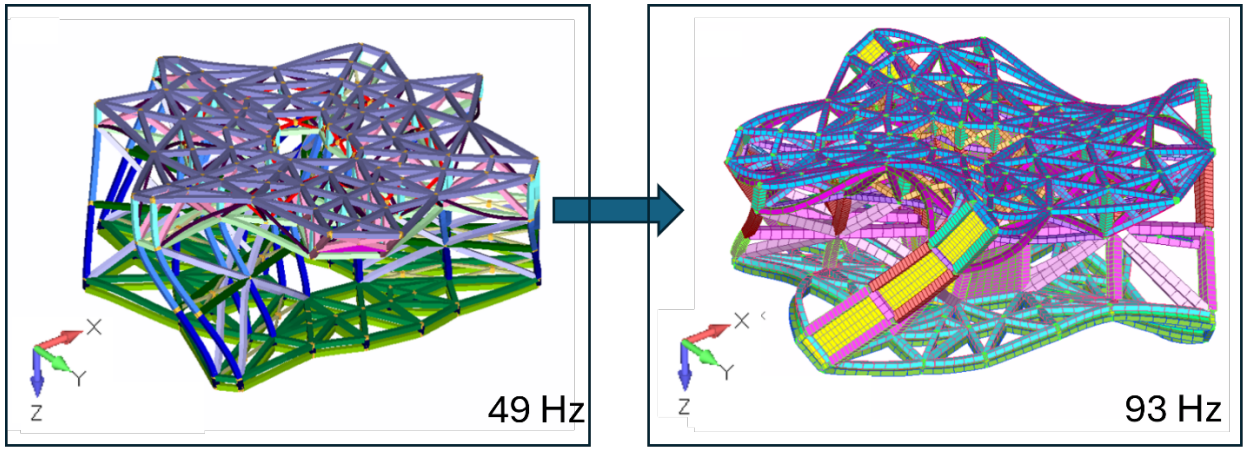

**Fig. 33** EAC2 IMS Frequency Progression.

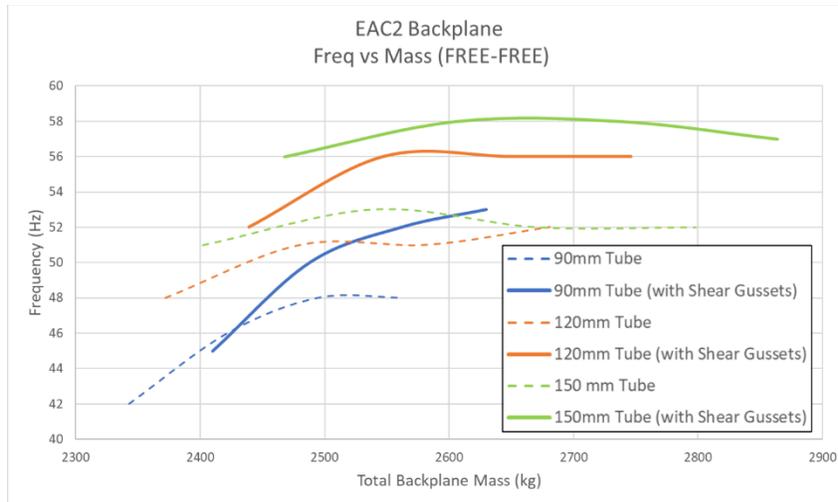

(a)

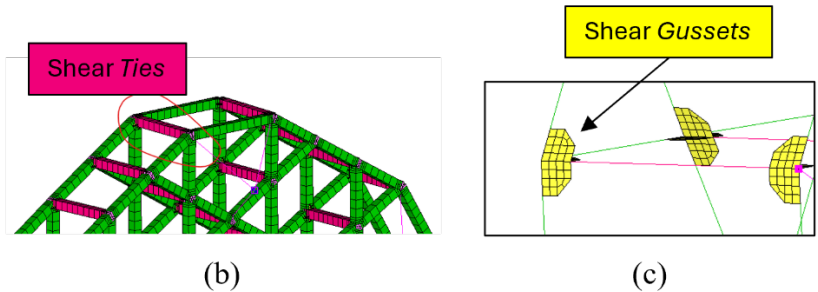

(b)                                    (c)

**Fig. 34** EAC2 PMBA: (a) Frequency Progression, (b) Shear Tie Design, and (c) Shear Gusset Design

### 2.3.5 EAC2 Thermal Configuration

The EAC2 thermal configuration built on the thermal design work from EAC1, specifically addressing several areas of weakness. The primary focus areas for EAC2 included assessing the DBS, refining instrument radiator accommodations, and resolving the CI thermal design challenges identified in EAC1. In the end, with the same environmental assumptions and passive cooling philosophy from EAC1, the EAC2 thermal design implemented sufficient changes to the instrument module configuration and parasitic heat load calculation methods to achieve a significantly improved thermal design.

EAC2 incorporated several similar features from EAC1: deployables to block direct solar impingement and indirect IR backloading from the solar arrays; the same temperature limits and dissipations for the SIs, despite EAC1 guidance to consider changes; and a DBS, like the MBA in EAC1, analyzed as a fully passive and "heated" system, responsible for maintaining the telescope radiative environment stability. It should be noted that the heated concepts assumed perfect boundary temperatures for actively controlled areas and that further work will be required to assess instability due to dissipated power perturbations in the heated areas as well as induced gradients.

The EAC2 overall architecture provided a few new concepts for thermal consideration. First, the DBS had a different deployment system which required structural panels rather than soft goods. Second, the instrument module organized the SIs in a radial pattern instead of a tower-like configuration, with the CI located at the top with a relatively unencumbered view to deep space. While this reflected a more optimal location for the CI, the other instrument radiators still required solar protection using the deployable shades to maintain sufficient sink temperatures. Figure 35 illustrates the overall EAC2 thermal architecture, with key components highlighted.

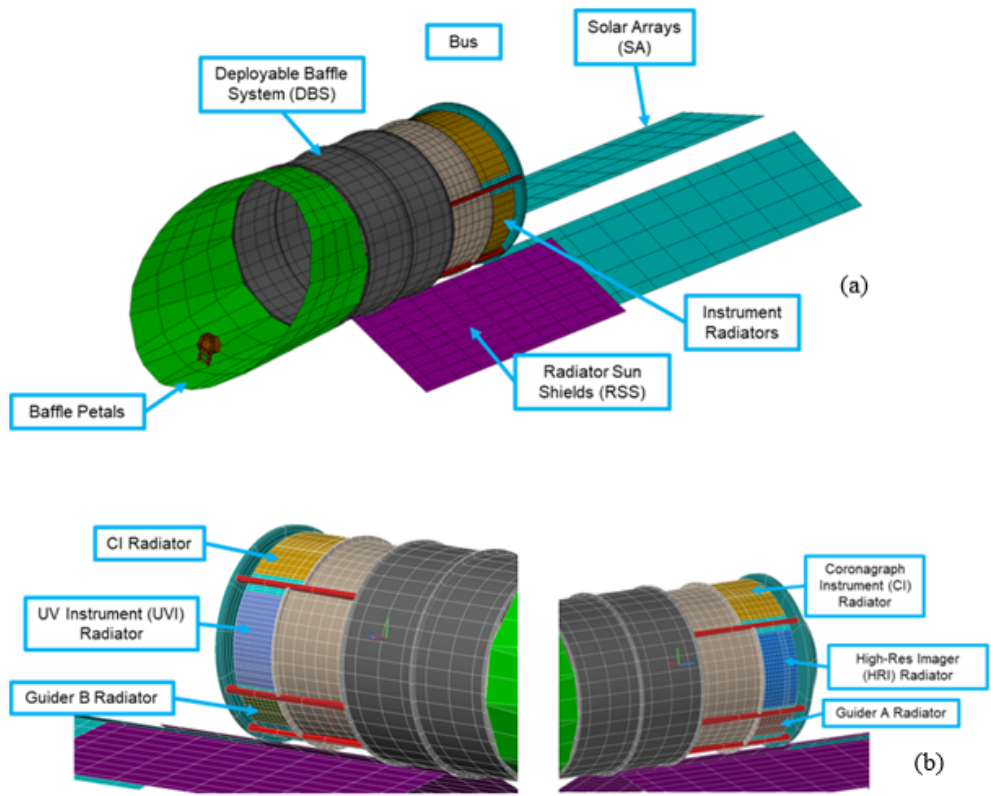

**Fig. 35** EAC2 Thermal Model Architecture: (a) Full Observatory and (b) Instruments & Radiators.

The HRI, UVI, and GI parasitics were based on a refined standard calculation, with scaling factors derived from the Roman Wide Field Instrument (WFI); again, room temperature components were assumed to have negligible parasitic heat loads. Given the EAC 1 issues with the CI thermal closure, the EAC2 CI thermal design included a staged radiator system and parasitics based on a generic relationship between overall support conductance, the $\Delta T$ involved, and the supported mass.[12] For EAC2, given the assumptions for the efficiency of the deployed sunshades, radiator designs closed for all SIs.

Exiting the EAC2 system architecture effort, the team identified several open items and recommendations. Key open items included omission of struts in the radiator view after deployment, inability to achieve assumed low emissivity values, and potential plume impingement from the thrusters onto the solar shields. Recommendations for future EAC efforts included a more nuanced understanding of all IR backloading paths onto the radiators and concurrent collaboration with the mechanical team on solar shield implementation for higher efficiency.

## 2.4 EAC3 Design Formulation

### 2.4.1 EAC3 Optical Telescope Configuration

Unlike EAC1 and EAC2, the optical design of the EAC3 observatory is centered (i.e. obstructed, "on-axis"), where the obstruction ratio was set to be less than 10 percent, as shown in Fig. 36. This minimal obstruction constrains the FOV to fit within the hole of the primary (M1), which is kept to approximately the same size as the shadow of the secondary mirror (M2). Figure 36 also shows the clustered fields for the instruments, set at each quadrant about the axis, having about ~0.25 degrees.

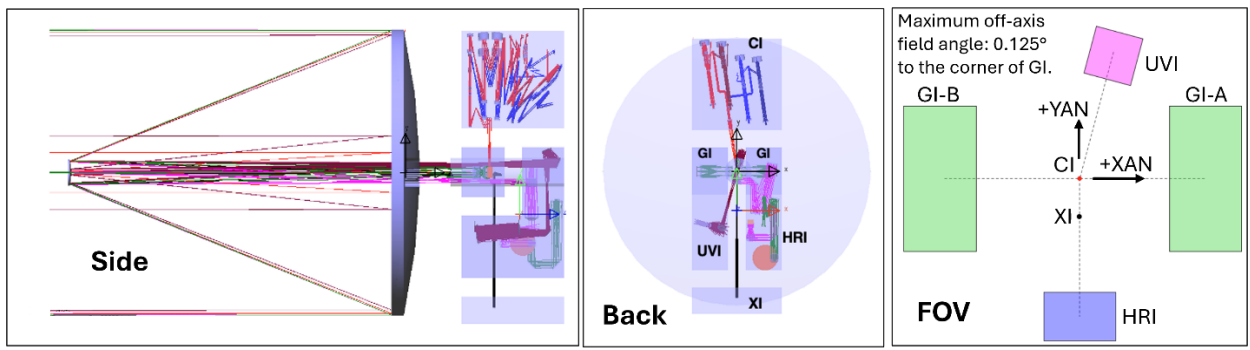

**Fig. 36** Centered EAC3 Telescope with Instruments, and Field of View plot (FOV).

Figure 37 illustrates the M1 segmentation and M2 support obstructions as well as shows the projected geometry in the entrance pupil plane. M1 has an 8 m diameter that consists of four rings of "keystone" ULE glass segments surrounding a central hole of 800 mm in diameter. The segments are separated by 8 mm physical gaps.

The EAC3 keystone segments have a maximum linear dimension of 1.8 m along the outer secant line. This is within the current maximum boule size for ULE manufacture.

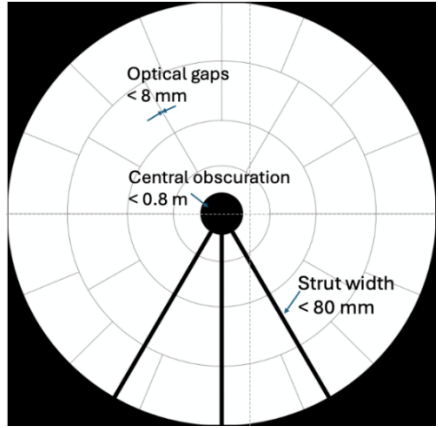

**Fig. 37** EAC3 Primary Mirror Geometry.

### 2.4.2 EAC3 Science Instruments

Similar to the EAC1 and EAC2 CI, the optical design of the EAC3 CI consists of four channels; two are separate polarization channels for the Visible wavebands and two are separate polarization channels for the NIR channels. Each channel incorporates its own Deformable Mirror (DM), Wavefront Sensor, and Integral Field Spectrometer (IFS). The optical layout of the CI concept, illustrated in Fig. 38, shows the packaging of all four CI channels into its allocated volume.

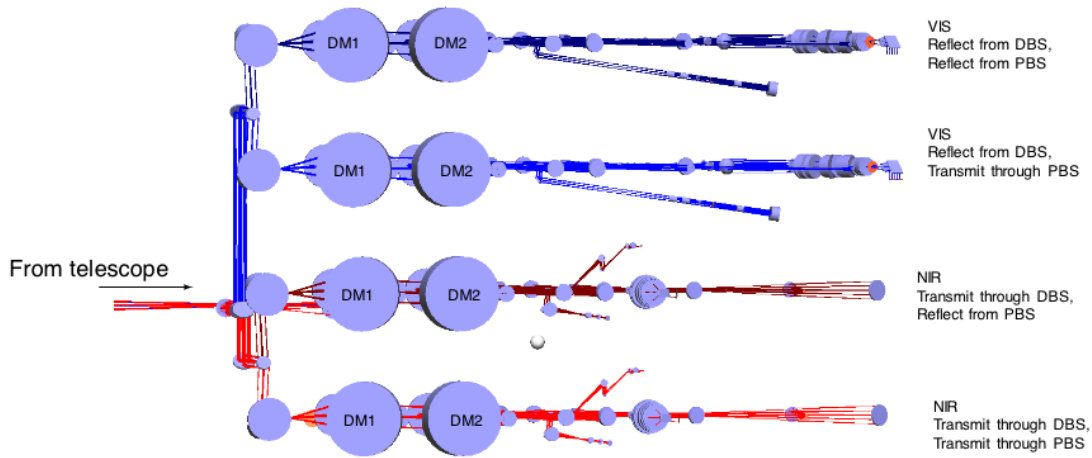

**Fig. 38** Side View Packaging of the Four Channels for the EAC3 CI.

Figure 39 illustrates the UVI optical layout. The UVI has the following four optical channels: 1) FUV Imager, 2) FUV MOS; with short wavelength grating modes, 3) FUV MOS with medium wavelength grating modes, and 4) NUV MOS; with longer wavelength grating modes. All four channels pass through the same MSA and then into their individual optics. Similar to EAC2, one of the FUV MOS grating modes was not included due to volume constraints but will be considered for future EACs.

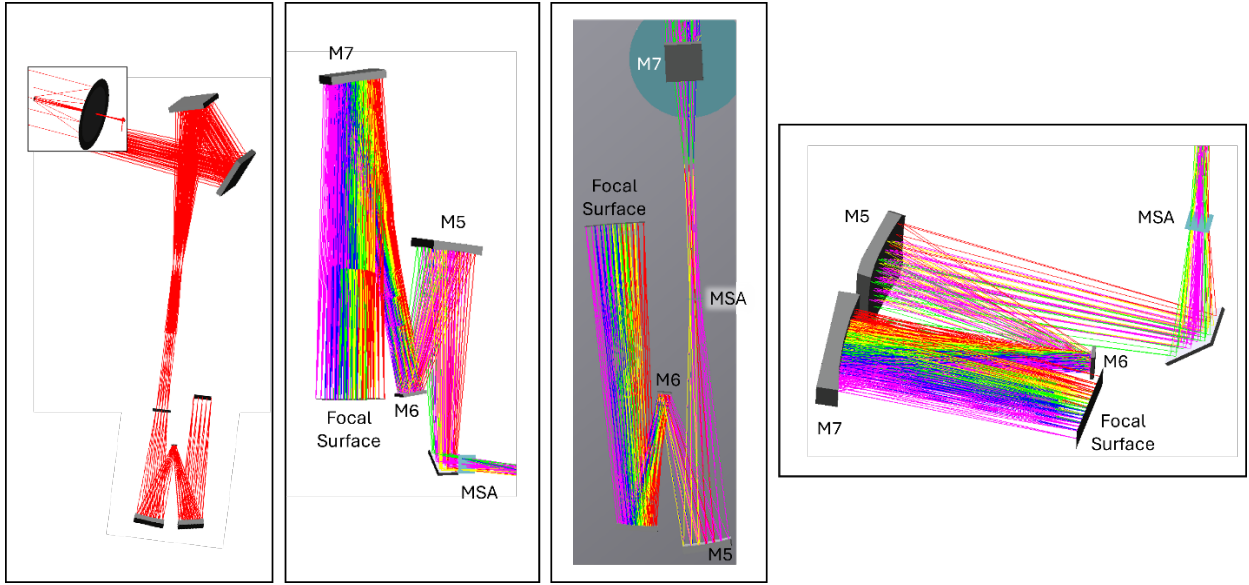

**Fig. 39** Optical Layout of the EAC3 UVI.

HRI provides UV / Visible and NIR imagery over a 3 x 2 arcmin FOV; Fig. 40 illustrates its optical layout.

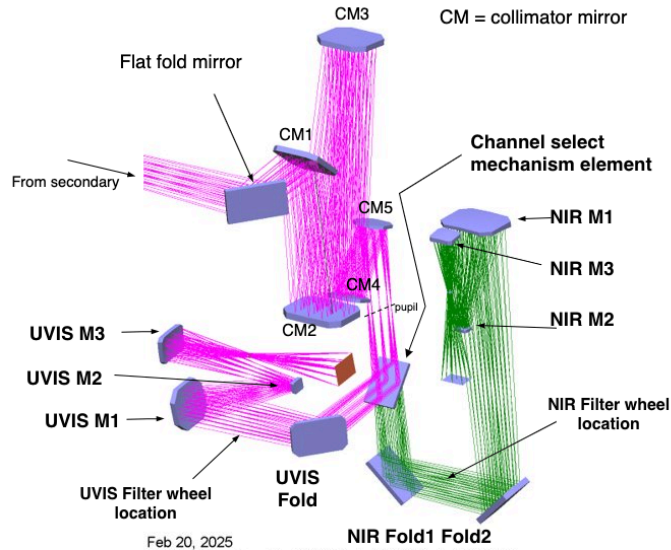

**Fig. 40** Optical Layout of the EAC3 HRI.

The two Guiding Instruments (GI-A and GI-B), provide the fine pointing signals to the observatory control system. These imagers operate in the waveband from 400 nm to 1000 nm over a 3 x 6 arcmin FOV, double that for EAC1 and EAC2; the telescope design for EAC3 optimized the guider area for possible Wavefront Sensing and Control (WFSC). Figure 41 illustrates their optical layout.

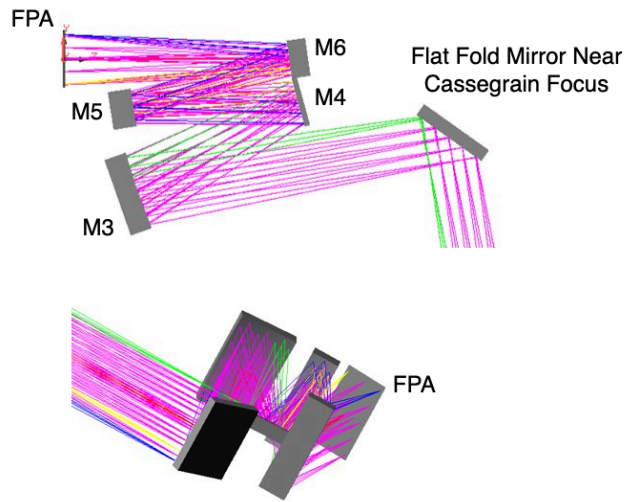

**Fig. 41** Optical Layout of the EAC3 GI.

### 2.4.3 EAC3 Mechanical Configuration

The initial concept for EAC3 is similar to EAC1 but focused on an 8 meter on-axis, large FOV telescope, which allowed a shorter PM-SM spacing. Further, the EAC3 stowed configuration considers three launcher options: NASA SLS, SpaceX Starship Standard, and Blue Origin New Glenn 9x4. Similar to EAC2, the EAC3 design effort benefitted from prior EAC lessons learned by positioning CI on the cold side of the observatory. Figure 42 shows the overall EAC3 design evolution, in part driven by the lessons learned.

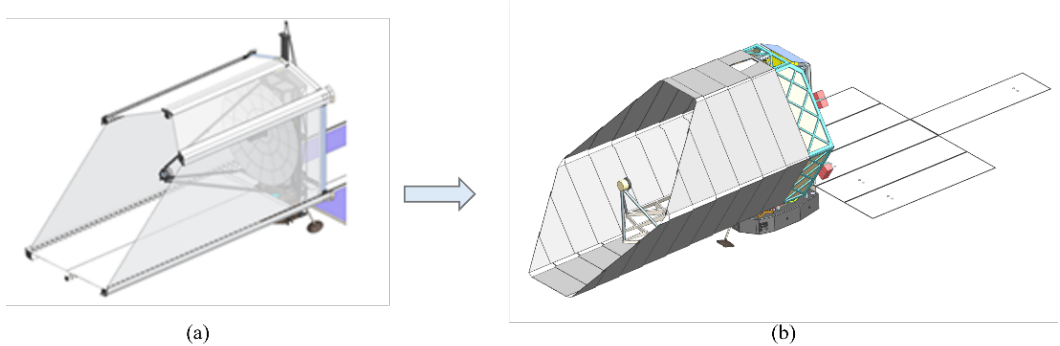

(a)                                                                          (b)

**Fig. 42** Evolution of the EAC3 Concept: (a) before and (b) after Design Evolution.

This evolution included the following major changes from the initial configuration:

- Adopted new DBS and SMSS designs; this included an updated baffle deployment strategy to SLiding Aluminum Panels (SLAP) as well as an updated SMSS design to reduce strut obscuration; note, future designs can consider a composite material to reduce mass

- Converted rear sunhades into blanketed solar panels, which eliminated large roll-out solar array deployment

- Positioned the CI on the cold side of the observatory to afford the CI radiators a complete cold view to space

- Revised the PM Backplane based on iterative work between design and structural analysis personnel

Figure 43 and Fig. 44 illustrate the EAC3 observatory physical deployed and stowed configurations, respectively.

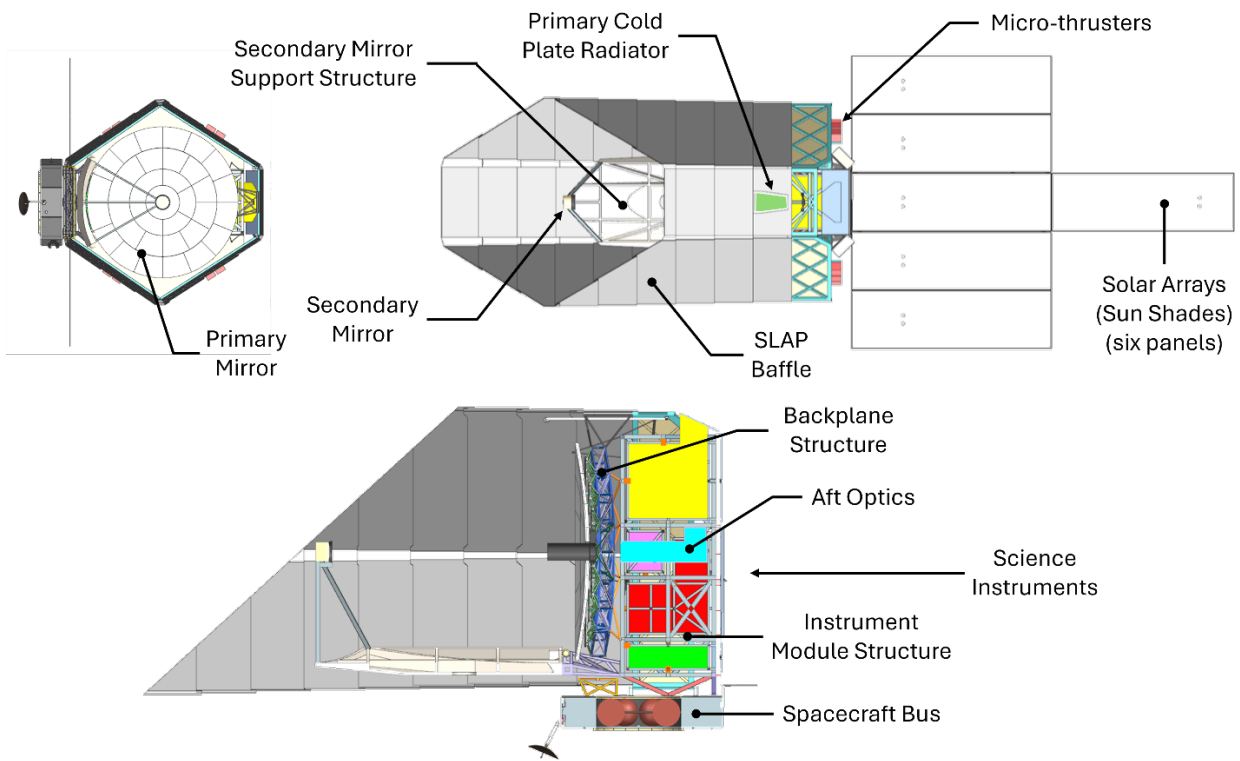

**Fig. 43** EAC3 Deployed Configuration.

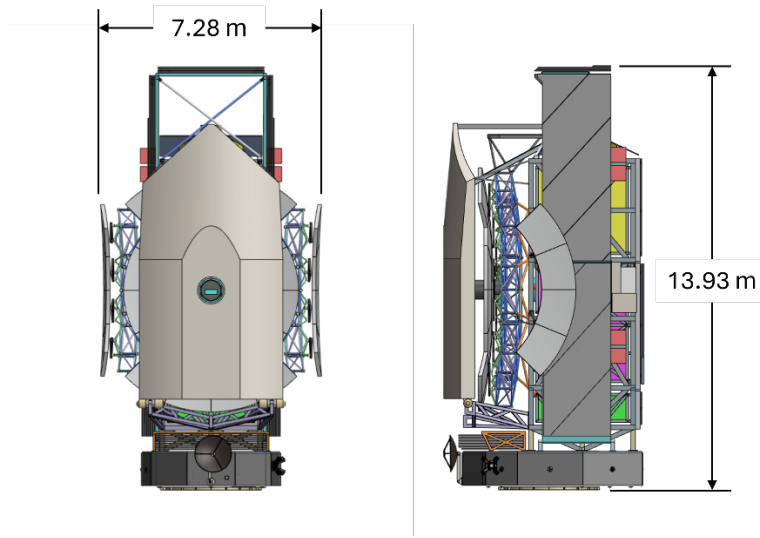

**Fig. 44** EAC3 Stowed Configuration.

EAC3 incorporates radiators into the instruments to facilitate servicing; this forces the SI pallets to act as launch primary structure with multiple latches to the IMS, so radiators remain part of the SI (on-orbit, extra latches release to provide near-kinematic support to SIs).

The EAC3 deployment sequence, as shown in Fig. 45, shares similarity with EAC1, though driven by a cable-pulley deployment mechanism given the difference between the SLAP Baffle construction and the EAC1 soft baffle. The deployment sequence includes: 1) Folded solar array panels and the HGA, 2) SMSS (immediately preceded by release of the Upper Strut Assemblies, which support the SMSS during launch), 3) the two PM wings, 4) the SLAP Baffle structure "squats" and, finally 5) a cable-pulley system deploys the baffle sliding panels.

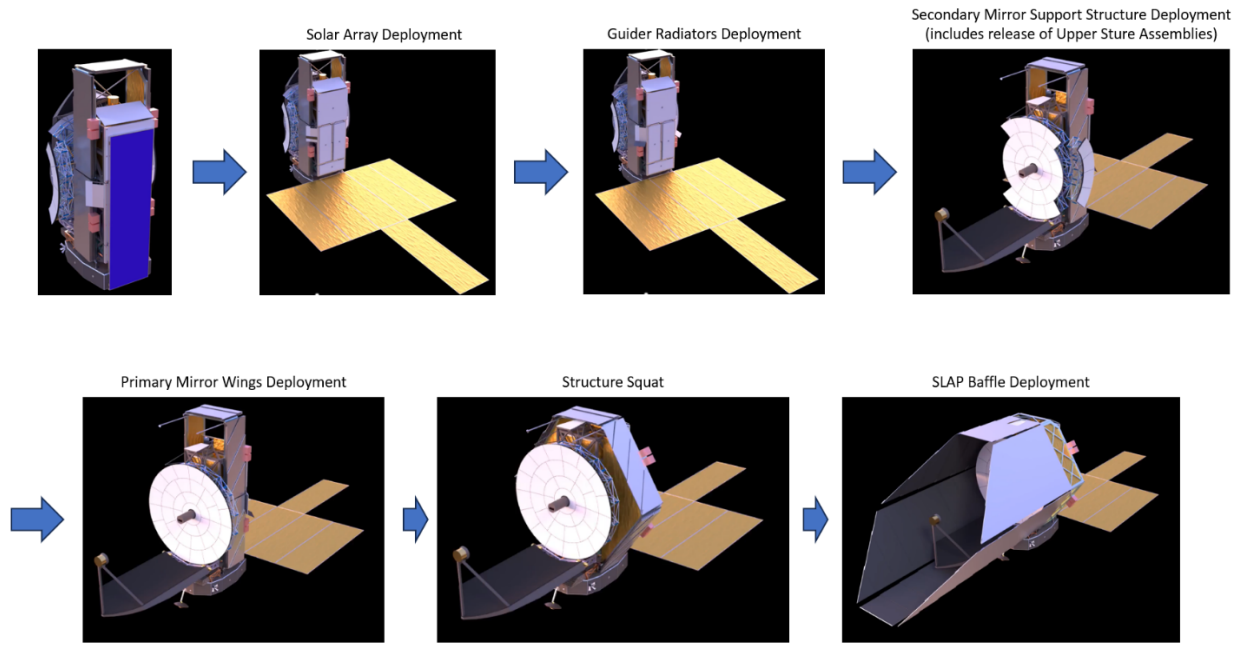

**Fig. 45** EAC3 Deployment Sequence.

### 2.4.4 EAC3 Structural Configuration

EAC3 structural modeling and analysis efforts were more limited in scope compared to both EAC1 and EAC2, with no system level FEM developed for system level analysis. With a similar overall architecture to EAC1, EAC3 modeling and analysis efforts focused on the SMSS, PM segments, and the DBS. Design goals were established for critical components and subsystems, as shown in Table 6; though, note predictions are available for a only a smaller set of components.

**Table 6** EAC3 Deployed Frequency Goals.

| Component | Frequency Goal, Low (Hz) | Frequency Goal, High (Hz) | Current Estimate (Hz) | Boundary Conditions / Comments |
|---|---|---|---|---|
| SMSS | 10 | 20 | 10 | Fixed based at IMS |
| Mirror Segment, First Flexible Mode | 200 | 300 | 240 | Free-Free Mirror |
| Isolation Systems | | | | |
| Spacecraft Isolator (Modes 1-6) | 0.5 | 1.0 | 0.51 | With lumped SC and OTE masses |
| Reaction Wheel Isolator (Modes 1-6) | 2.5 | 6.5 | 3.1 | Fixed base with RW and housing mass included |
| Spacecraft | | | | |
| DBS / SLAP | 0.07 | 0.40 | 0.10 | Fixed base |

Notably, EAC3 has an on-axis optical design which requires a centered SM, a significant deviation from prior EACs. While the observatory scale is too large for a JWST tripod style SMSS, EAC2 experience demonstrated the feasibility of a curved and panelized design; EAC3 simply required a cantilevered SM placed at the tip. The team developed this panelized concept and optimized to achieve a 10 Hz minimum frequency, while minimizing mass. Figure 46 shows the optimized first mode (bending mode shape).

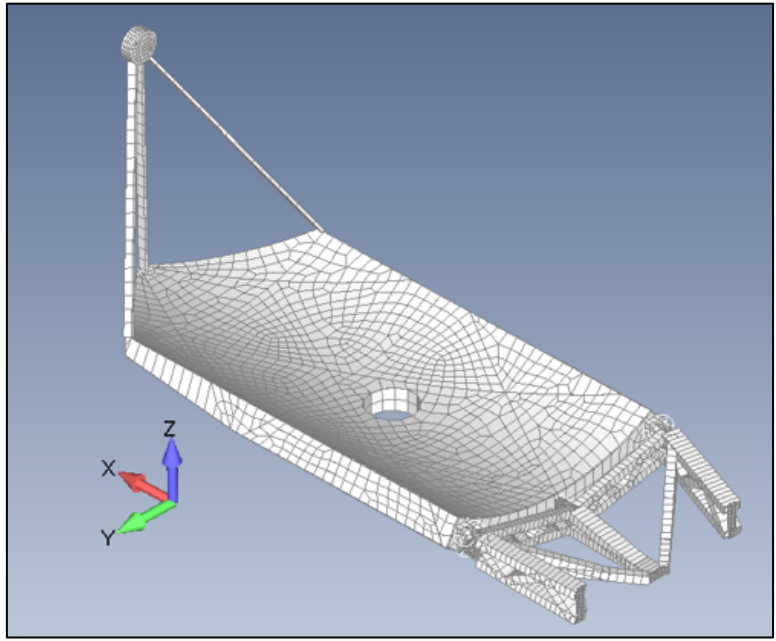

**Fig. 46** EAC3 SMSS First Mode after Optimization.

The PMSA mirror segments were constructed using the Arnold Mirror Modeling[13] software, similar to prior EACs. Free-free boundary conditions were applied to assess the overall segment stiffness; Fig. 47 shows the resonances of the center segment (240 Hz) and the resonance of the remaining three segment types (all ~375 Hz).

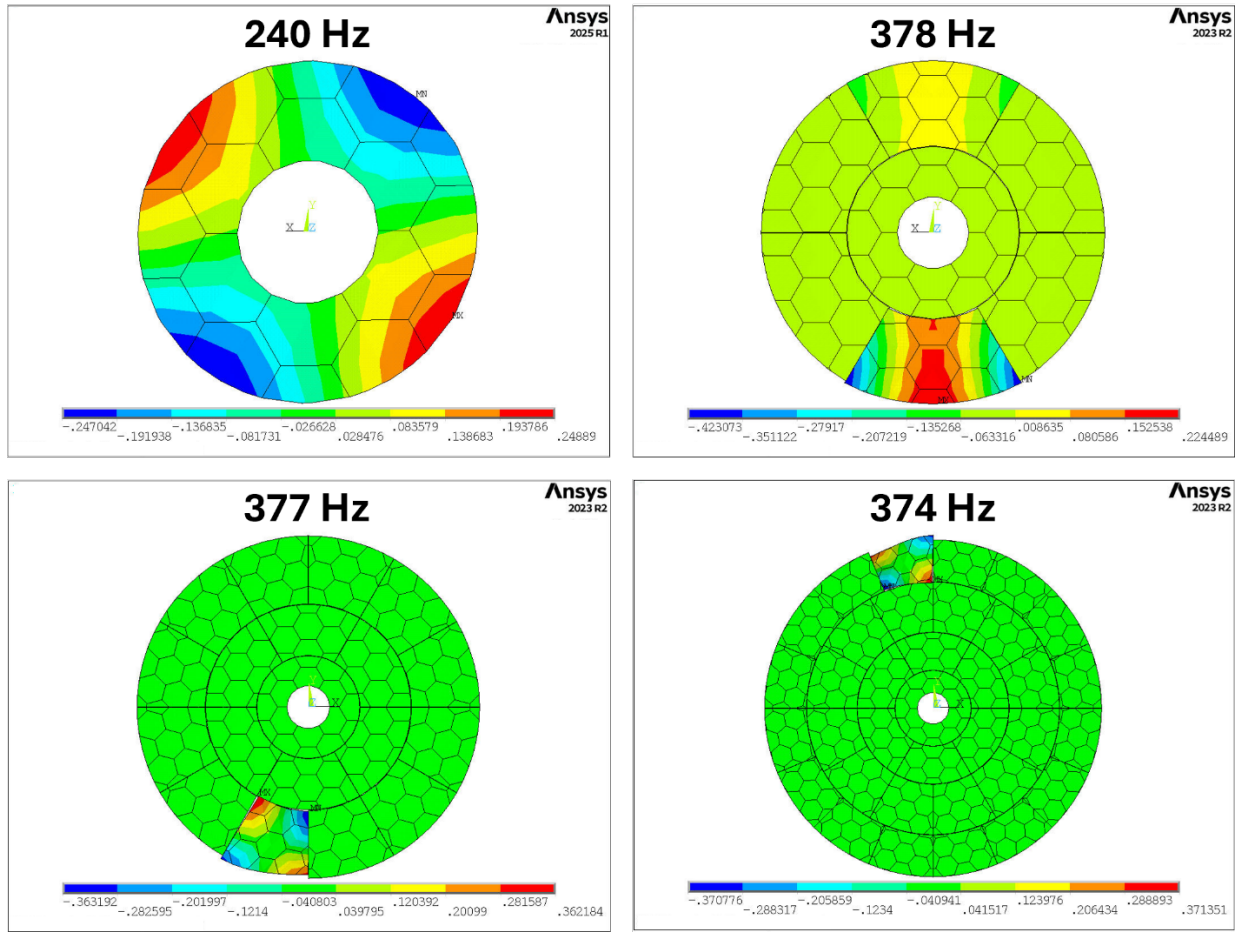

**Fig. 47** EAC3 Keystone Segment Resonances.

EAC3 PMBA design and analysis took a different approach; initial design included a starting PMBA strut layout provided adequate nodal points to receive the 35 individual PM segments as well as a limited 0.5 m backplane depth. Following, five candidate beam sections were chosen based on previous EAC experience and fed into a 3,000 trial Monte-Carlo analysis, which computed mass and first mode frequency for randomly chosen candidate beams at 353 individual locations. The top 2% of results (all cases above 46 Hz) were saved and used as seeding cases for an optimization run; the final configuration had a mass of 2,238 kg, with an additional 2,420 kg to account for PMSAs and associated thermal control hardware.

The EAC3 SLAP Baffle is similar to EAC1, though smaller overall and implemented with telescoping solid panel sections instead of membrane (although smaller, its 3,690 kg predicted mass approaches that of EAC1). The SLAP Baffle lowest predicted frequency is 0.1 Hz, though is relatively small in terms of modal mass participation. The larger participation modes fall within the isolation system frequency range of 0.5 to 1.0 Hz; this interaction between the SLAP Baffle and isolation system modes is not desirable and will be studied should an EAC3 type design be selected for future analysis work.

### 2.4.5 EAC3 Thermal Configuration

The EAC3 thermal configuration included elements from both EAC1 and EAC2. From EAC1, EAC3 preserved both the instrument module (though with different instrument locations) and the overall baffle shape (though with soft goods membrane replaced by the more rigid panels in the SLAP Baffle). Notably, the CI included increased electronics power and an additional IR detector, which challenged the thermal design. Further, the Roll Out Solar Arrays (ROSAs) from EAC2 were replaced with blanketed, solid solar panels.

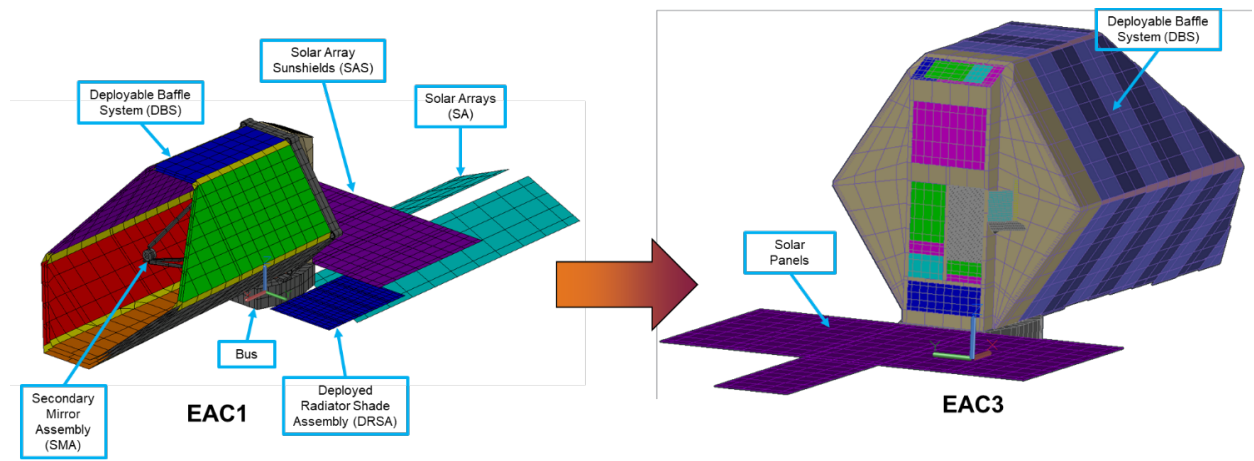

**Fig. 48** EAC3 Thermal Model Architecture (EAC1 shown left for reference).

The instrument layout changed significantly from the EAC1 configuration; in response to a key EAC1 lesson learned, the EAC3 thermal architecture relocated the CI to the top of the instrument "tower" – this prioritized a location for the CI radiator with a direct, clean view to deep space, making passive cooling for the 65 K detectors more feasible. While the CI heat rejection loads and parasitics increased, the radiator design allowed for closure.

In lieu of dedicated, deployed radiator solar/IR shades, the solar array panels doubled as shades. Despite the new solar blockers, the Guiders required additional, local thermal shading and the HRI used directional vanes to reduce the sink temperature for the 100 K detectors.

### 2.5 Attitude Control System

The Attitude Control System (ACS) architecture design for HWO is driven by the CI's stringent image quality requirements. The ACS system must provide an ultra-stable, low-disturbance pointing platform to facilitate state-of-the-art instrument-level LOS and wavefront stability, while also providing sufficient agility to enable observing efficiency goals for Roman-like coronagraph ConOps and general astrophysics observations. For each EAC, the study has formulated and maintained several distinct ACS architectural concepts. Here, we focus on 1) a conventional "reaction wheels only" ACS architecture that utilizes a suite of reaction wheels (RWA) to actuate both large maneuvers and science observations, and 2) a "hybrid" ACS architecture that employs RWAs for large maneuvers and low-disturbance micro-thrusters for precision pointing observations. The "RWA Only" ACS architecture is paired with conventional dual-stage passive isolation at the payload and RWAs. Preliminary jitter analyses have introduced active payload isolation to the RWA Only case, described in Sec. 4.2.1.4; future studies will also consider active RWA pallet isolation.

### 2.5.1  *ACS Architecture Overview*

The HWO instrument LOS control architecture in Fig. 49 includes an ACS control loop (outlined in red in the figure) that uses sensing data from Star Trackers or Stellar Reference Unit (SRU), Inertial Reference Unit (IRU), and Fine Guidance Sensor (FGS) to control the RWAs and/or micro-thrusters. Slew maneuvers are performed by RWA actuation using measurements from the SRU/IRU. FGS is used during science observations for precise attitude control. In the "Hybrid" ACS architecture implementation, pointing control authority is transitioned to micro-thrusters while the RWAs are spun down to zero speed to minimize jitter contributions. The "RWA Only" ACS architecture removes the micro-thrusters from this control structure and instead operates the RWAs within a quiet speed range during observations. A wheel speed bias maneuver is performed to maximize the time the RWAs remain within this region while momentum is accumulating from solar radiation pressure.

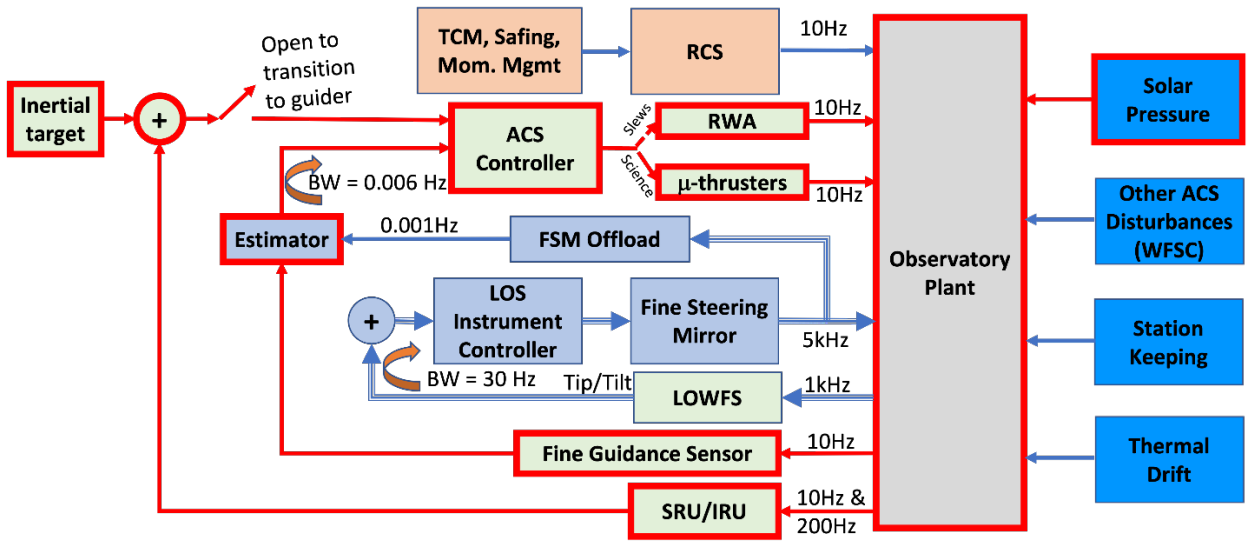

**Fig. 49** ACS Architecture (boxes outlined in red) as Part of HWO Instrument Line-of-Sight Control; light green represents ACS control, light blue represents instrument level control, orange represents orbital maneuver system (RCS), and dark blue to right represent external disturbances.

Each SI will have its own Instrument Line-of-Sight Control and FSM, as required by its pointing requirements. Each FSM is controlled by either the Low Order Wave Front Sensor (LOWFS), as in the case of the CI, or by commands from the GIs and from the ACS.

The ACS architectural options are applied to each of EAC 1-3 to 1) examine their accommodation requirements 2) assess their ACS-level pointing performance and drivers 3) and identify challenges and operational considerations for future EACs. Here we describe the formulation process of the ACS architectures for EAC1, which focuses on the selection and sizing of both the micro-thruster and RWA configurations.

### 2.5.2 EAC1 Micro-thruster Configuration and Sizing Study

Figure 50 shows the baseline micro-thruster configuration on EAC1. In the Hybrid ACS Architecture, the micro-thrusters are tasked with 1) preventing momentum accumulation on the RWAs during slew maneuvers and 2) maintaining ACS precision pointing during science in the presence of external disturbances. Given the Solar Radiation Pressure (SRP) torque is the dominant external disturbance on HWO, the micro-thruster geometry and sizing is chosen to handle the worst-case SRP torque for all allowable Sun directions, based on EAC1 mounting location and plume impingement constraints. A symmetric micro-thruster geometry is chosen to provide arbitrary torque without imparting a net force on the observatory.

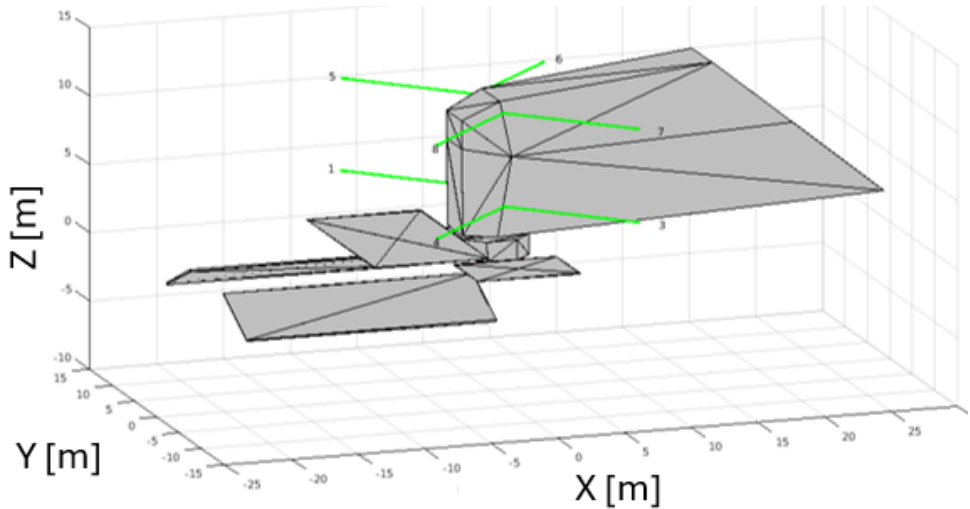

**Fig. 50** EAC1 Observatory Geometry Model and Micro-thruster Configuration.

The SRP torque for EAC1 is estimated by using a simplified geometry model derived from the EAC1 mechanical drawings, as shown in Fig. 50. The Sun direction is then varied over the entire Sun Field-of-Regard (±45º Sun Pitch, ±22.5º Sun Roll) to compute the SRP torque at each Sun direction as shown in Fig. 51. For EAC1, the SRP torque is computed using a standard "N-plate" torque model with surface reflectivity set uniformly to 100% specular to conservatively bound the SRP torque (see Sec. 4.1.1.2). The EAC1 center-of-mass (CM) location is set using the EAC1 mass properties (see Sec. 4.1.1.1), and the Center-of-Pressure (CP) offset is set to +1 meter along the observatory X axis to study SRP torque sensitivity to CM-CP offset; certain assumptions are made to specify this offset, including the assumption that the solar arrays are not constrained by power and lengths can be changed to shift the CP. The micro-thrusters are modeled as the Busek Colloid Micronewton Thrusters (CMNT) flown on the LISA Pathfinder Space Technology 7 (ST7) demonstration in 2015, as described in Sec. 4.1.1.4.

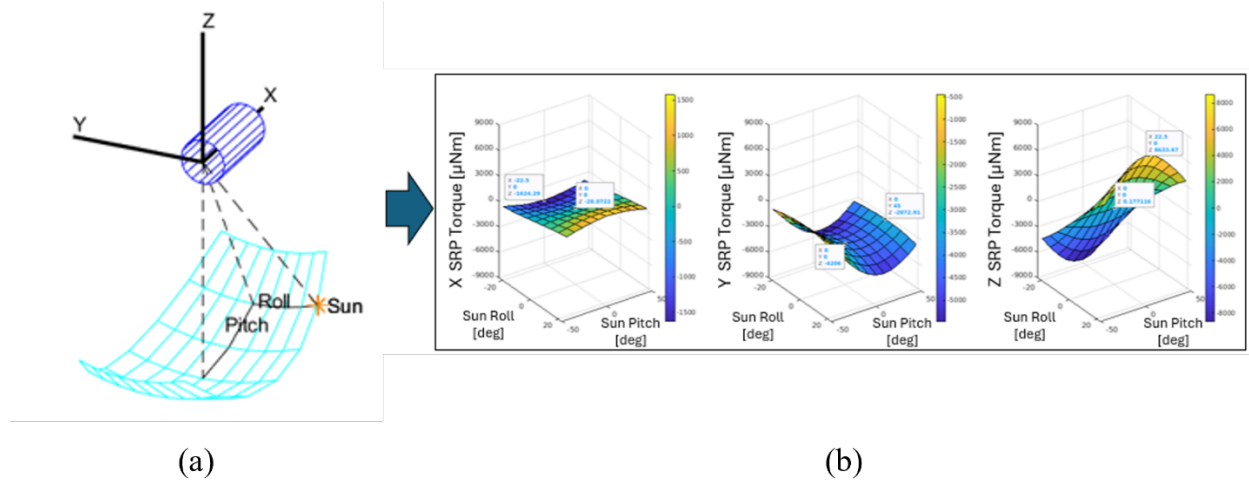

<div align="center">(a)                   (b)</div>

**Fig. 51** (a) Sun Angle Definitions (b) SRP Torque vs. Sun Angles (with 1 meter +X CM-CP Offset).

Figure 51 illustrates the variation in SRP torque as a function of Sun angles for EAC1. The SRP torque about the observatory Y-axis is 4,200 µNm when the Sun is at the solar array normal (0º Sun Pitch and 0º Sun Roll) due to the +1 meter CM-CP offset at this direction, and it varies strongly as the CM-CP alignment changes with Sun Pitch. The SRP torque about the observatory Z-axis is near zero when the Sun is at the solar array normal, but reaches 8,600 µNm at the maximum Sun Roll angles. This large Z-axis SRP torque drives the micro-thruster sizing on EAC1 as the largest over the EAC 1-3 configurations; this is due to the long length of the EAC1 MBA, which creates an asymmetric observatory geometry at this Sun orientation.

The micro-thrusters for EAC1 are each sized to about 700 emitters to accommodate the estimated SRP torque, when assuming a scaled version of the Busek CMNT reference model (see Sec. 4.1.1.4 ). EAC1 micro-thruster fuel use is estimated over a 10-year mission by assuming equal time is spent at all allowable Sun directions. Using the Busek CMNT nominal $I_{sp}$ of 240 sec, the estimated fuel is 520 kg, split evenly between the eight fuel tanks co-located at the micro-thruster locations. The management of SRP disturbance torque on subsequent EACs will continue to drive micro-thruster sizing and fuel use for the Hybrid ACS Architecture, as well as RWA momentum accumulation for the RWA Only ACS Architecture. The large projected area of HWO (600 m$^2$ for EAC1, about three times the JWST sunshade area) makes SRP torque sensitive to any CM-CP offsets that occur due to observatory asymmetries from different Sun directions.

*2.5.3 EAC1 RWA Configuration and Sizing Study*

Each ACS architecture option uses an eight RWA pyramid configuration as shown in Fig. 52, oriented for equal agility in the X, Y, and Z body axes. The RWA pyramid is defined by three angles $\alpha_1$, $\alpha_2$, and $\beta$, where $\alpha_1$ and $\alpha_2$ are the angular offsets (10° and 30° respectively) from the Y axis in the X, Y plane and $\beta$ is the tilt angle (34.1°) towards Z from the X, Y plane. The describing angles for each architecture are derived from the ratios of the primary axes' moments of inertia such that each axis has the same maximum rotational acceleration capability.

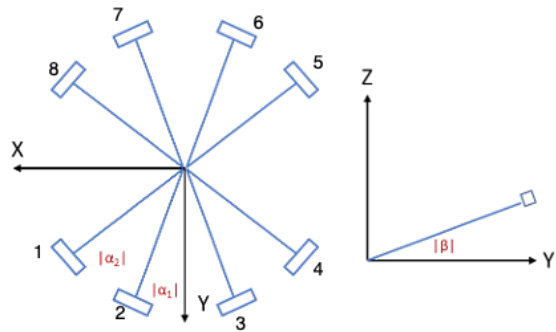

**Fig. 52** Reaction Wheel Pyramid.

Reaction wheel sizing for the Hybrid ACS architecture is mainly driven by slew agility with the assumption that the micro-thrusters are always continuously countering eternal disturbance torques such as solar radiation pressure. A trade to optimize slew profile duration by comparing maximum momentum and available torque along a constant power curve has been performed and is shown in Fig. 53. These two contour plots show the time in minutes to complete the slew profile for a 90-degree pitch and 45-degree roll against the maximum per wheel momentum and torque. The dashed blue lines represent the momentum and available torque along a constant power curve for the 8-wheel, 12-wheel, and 16-wheel configurations. The 8-wheel configuration power curve is bounded by performance parameters of 0.2 Nm at max 250 Nms and 0.4 Nm at max 125 Nms with the circle marking the optimal slew time performance. The power curves representing a 12-wheel and 16-wheel configuration show comparative performance increase of additional wheels by scaling the per wheel momentum and torque by 1.5 and 2.0, respectively. The dashed lines represent the momentum and available torque along a constant power curve, with the circle marking the optimal slew time performance. Power curves representing a 12-wheel and 16-wheel configuration show comparative performance increase of additional wheels by scaling the per wheel momentum and torque by 1.5 and 2.0 respectively. As the optimal point along the power curve is dependent on slew direction and length, a baseline maximum torque of 0.2 Nm and 250 Nms momentum for each reaction wheel in the 8-wheel configuration is used for the analysis presented in Sec. 4.1.

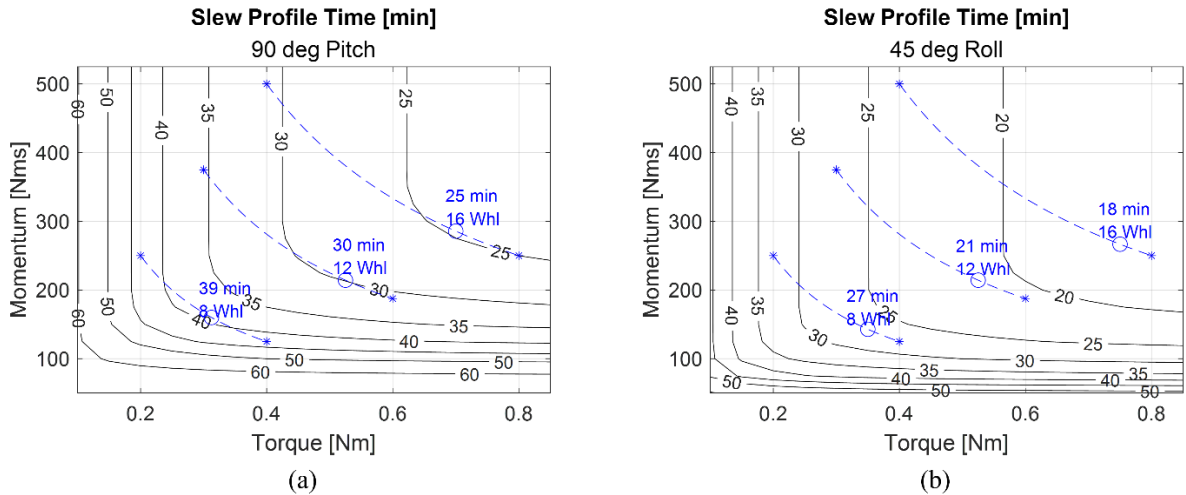

**Fig. 53** Reaction Wheel Slew Performance: (a) 90 Degree Pitch and (b) 45 Degree Roll.

## 2.6 Findings from EAC Designs

Through the EAC design and analysis exercises, the SET identified several challenges, many of which are common across the first three EACs; these findings will guide the SET in developing the next set of EAC designs. The top-level findings are summarized below:

<u>Optical Design</u>

a. A four-channel Coronagraph Instrument is estimated to require a volume of ~18 m³, a mass of ~3,000 kg, power dissipation of ~2,000 W, and thermal management to cool multiple NIR detectors to 50-65 K. Coronagraph architectures exceeding four channels will pose significant design challenges.

b. UVI drives the OTE prescription, including focal length, for all three EACs. For EAC3, the PM and SM are optimized to provide a flat field and minimize wavefront error at the MSA plane in UVI.

c. Polarization considerations for the CI constrain the PM-SM distance to a minimum length (or speed of the primary mirror), keeping the SM structure at 15 meters or longer from the PM.

d. The size of the PM segments are recommended to be less than 1.8 m to keep a standard boule size, and avoid ULE facesheets with multiple cores, which may exacerbate ULE uniformity concerns and require an extensive schedule for developing a new process. ULE CTE uniformity will be studied through IM to understand the impact on optical stability.

  e. Stray light protection via scrapers and vanes will be studied for future EACs; addition of these components will further increase the baffle design complexity.

Mechanical and Structural

  a. Launch vehicle capabilities have driven and will continue to limit observatory mass and volume requirements. The goal remains to launch all primary mission components in a single launch, but future EACs will also consider two-launch configurations to assess potential benefits and cost implications.

  b. Baffle size is driven predominantly by the FOR (no light into the baffle) and MMOD off of the secondary mirror, though a number of other effects, such as SM position, also drive the size.

  c. The current displacement of the Center of Mass from the Center of Pressure presents challenges for angular momentum control. Several options are under consideration to reduce the CP-CM offset to less than a meter, per ACS design assumptions.

  d. Off-axis telescope, plus desire for minimizing bounces and polarization for UV light significantly drive SI placement with CU and UVI MOS taking precedence.

Thermal and Thermal-Mechanical

  a. A deployable baffle capable of providing thermal stability as well as micro-meteorite and orbital debris protection is a significant challenge.

  b. Solar impingement and solar arrays are major barriers to instrument radiator heat rejection. Given the power dissipations and thermal needs of the CI, provisions should be made to locate this instrument with the best possible view factors to cold space.

  c. Thermal isolation and cryogenic heat transport: To achieve passive cryogenic cooling for the detectors, parasitics will drive radiator size, requiring both cutting-edge thermal isolation and the capabilities to effectively move heat at cryogenic temperatures over distances with low $\Delta T$.

  d. The large FOR (Field of Regard) assumption challenges the thermal design, requiring large sunshades for instrument radiators.

## Servicing Accommodation

a. Solar array and radiator shades may have to move out of the way to allow instrument access, which might require motorized hinges, rather than one-time spring driven deployments, to allow access.

b. Latches to secure and align the instruments will introduce structural non-linearities that will be problematic for a structure that needs to achieve "picometer-level stability". Adequate latch preload is necessary to maintain stability.

c. The disconnection and reconnection of cryogenic heat straps robotically will present design challenges as well as ΔT losses in the conductive paths that must be addressed.

d. Primary structure for launch loads will of necessity close off instrument compartments. SI radiators must be outside of this structure to provide efficient view to space. For servicing, such structure with radiators must be removed and replaced, with connections to the radiators remade, or be part of the new instruments when installed, while also restabilizing the SI compartment.

## Mass and Power

a. For EACs 1–3, the design fidelity is sufficient for identifying technical challenges but not detailed enough for accurate mass estimates. Nevertheless, the SET conducted a mass estimate exercise by scaling similar components from JWST, using structural analysis to provide mass predictions where models are available, and incorporating study information from LUVOIR. While this is not a true bottoms-up estimate, it provides a rough order of magnitude for mass comparison against launch vehicle lift capabilities. The EAC1 mass estimate is approximately 23,000 kg, while EAC2 and EAC3 are around 36,000 kg, with each major element (OTE, IGA, SCE, and MBA/DBS) exceeding 5,000 kg. More detailed work, including a comprehensive Mass Equipment List (MEL), will be performed to ensure adequate mass margin is maintained against launch vehicle lift capabilities.

b. Similarly, a rough power estimate was generated based on telescope heater power, segment control hardware (laser metrology, edge sensors, and actuators for 19 PM segments), worst-case assumptions for heating the entire baffle, a CI estimate, LUVOIR information for other astrophysics instruments, and scaled spacecraft information from Roman. The total with 25% margin is approximately 16 kW. This estimate is primarily intended to ensure adequate solar array sizing. A detailed Power Equipment List (PEL) will also be created for future EACs to ensure adequate power margin.

## 3   EAC1 System Modeling

After developing the EAC1 design, the SET effort was divided into two parallel tasks: the design team focused on maturing EAC2 and EAC3 designs and performing dedicated design analyses, while the system analysis team developed physics-based models to support the integrated modeling (IM) pipeline, which enables system-level performance assessments. Similar to JWST and Roman, HWO will rely on IM for final verification prior to launch, as end-to-end system testing to the required picometer stability will be impossible on the ground. Hence this IM process will stand for the duration of the project lifecycle with increasing fidelity as designs are refined and sub-system models are validated against test data.

The HWO TMPO recognizes the importance of establishing the IM pipeline during the pre-formulation phase as an initial investment rather than risking mission-critical schedule delays by building this capability during the implementation phase. More importantly, many of the JWST and Roman experienced analysts are still available and willing to transfer their knowledge and expertise to the next generation of HWO engineers. This is the right time to develop and extend the IM pipeline and provide an even stronger bridge between science and engineering analysis, allowing more efficient requirement flow-down from science and optical metrics to other engineering disciplines such as thermal, mechanical, control systems, etc.

The first IM cycle for HWO using EAC1 provided the opportunity to "pipeclean" the process, which entailed creating and verifying each of the discipline models and their respective data interfaces in the various threads of the pipeline. The EAC1 models, while very detailed for this mission phase, do not incorporate all features required for high-fidelity performance verification. However, they are powerful tools for understanding design sensitivities and providing comparative studies. Through this exercise, we are encouraged to demonstrate that the results are within reach of our stability goals, though the SET recognizes that additional error sources and imperfections need to be incorporated into future analyses. As expected, significant challenging work remains for system analysis, but the team has established a strong foundation to move forward.

### 3.1 Introduction to HWO Integrated Modeling Pipeline

Integrated Modeling (IM) on HWO builds upon the capabilities successfully demonstrated on JWST[14] and Roman.[15,16] IM is the analytical process by which multi-disciplinary engineering models of the Observatory system are exercised in a variety of sequences depending on the end-to-end performance to be estimated; this is what we refer to as the multi-thread, cross-disciplinary system modeling pipeline. Each discipline step in the thread requires a discipline specific model which can be exercised using either commercial or custom developed analysis tools. All disciplines models are consistent with the mechanical geometry, the optical design, and the coordinate systems defined in the computer-aided design (CAD) model. Component and data interfaces between the models are strictly defined to enable automated data flow through the pipeline across different analysis teams and without the need for users in the loop. The main branches of the HWO pipeline, as shown in Fig. 54, are the ACS / LOS / Jitter thread at high temporal frequencies and the thermal drift thread at low temporal frequencies, also commonly referred to as the Structures, Thermal, Optical Performance (STOP) thread. Both threads can then be fed through the wavefront sensing control and diffraction analyses to estimate nonlinear coronagraph e-fields, raw contrast, and contrast stability which are compared to the top-level coronagraph requirements in the FRN budget (see Sec. 1.2).

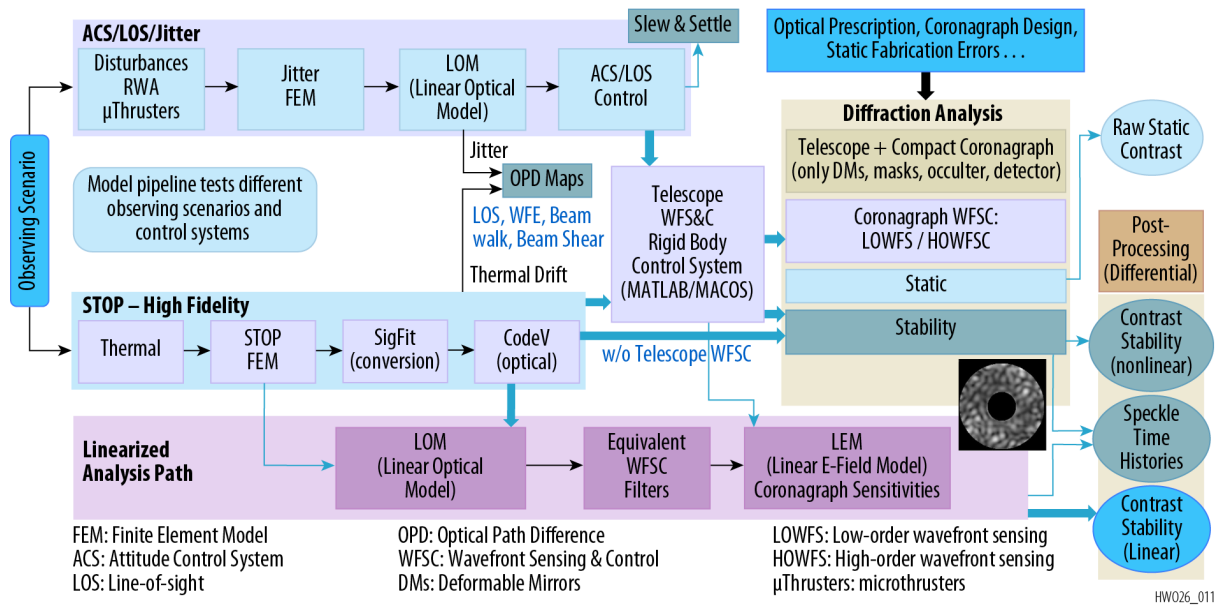

**Fig. 54** Integrated Modeling Pipeline

The ACS/Jitter thread is exercised by inputting ACS mechanism disturbance models, which for HWO are still under investigation and include RWAs and micro-thrusters. The disturbances are injected into a state-space dynamics model derived from the modes and mode shapes of the FEM. The outputs include six degrees-of-freedom rigid body motion of all optics and surface error on the primary and secondary mirror to capture bending-induced figure errors. Coupling the dynamics analysis outputs with the linear optical model (LOM) generates engineering performance metrics that support error budget development: LOS, WFE, beamwalk, pupil shear, and optical path difference (OPD) maps. This process is combined with the ACS/LOS simulation to evaluate slew/settle performance and feeds through the optical control system, such as the FSM, to be included in the diffraction analysis assessments.

For STOP, observing scenarios (OS) define the sequence of roll and pitch maneuvers of the observatory which affect the thermal environment and stability, as described in Sec. 1.2. A Thermal Desktop analysis is performed to evaluate temperatures throughout the observatory either for a full transient analysis of the entire OS or a pseudo-steady state analysis on informed field of regard points for fast turnaround trade studies. A link is established between the thermal model nodes and structural model nodes, and, using this link, temperatures are mapped from the thermal analysis to the structural model nodes for the subsequent structural analysis. The structural run uses thermal loading from the mapped nodes and is analyzed via NASTRAN, resulting in displacements representing all the optics in the system, similar to jitter outputs, – rigid body motion for all optics and detailed surface motion (at this point in time) for the primary and secondary mirrors. From the structural analysis output, there are two parallel path options for optical analysis: One path is continued MATLAB processing of the data using the LOM. The other parallel path is performing a full ray-trace analysis within Code V. Both paths can output WFE, LOS, beamwalk, pupil motion, and OPDs.

Following the jitter and STOP threads, the analysis includes the application of various control systems, optical diffraction, and the coronagraph. The different control architectures include combinations of rigid body telescope segment correction, FSM LOS correction and deformable mirror figure correction, which are described in more details in Sec. 4.3.3 and Sec. 4.3.4. For the initial IM pipeline, the telescope Rigid Body Control System (RBCS) takes the uncorrected rigid body motion of the primary mirror segments and the secondary mirror and runs the data through a control model that includes 4 pm rms per leg of metrology noise. This model applies corrective actuation for each optic and returns residual rigid body motion that is re-applied to the optics in the form of surface error for the primary mirror segments, to be passed on to the diffraction model.

The last model in the pipeline, diffraction and coronagraph using PROPER, produces a set of images called a speckle time series and calculates the change in the mean normalized intensity for a variety of annular regions around the target star. The goal is to achieve a sufficiently low mean normalized intensity to observe an Earth-sized planet around its star within the habitable zone.

Because these analyses are nonlinear and computationally intensive, a linearized analysis path can be exercised for rapid turnaround of design trades. The IM pipeline was initially developed for the coronagraph channel; for other HWO instrument channels IM will use essentially the same observatory models and analysis threads as for the coronagraph but will propagate through each instrument's individual optical path to evaluate instrument specific metrics at the detectors such as Encircled Energy (EE), EE Stability, and Strehl Ratio.

Prior to generating performance results, the models go through a rigorous verification process to ensure that the pipeline is computationally ready to generate end-to-end simulations, that there are no inconsistencies across the disciplines and interfaces, and that the models generate the expected results for a pre-determined set of analysis cases.

*3.2 Model Descriptions*

Examples of the various EAC1 discipline models are shown below in Fig. 55. While each discipline model appears to be different, they are in fact all consistent with the baseline geometry defined in the CAD and Optics models; they are also consistent with the overall materials and properties definition. However, only components which impact the engineering discipline performance are modeled in each step of the pipeline. For instance, while the thermal model includes the baffle, spacecraft, and sunshade (all components which drive thermal stability), the STOP model only uses temperatures along the telescope and instrument payloads to determine the stability of the alignment and wavefronts.

Sections 3.2.1, 3.2.2, and 3.2.3 below cover the optical, structural, and thermal models, respectively, which form the backbone of the IM STOP analysis process. Additionally, Sec. 3.2.4 describes the optical control architecture and modeling approach, necessary given the extensive control necessary to achieve HWO stability. The appropriate sections in Sec. 4 provide detail on other models, such as ACS, jitter, and diffraction used for specific types of analysis.

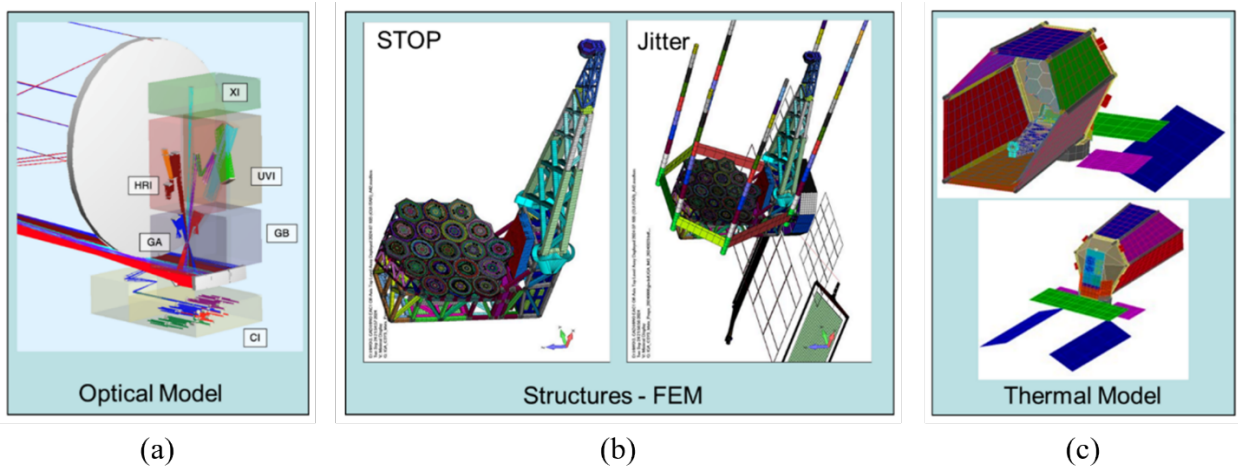

<div align="center">(a)        (b)        (c)</div>

**Fig. 55** Graphical Representation of the various Discipline Models Stitched together in the HWO Integrated Modeling Pipeline: (a) Optical Model, (b) Structural Model, and (c) Thermal Model.

### 3.2.1 Optical Model: Code V and Linear Optical Model

The EAC1 optical design is modeled in Code V software, including surface geometries, materials, and aperture definitions, to perform ray tracing, aberration analysis, and image quality performance evaluation. The optical models are used to support two parallel STOP analyses – one through the Linear Optical Model (LOM) and another, through raytracing (RT). The LOM is a MATLAB structure and provides a quick method for trade studies and optical performance predictions.

A set of linear optical models (LOM), generated from the nominal optical designs using rigid body motion sensitivities of all the optics in the imaging path through the CI, are implemented for system level image quality analysis; also included are component figure shape sensitivities for the optics in the telescope. The LOMs output image motion and wavefront error for a given set of rigid body motions and figure change inputs.[17] The output wavefront error maps can be fit to polynomials, such as Zernike circular polynomials,[18] to further parameterize the image quality degradation.

OPD maps generated by either the LOM or the RT approach are stored and passed through a full diffraction analysis to obtain contrast results. The diffraction analysis is a slower process and is primarily used for more accurate predictions. Alternatively, outputs from the LOM or RT can be parametrized in statistical terms including mean, delta-mean, variance, and delta-variance RMS WFE, which can be converted to contrast results using linear sensitivities. This approach is routinely used for trade studies and error budget allocations.

*3.2.2  Structural Model*

This section will describe the FEMs used to support integraged modeling analyses. Depending on criticality to optical performance, and drawing from recent experience on JWST and Roman, the FEMs vary in detail and expended effort.

Before discussing component modeling, it is prudent to present the overarching approach to setting system frequency goals; this discussion can be split into spacecraft and telescope:

- In general, the telescope was designed to be as stiff as possible (i.e. with the highest natural frequency), to make the payload vibration isolation system as effective as possible. Further, the telescope modes are strategically spaced in frequency; having telescope vibration modes that do not significantly couple helps to simplify understanding of the micro-vibration in the frequency domain. The lowest resonance, as expected, is the 15 meter long SMSS (10.74 Hz). This is followed by resonances of the PM as a rigid body (20 Hz and up). Next are the first flexible modes of the deploying "wings" at about 40 Hz. Finally, the lowest mirror segment frequencies, the "tip/tilt" modes, start at approximately 50 Hz.

- On the Spacecraft side, the vibration isolation system lessens the criticality of many of the spacecraft resonant frequencies. The most critical spacecraft side resonances are the low frequency deployable structures due to their influence on the ACS slew/settle time.

The remainder of this section goes into more detail for some of the key subsystems.

<u>Spacecraft + Deployables + Micro-thrusters + RWA</u>: Less modeling precision was used on the Spacecraft and deployable models compared to the RWA. In general, the Spacecraft was modeled as realistic with respect to expected mass and required size and focused on a feasible structure that could withstand the expected launch vibration environment (see the Spacecraft side FEM in Fig. 56 below). The deployables were matched to required size estimates with the analysis aimed at providing a reasonable prediction of natural frequencies. The DBS on EAC1, quite large and cantilevered off the Spacecraft, presents the lowest first mode prediction at just 0.046 Hz; this lowest deployed system frequency is critical for the ACS system design.

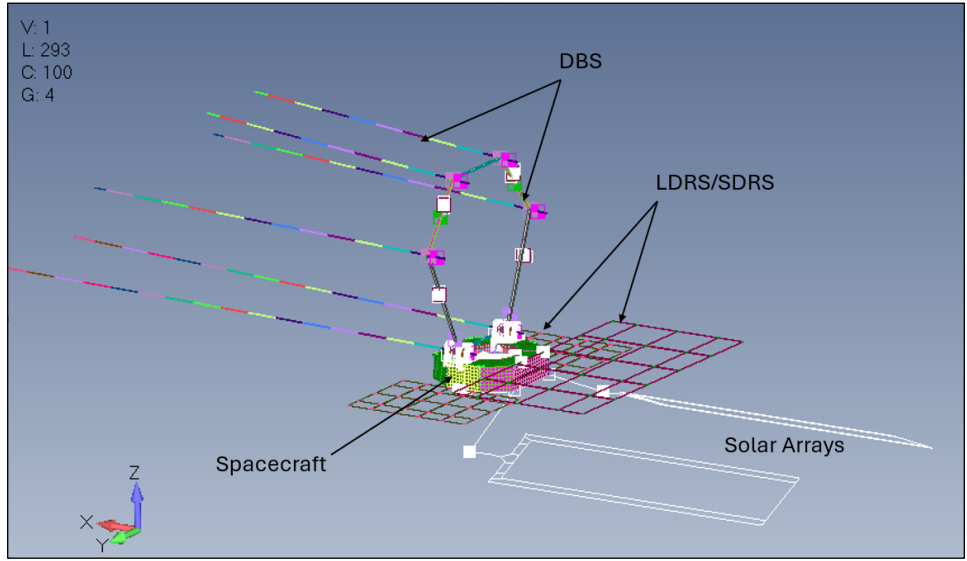

**Fig. 56** Spacecraft and Associated Deployables.

The RWA model (eight reaction wheels with associated isolation systems) borrows from Roman as an appropriately detailed model is critical for accurate micro-vibration analysis. Passive isolation was applied to each wheel and sized to achieve isolation system frequencies (six per wheel) between 3.5 and 6.5 Hz. This baseline frequency range is varied in trade studies and is based on Roman and JWST experience. The Micro-thrusters required little specific FEM modeling other than identifying mass and location in the model.

<u>Payload Isolation System</u>: The EAC1 architecture is in many ways similar to JWST, which used a passive isolation system between the spacecraft and payload. Rather than expending resources with design of an actual EAC1 isolation system, modelers used an oversimplified "spring" model where with an appropriate stiffness simply tuned to achieve system isolation frequencies between 0.5 and 1.0 Hz. This simple representation, although challenging to achieve in hardware, enables trade studies in model space for the isolation system.

<u>PMSA + PMBA</u>: Fig. 57 shows the PMSA model(s). The mirror substrate itself has a rather high first mode (in the free-free boundary condition) with a goal of achieving 400 Hz. The mirror, including reaction structure and supporting actuator struts, has a predicted first mode (fixed base boundary conditions) of 50 Hz; the 50 Hz modes (called segment "tip/tilt" modes) are expected to show up in subsequent jitter analyses. Extensive effort was expended on accurately modeling not only the mirror segment itself (ULE glass with precise thermal expansion coefficients), but also modeling the structure connecting the PMSAs to the backplane (RTV mount pads, invar mounts, reaction structure, actuator struts with flexure ends). The PMSA model is likely the most critical model for STOP assessment.

Figure 58 shows the PMBA model. It requires significant modeling detail to capture the local behavior of such a structure. For example, the bonded joints require high-fidelity to capture through-the-thickness behavior. The EAC1 model is simplified, being composed of primarily of beam and plate elements, with only a single layer of solid elements representing the RTV bond line. Nonetheless, it is appropriate for EAC cycle analysis, as it will capture global structural behavior such as mirror response to bulk temperature gradients and mirror on mount modes.

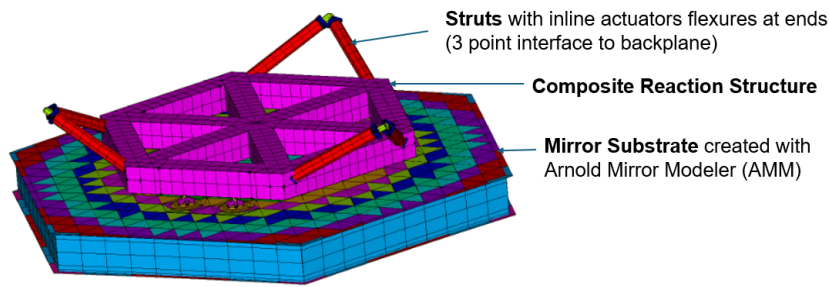

**Fig. 57** Primary Mirror Segment Assembly (PMSA) FEM.

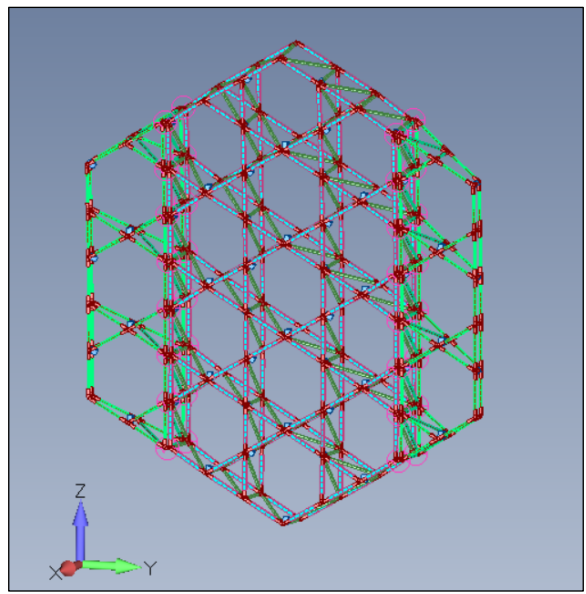

**Fig. 58** Primary Mirror Backplane Assembly (PMBA) FEM.

IMS: The Instrument Module Structure serves as the primary structural component of the telescope. It holds the PMBA, supports the deployed SMSS, encloses the instruments, and is the primary interface to the Spacecraft. The IMS stiffness and strength are important for both launch (stowed) and for operational (deployed) configurations. Similar to the JWST PMBSS structure, it is locked to the Spacecraft for launch but floats on a very soft passive isolation system in operation. Figure 59 shows the IMS FEM.

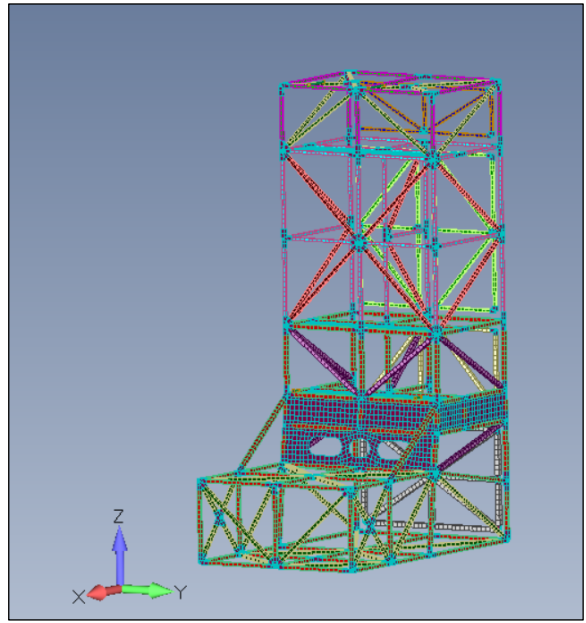

**Fig. 59** Instrument Module Structure (IMS) FEM.

<u>SMSS + Secondary Mirrors</u>: The 15 meter long Secondary Mirror Support Structure (SMSS) supports the Secondary Mirror Assembly (SMA). Significant effort was invested in designing and analyzing an optimized truss structure with a minimum goal of 10 Hz; the SMSS presents the lowest natural frequency (10.7 Hz) on the telescope, as shown in Fig. 60.

Detailed models were constructed for the SM (located at the end of SMSS), M3, and M4 (mounted in the AOS inside of the IMS). Similar to the PMSA model, extensive effort was undertaken to model the ULE glass, RTV Mount Pads, Invar mounts, and Titanium flexures. Figure 61 shows these detailed mirror models.

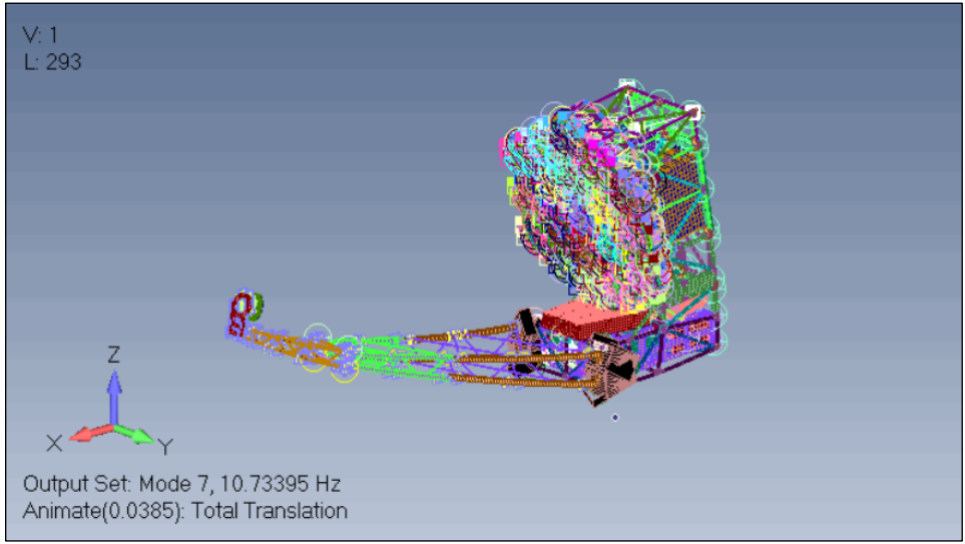

**Fig. 60** Lowest Telescope Side Vibration Mode at 10.7 Hz.

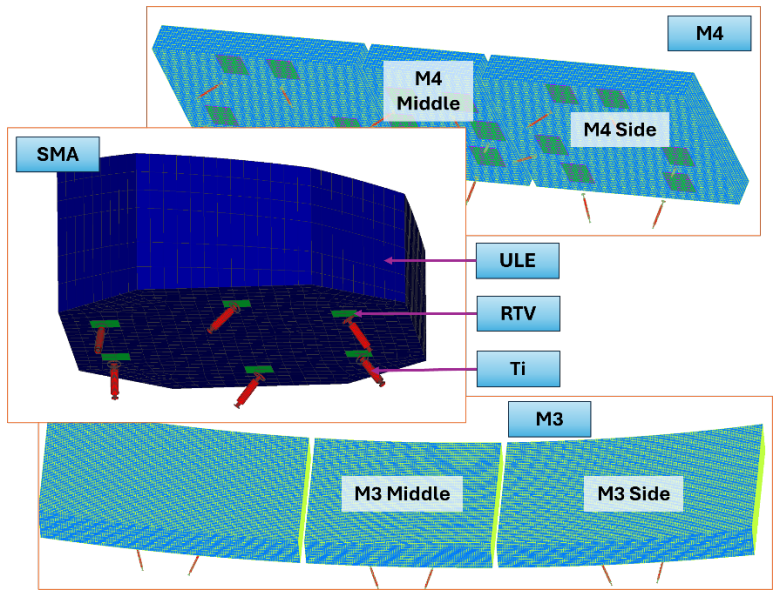

**Fig. 61** FEM of the SMA, M3, and M4.

<u>CI</u>: The EAC1 coronograph model (Fig. 62) mounts to the inside of the IMS using a kinematic 3-2-1 constraint system reminiscent of the HST instrument "A-B-C" latch system. This kinematic interface, while minimizing thermo-elastic interaction with the IMS, is a contributor to a lower than desired first natural frequency of 15.7 Hz; future EAC CI designs are expected to have much higher frequencies as the design concept matures.

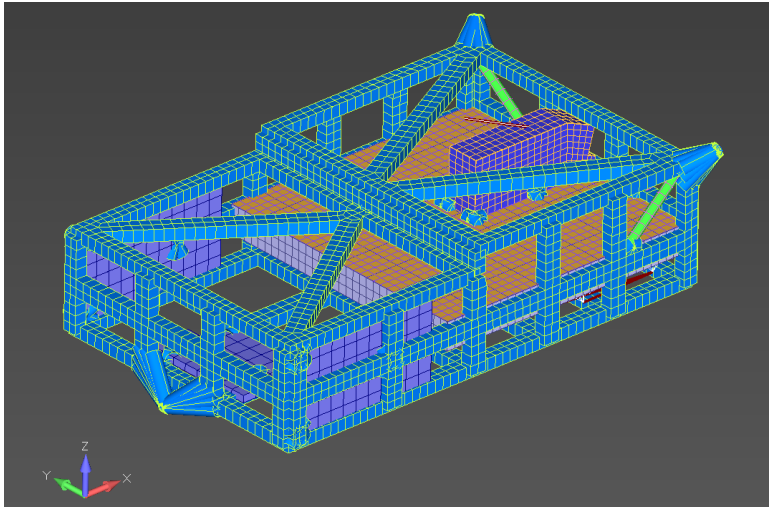

**Fig. 62** Coronagraph Instrument (CI) FEM.

### 3.2.3  Thermal Model

The EAC1 detailed thermal model was developed collaboratively with the mechanical team (CAD model) and the structures team (structural FEM). For key observatory subsystems, particularly the telescope, the thermal model was specifically designed with the structural FEM in mind to ensure seamless temperature mapping; all other observatory components were developed directly from the CAD model without specific FEM mapping requirements.

Thermal model development followed an iterative approach, with refinements driven predominantly by evaluation of thermal model output, temperature mapping onto the structural FEM, and downstream optical results from the IM pipeline. Targeted sensitivity studies further identified a number of model improvements as well as uncovered potential modeling errors; unlike many prior programs, HWO's stringent stability requirements mean even minor thermal shorts and radiative surface overlaps could significantly impact overall performance. This iterative process not only generated valuable lessons learned but also contributed to the development of broader modeling guidelines and standardized processes for future modeling efforts.

Several key model enhancements of note from this iterative process included:

- Baffle optimization: Remodeled baffle-related components to ensure light-tight performance
- Thermal protection: Added insulation to prevent solar heat entrapment
- Radiation analysis refinement: Created distinct radiation analysis groups to eliminate unintended solar impingement

- Geometric simplification: Streamlined CAD geometry (including removal of launch struts)
- Assumption updates: Modified key parameters, including MLI emissivity ($\varepsilon^*$) and optical properties

The model inherits thermophysical and optical properties from the Roman database; the only exceptions are those provided by the coronagraph team or by glass vendors, specifically for use on the PM segments. The model implements temperature-dependent properties for materials used across a broad range of temperature environments; for example, M55J is used throughout the payload on the PMB, IMS, and SMSS, each of which operate at different temperature ranges. The ULE used for the PM segments reflects one notable exception–as the system design locks the PM segments into a narrow temperature band, temperature-dependent properties are not necessary. For reference, the ULE thermophysical property assumptions are:

- Conductivity k = 1.31 W/m/K
- Specific Heat cp = 767 J/kg/K
- Density $\rho$ = 2,210 kg/m3

The most important thermal property assumptions are those used to model the PM's radiative environment; the deployable baffle, heater plate control surfaces that surround the mirror segments, and the mirror properties themselves represent the driving components, in this context. For critical surfaces, identified in Table 7, properties are defined for emissivity, transmissivity, and reflectivity in both IR and UC spectrums, where appropriate.

**Table 7** Key Optical Properties Locations.

| Optical Property Name | Section | Location |
|---|---|---|
| HWO_BlkKapton | | Inner Payload-Facing Layer |
| HWO_VDA_200K | Deployable Baffle | Middle Baffle Layers |
| HWO_MBA_OuterLayer | | Outer Space-Facing Layers |
| PMSA_EAC1_emiss_03 | PM Segments | Front Face |
| PMSA_EAC1_ULE_Corning | | Back Face, Side, Interior |
| PMSA_EAC1_emiss_80 | PM Heater Plates | Optics-Facing Bottom Plate |
| HWO_BlkKapton | | Optics-Facing Side Plate |

Figure 63 shows the cavities in which the PM segments are nestled; these heater plates are control surfaces, each with multiple control zones behind each segment. There are also Heater Plates surrounding the outer perimeter of the segment assembly, these eliminate the large gradient formed in the outer segments caused by a direct view between the segment sides and the MBA and the MBA containment boxes. Not shown in the figure, the PM backplane is modeled as composite beams and metallic fittings with geometry defined by symbols so dimensions can be modified easily without impacting alignment. Figure 64 shows these PM control surfaces within the context of the deployable baffle and greater EAC1 architecture.

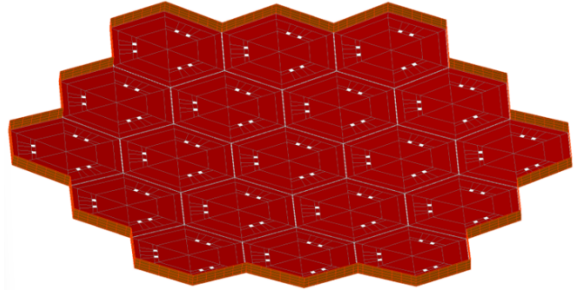

**Fig. 63** PM Radiative Bottom & Side / Perimeter Control Surfaces.

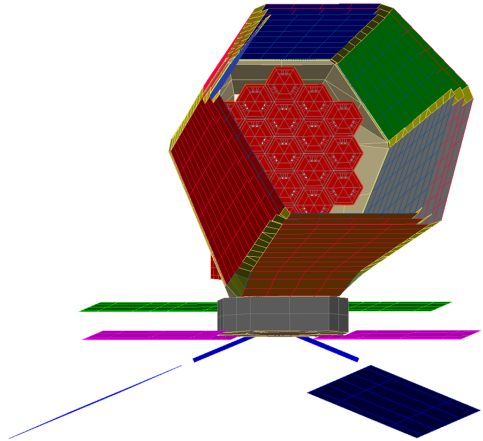

**Fig. 64** PM Control Surfaces & Deployable Baffle Within EAC1 Model.

Figure 65 shows the IGA thermal model with callouts for materials/properties. The IMS is similar to the PM backplane with modeled composite beams and fittings. The instruments are blanketed and held to operational temperatures; Fig. 66 shows the CI thermal model in more detail. Of note, given its importance, the CI model went through a thorough mass scrub to ensure alignment with the MEL distribution.

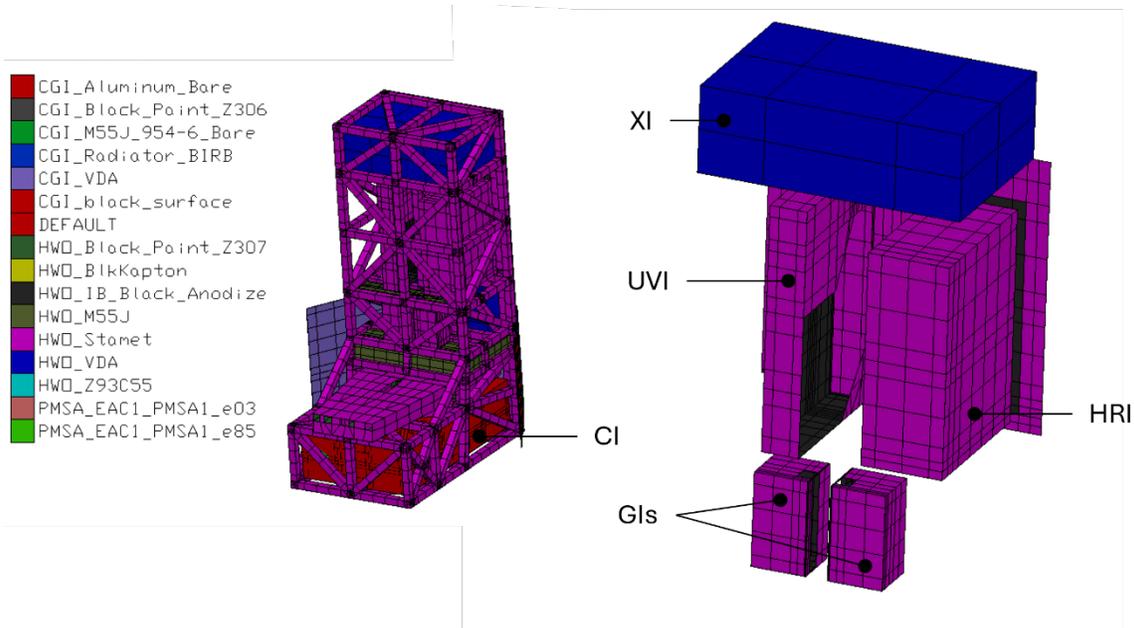

**Fig. 65** IGA Thermal Model Overview.

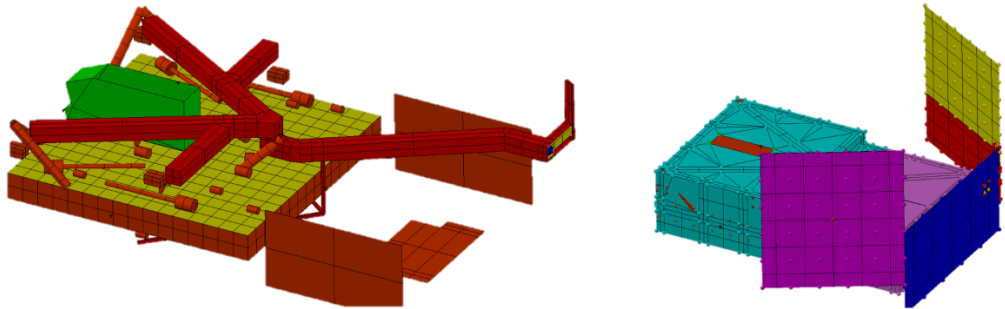

**Fig. 66** CI Thermal Model Overview.

Given the model's importance within the IM STOP analysis process, the model was carefully configured for easy and reliable reconfiguration to support the full suite of cases needed by IM; this was done predominantly by implementing control parameters, easily modified without changing model geometry. Further, the team implemented a number of recent lessons learned from Roman by including several specialized tools to aid post-processing and mapping (see Sec. 4.3.1 for more detail).

### 3.2.4 Optical Control Architecture and Modeling Approach

In order to achieve the picometer-level wavefront stability necessary for coronagraph science objectives, multiple layers of optical control loops must be applied to suppress line-of-sight (LOS) and wavefront (FW) errors to necessary levels, as defined by the error budgets. Three types of optical control are considered, as shown in Fig. 67: alignment control of the PM segments and SM (as represented by green), LOS control of the telescope and of the coronagraph instrument (as shown by magenta), and wavefront control using the Deformable Mirrors (DMs) within the coronagraph (as shown in blue). Each type of control has the potential for nested, inner-control loops to take advantage of higher temporal bandwidths and better rejection of thermally induced errors, that work in concert with the outer-most loops with more direct sensing to correct for non-common path errors on longer timescales. For reference, Fig. 67 also notes the control system temporal bandwidth range and spatial frequency correction capability.

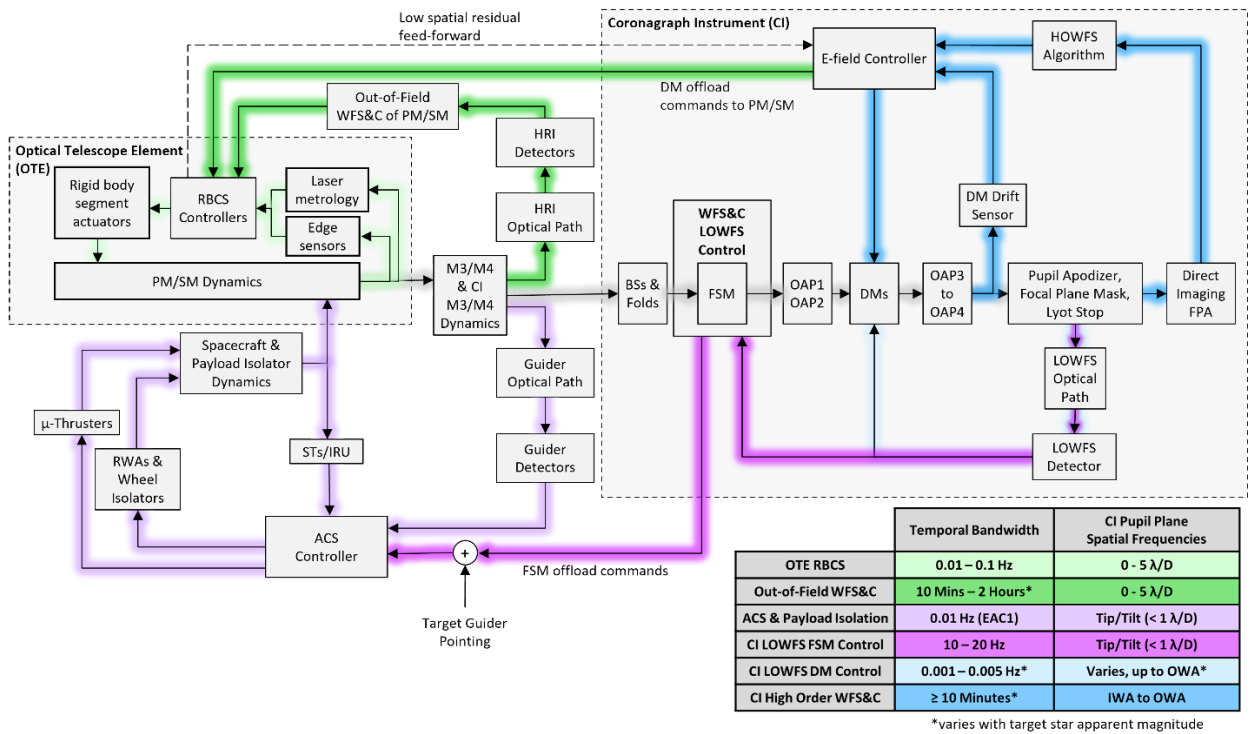

**Fig. 67** Coronagraph and Telescope Optical Control Block Diagram.

As described in Sec. 0, the EAC1 design contains a PM consisting of 19 hexagonal segments and a monolithic SM, with each of the mirrors capable of actuation in all 6 rigid-body degrees-of-freedom. The Rigid Body Control System (RBCS) uses laser distance gauges and edge sensors to measure the relative alignments of the Primary Mirror segments and Secondary Mirror and is responsible for maintaining mirror alignments (the commanded pose) relative to the optical path through the M3 mirror, in the presence of thermally driven deformations of the telescope structure. A more detailed description of the RBCS, and its performance modeling, is covered by Tesch et al.[19] Initial estimates of the optimal pose may be determined during telescope initialization, with refinements derived through outer-loop control, based either on imaging from other instruments within the observatory, such as HRI, or on feedback from the CI's high-order wavefront sensing. Further investigations of these outer-loop methods are a topic for a future paper.

Observatory LOS is primarily controlled by ACS with additional correction from the Fast Steering Mirror (FSM) in each instrument channel. During observation, ACS uses guide stars on dedicated Guider Instruments to provide pointing measurements. The residual pointing stability error after ACS corrections, combined with pointing knowledge errors of the CI LOS relative to the guiding instrument, must be low enough to ensure placement of the coronagraph target star within the capture range of the Low-Order Wavefront Sensor (LOWFS). In addition to lower-temporal-bandwidth wavefront sensing, the LOWFS also generates high-bandwidth pointing feedback for the coronagraph's FSM, which centers the target star on the coronagraph's focal plane mask. The closed-loop bandwidth of the FSM control is at least 10 Hz, to provide adequate rejection of thermally driven pointing errors, as well as low-frequency structural jitter. Knowledge of FSM corrections is provided to the ACS as an offload command, effectively shifting the desired registration of the guide star field on the guiding instrument; this preserves FSM mechanism range and reduces beamwalk effects within the bandwidth of the ACS loop. LOS stability errors outside of the ACS or ACS + FSM control bandwidths are mitigated by vibration isolations systems, which are described in Section 4.2.

High spatial-frequency wavefront errors are corrected through control of the two Deformable Mirrors (DMs) in each coronagraph channel. After establishing sufficiently low raw contrast through initial Electric Field Conjugation (EFC) iterations, contrast stability will be maintained during target observations through a combination of two control loops. First, LOWFS is primarily responsible for rejection of thermally driven optic surface figure changes due to observatory pointing maneuvers; a possible LOWFS implementation based on Zernike wavefront sensing is described and analyzed in more detail in Section 4.3.2 and in Tesch et al.[19] Second, high-order wavefront sensing (HOWFS) is responsible for correcting residual wavefront errors at spatial frequencies beyond what LOWFS can detect, out to the Outer Working Angle (OWA) of the coronagraph; these errors arise from sources such as beamwalk, optic surface roughness, machining imperfections, and non-common path errors after the LOWFS optical pick-off. One possible implementation of HOWFS would be the Dark Zone Maintenance (DZM) algorithm, as described by Redmond et al.[2]

When modeling optical control in response to disturbances from predicted thermal and structural environments, control loops that can couple with the environmental model, or with other loops, are modeled jointly with those models. For example, the time domain jitter analysis, described in Sec. 4.2 includes models of the ACS, telescope RBCS, and CI FSM control (though currently omits the temporally-slower WFS&C controls with bandwidth significantly below the first flexible-body mode of the observatory). When control loops are not expected to couple with the environment or with other loops, these loops are modeled separately and applied to simulated observation time-series sequentially. For structural-thermal-optical simulations as described in Section 4.3, using the pipeline as described in Section 3.1, the RBCS inner-loop control model is applied first, since the local PM and SM metrology sensors are not sensitive to DM or FSM actuations inside the coronagraph, followed by pointing control corrections which are largely insensitive to DM actuations due to temporal bandwidth separations, and lastly LOWFS control which is sensitive to all other control loops. This sequential approach both simplifies model implementation and provides insight into the performance benefits of each loop. Additional implementation details are covered in Section 4, and by Tesch et al.[19]

## 4 EAC1 System Analysis

### 4.1 Attitude Control System Analysis

The EAC1 IM effort includes implementation of each ACS architecture option in a time-domain ACS performance simulation. Each architecture is exercised using Coronagraph Observing Scenario 1 (OS-1), which contains repeating ±45 deg roll maneuvers followed by 1.75 hour observation periods at the target star 47 Ursae Majoris. ACS performance analysis simulates a single roll maneuver and observation sequence to 1) quantify the ACS-level pointing stability at the FGS, 2) evaluate observatory agility (the time to reach steady-state pointing stability), and 3) identify the driving ACS error sources to inform pointing error budgets. The time-efficient ACS simulation is used to establish an operational baseline for each ACS architecture and investigate any interplay with high-frequency jitter mitigation strategies. For the RWA Only ACS Architecture, simulation results inform the frequency domain jitter analysis on the RWA speed range required to execute the OS-1 scenario.

ACS pointing stability is evaluated at the FGS sensor location, before any instrument level compensation. The working "desirement" for ACS-level pointing stability is < 4 mas, 1σ over the observation period, as tabulated in Sec. 1.2 (this is further reduced to < 0.1 mas, 1σ by the CI FSM, which is assessed in the ACS time-domain jitter simulation). Observatory agility is evaluated as the time required to slew to a new target and reach the desired ACS pointing stability, as evaluated over a shorter time window determined by the ACS bandwidth (currently set to 3 minutes). The ACS pointing stability is then evaluated as the 1σ pointing stability over the entire observation duration (1.75 hours for OS-1).

### 4.1.1 ACS Model Descriptions

#### 4.1.1.1 ACS Observatory Dynamics Model

The ACS time-domain simulation employs a state space observatory dynamics model derived from the EAC1 FEM, reduced to capture ACS relevant flexible modes (300 modes up to 10 Hz). The lowest frequency modes include the 0.046 Hz MBA and 0.063 Hz solar arrays, which are modeled with 0.25% structural damping. Micro-thruster and RWA input grid locations reflect the baseline configuration for EAC1. Motion at the IRU output grid location is used as a reference for FGS motion for this analysis. The dynamics model includes the dual passive isolation configuration: 0.5-1.0 Hz (Payload) and 3-6.5 Hz (RWA). Additional isolation configurations are assessed with separate observatory dynamics models. Figure 68 summarizes the EAC1 observatory mass properties.

$$I_{EAC1} = \begin{bmatrix} 366915 & 216 & -42195 \\ 216 & 786039 & -388 \\ -42195 & -388 & 608702 \end{bmatrix} kgm^2$$

$$m_{EAC1} = 23967 \text{ kg} \qquad CM_{EAC1} = \begin{bmatrix} 1.07 \\ 0 \\ 4.40 \end{bmatrix} m$$

**Fig. 68** EAC1 Observatory Mass Properties.

#### 4.1.1.2 Solar Radiation Pressure (SRP) Torque Disturbance Model

The ACS simulation uses an "N-plate" SRP model to compute the total SRP torque disturbance on the observatory. The N-plate model, shown in Fig. 69 represents the EAC1 observatory geometry as triangular plates with defined area, reflectivity, orientation, and location relative to the observatory CM. The torque contribution from each plate is then added to compute the net SRP torque for any Sun Roll angle and Sun Pitch angle in the observatory body frame, as defined in

Fig. 50. The operational sun constraints for EAC1 are set at ±45º Sun Pitch and ±22.5º Sun Roll. Surface reflectivity is set uniformly to 100% specular for EAC1 to conservatively bound the SRP torque magnitude. Small changes in SRP torque magnitude due to solar flux variation are also modeled; the variation is implemented as a 0.1% amplitude variation over a 5-minute period based on data from the VIRGO Experiment on the Solar and Heliospheric Observatory (SOHO).[20]

**N-plate model**

$$F_{SP} = -PA\cos\theta\left[(1-C_s)\hat{s} + 2(C_s\cos\theta + \tfrac{1}{3}C_d)\hat{n}\right]$$

where P — is the solar radiation momentum flux at L2
A — is the area of the spacecraft face to be analyzed
$C_s$ — is the surface coefficient of specular reflection
$C_d$ — is the surface coefficient of diffuse reflection
$\hat{s}$ — is the unit vector from the spacecraft to the sun
$\hat{n}$ — is the unit vector normal to the spacecraft face
$\theta$ — is the angle between $\hat{s}$ and $\hat{n}$

**Fig. 69** "N-Plate" Solar Radiation Pressure (SRP) Torque Disturbance Model.

### 4.1.1.3 Reaction Wheel Model

The ACS simulation implements a high-fidelity reaction wheel model with a maximum output torque of 0.2 Nm and a maximum stored angular momentum of 250 Nms at 6,000 rpm. The model includes key torque error sources such as wheel drive electronics command quantization, bearing drag effects (Coulomb, viscous, and Dahl friction with a LuGre formulation), and torque ripple and cogging. Static/dynamic imbalance and bearing eccentricity EFT tones are omitted here, but are included in the high-frequency time domain jitter simulation. The wheels are assumed to be capable of utilizing the full ±6,000 rpm range subject to sufficient bus voltage, indefinite operation at zero speed, and repeated power cycling without wheel health degradation. For micro-thruster pointing operations, various wheel management options are considered, include parking the wheels at low speed or powering them off entirely (the baseline operational approach for the Hybrid ACS Architecture on EAC1).

### 4.1.1.4 Micro-Thruster Model

The ACS simulation uses the Colloid Micro-Newton Thruster (CMNT) as a representative model for micro-thruster performance on EAC1. The CMNT is an electrospray micro-thruster developed by Busek Co. Inc. in collaboration with JPL and flown in 2015 on the LISA Pathfinder Space Technology 7 (ST7) demonstration.[21] CMNTs apply a high electric potential to conductive charged liquid propellant at the end of a hollow needle emitter to accelerate a beam of charged droplets and generate thrust, as illustrated in Fig. 70.

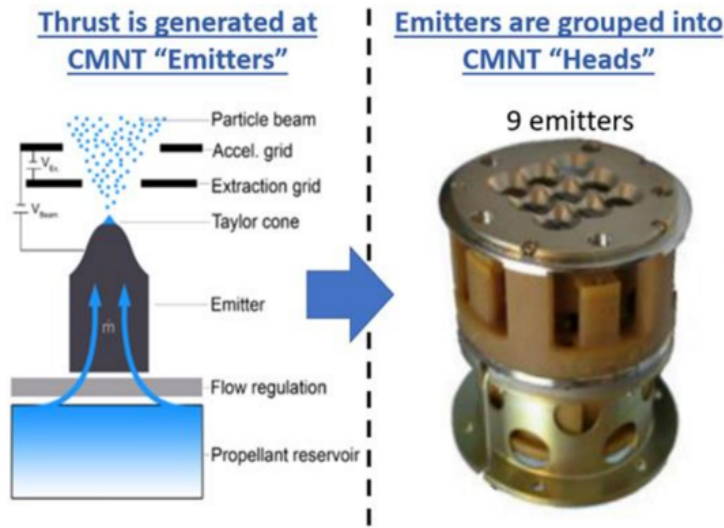

**Fig. 70** Busek Colloid Micro-Newton Thruster (CMNT) Emitter (left), CMNT Thruster Head with 9-emitter Configuration flown on LISA Pathfinder (right).

CMNT performance models for thrust and fuel consumption rate were validated on LISA Pathfinder over a wide range of conditions. CMNT thrust is modulated at 10 Hz by a Digital Control Interface Unit that adjusts beam voltage and current in response to thrust commands. The CMNTs on LISA Pathfinder operated over a thrust range of 0.5 to 3 μN per emitter, demonstrating ≤0.1 μN thrust resolution, ≤0.1 μN/$\sqrt{Hz}$ thrust noise, <10 sec full thrust range response time, and a nominal specific impulse of 240 sec. The propellant is a high-density liquid (1.53 kg/L) stored at low pressure (4 atm). Individual CMNT fuel tank / micro-thruster pairs are arranged as modular units across the observatory to accept ACS thrust commands. The CMNT plumes have a required half angle of <35 deg for 95% of the beam current, and typical performance of <23 deg. At higher thrust commands, the emitter particle beam widens and impinges on the electrodes, leading to potential electrical shorting and driving CMNT lifetime. Maximum thrust is therefore scaled by grouping multiple emitters into CMNT thruster heads; Fig. 70 shows the 9-emitter thruster head configuration flown on LISA Pathfinder. Each micro-thruster head on EAC1 is scaled to 700 emitters based on EAC1 SRP disturbance torque estimates (see Sec. 2.5.2).

The CMNT peaked at TRL 7 against LISA requirements when flown on LPF; its design has since been improved for the long lifetime required for NASA missions, such as HWO. "Colloid Thruster Life Testing and Modeling" has been funded through the 2022 NASA Strategic

Astrophysics Technology (SAT) Awards, which is conducting long duration CMNT hardware testing to validate the current design and lifetime models from TRL 4 to 5 for HWO and validate the thruster design and scalability to any thrust level.[22]

### 4.1.1.5 Stellar Reference Unit (SRU) and Inertial Reference Unit (IRU) Models

The SRU hardware model simulates the performance of the EAC1 star tracker assuming a three optical head configuration with a noise equivalent angle (NEA) of 1.7 arcsec RMS. The IRU hardware model simulates the performance of the IRU with an angle random walk of 0.0002 deg/$\sqrt{Hr}$. The STA and IRU measurements are both sampled at the ACS rate of 10 Hz and are used only during ACS slew maneuvers.

### 4.1.1.6 Fine Guidance Sensor (FGS) Model

The Fine Guidance Sensor (FGS) is a critical subsystem in space observatories, as it provides the high precision measurement of the telescope line-of-sight needed to achieve and maintain stable pointing. In this study, the FGS operates at fine angular resolutions (sub-mas) and high update rates (10 Hz), directly supporting the ACS control loop during the science observations.

The ACS simulation currently uses a simplistic additive noise model that is meant to conservatively bound the FGS measurement noise due to star centroiding error. This FGS noise model is currently set to bound a measured angular error of 2 mas, 1σ for observatory tip/tilt, and 100 mas, 1σ for observatory roll. Since FGS measurement noise is expected to drive the performance of the Hybrid ACS architecture, a higher fidelity model has been developed in parallel to further evaluate the performance of the EAC1 FGS configuration using known star fields at OS-1 observations. This simulation confirms that the noise equivalent angle assumptions (tip/tilt of 2 mas and roll of 100 mas) can be achieved. Future work will focus on evaluating more scenarios and exploratory cases, as well as directly integrating the high-fidelity FGS model within the ACS simulation and estimation.

### 4.1.2 ACS Modes and OS-1 Operational Timeline

Figure 71 shows the operational timeline for a typical target repoint using the Hybrid ACS Architecture. The timeline illustrates an example slew from a reference calibrations star to the target observation star. After the conclusion of the slew, the attitude control system begins coarse pointing to manage momentum and reduce error using reaction wheels. The system then hands over pointing control actuation from reaction wheels to micro-thrusters for fine pointing during observations. The RWA Only ACS Architecture's timeline is the same but will instead maintain reaction wheel control during fine pointing observations. Table 8 lists the ACS actuators and sensors used within each ACS mode.

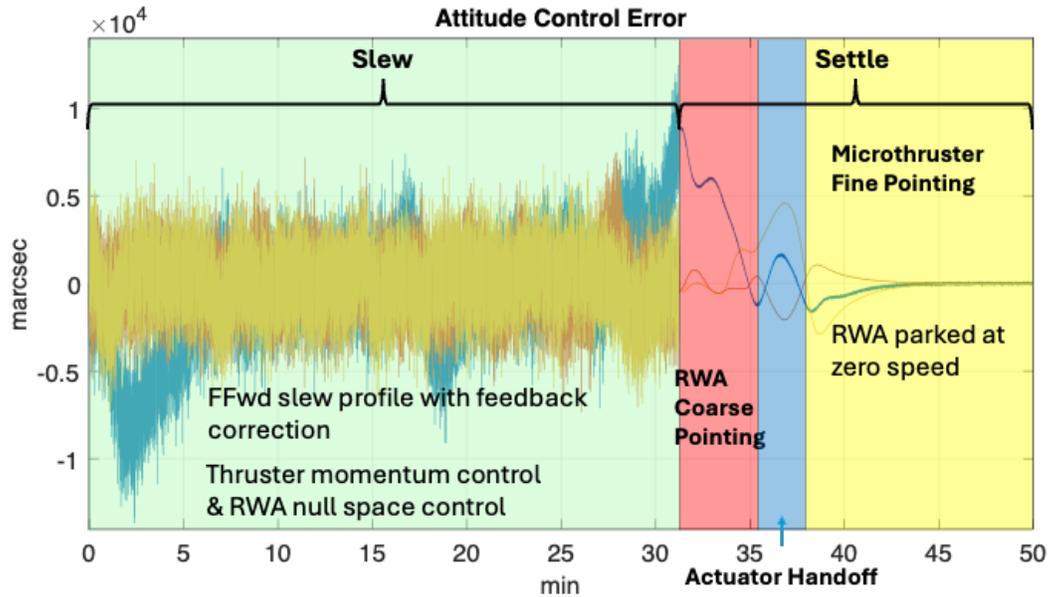

**Fig. 71** Example Target Repoint Operational Timeline – Hybrid ACS Architecture.

**Table 8** ACS Actuators and Sensors by Mode – Hybrid ACS Architecture.

| Mode | RWA | Micro-Thrusters | Star Tracker | IRU | Fine Guider |
|---|---|---|---|---|---|
| Slew | ✓ | ✓ | ✓ | ✓ | ✗ |
| Coarse Pointing | ✓ | ✓ | ✗ | ✗ | ✓ |
| Actuator Handoff | ✓ | ✓ | ✗ | ✗ | ✓ |
| Fine Pointing | ✗ | ✓ | ✗ | ✗ | ✓ |

The slew control system consists of a quaternion and rate trajectory tracking feedback controller following an onboard generated slew profile with reaction wheel feed forward torque commands. The slew profile shown in Fig. 72 was generated using a piecewise smooth $C^2$ function designed to avoid excitation of flexible modes by gently ramping up acceleration. The characteristics and duration of each segment of the piecewise function are determined by the slew angle, direction, and reaction wheel momentum and torque limitations. The duration of the ramp acceleration segment is designed to be at least twice the period of the lowest structural mode to avoid excitation.

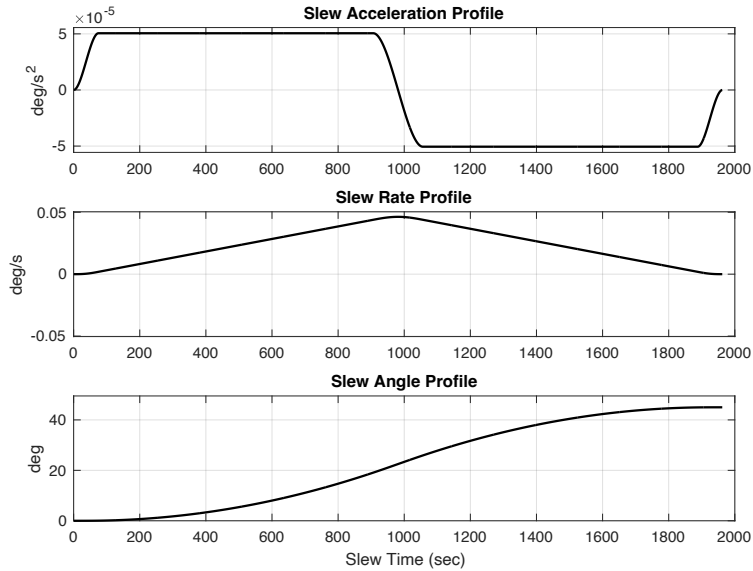

**Fig. 72** 45° Roll Slew Command Profile Completed in 32.7 Minutes.

Coarse pointing mode utilizes reaction wheels to reduce control error post slew maneuver while micro-thrusters continuously unload momentum from solar radiation pressure. A reaction wheel speed controller is used to bias the wheels into the null space for a zero-torque actuator handoff sequence. The actuator handoff maneuver is performed by ramping down all the reaction wheels simultaneously with the wheel speed controller along a null vector while control error actuation authority shifts from the wheels to the micro-thrusters. The goal is a zero-torque handoff to reduce excitations of flexible modes and minimize pointing stability settling time. The RWA ACS Only Architecture does not utilize the coarse pointing and actuator handoff maneuver and will instead transition into fine pointing post slew.

The actuator handoff maneuver is performed by ramping down all reaction wheels simultaneously with the wheel speed controller along a null vector while control error actuation authority shifts from the wheels to the micro-thrusters. The goal is a zero-torque handoff to reduce excitations of flexible modes and minimize pointing stability settling time.

During Hybrid ACS Architecture fine pointing mode, control error and continuous momentum unload is performed solely by micro-thrusters. The reaction wheels are spun down to zero speed to minimize jitter disturbance. RWA Only ACS Architecture biases the wheel speeds to avoid zero-speed crossings during fine pointing observations.

### 4.1.3 ACS Architecture #1: RWA only Performance Analysis

An example OS-1 roll maneuver and coronagraph target observation is simulated to demonstrate the operational timeline and evaluate the performance of the RWA Only ACS architecture. Slew and settling time is assessed, and pointing stability is evaluated over the 1.75-hour observation period. Figure 73 presents the results with all ACS error models enabled. The operational baseline assumes zero initial system momentum following an RCS thruster momentum unload prior to the slew, consistent with Roman coronagraph ConOps. The maneuver and observation are executed using reaction wheels, with wheel speeds biased during the momentum unload to avoid zero-speed crossings during the observation. The observations duration is bounded by wheel speed drift due to Z-axis solar radiation pressure torque at the high Sun roll angle.

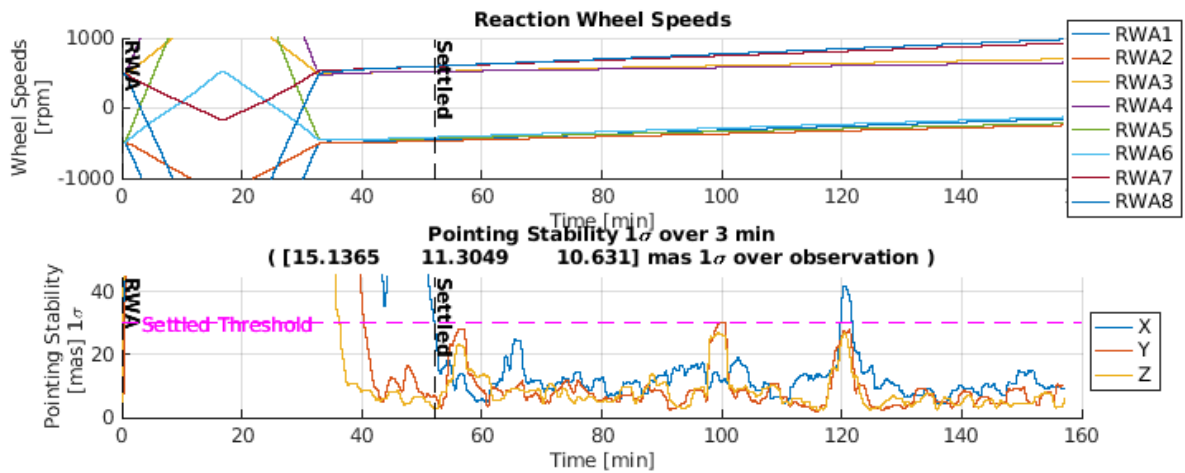

**Fig. 73** OS-1 Roll Maneuver and Target Observation, RWA Only ACS Architecture.

For the RWA Only ACS configuration, steady state pointing stability is 11 mas, 1σ in tip/tilt and 15 mas, 1σ in roll. The roll maneuver slew and settle time is 52 minutes for the baseline RWA sizing. Settling is defined when the pointing stability over 3 minutes reaches 30 mas, 1σ. However, a true steady state stability is not reached, as performance varies with changing wheel speeds, which drift at approximately 160 rpm per hour due to solar radiation pressure. The resulting wheel speed range is used to inform the high-frequency jitter analysis on the maximum wheel speed to analyze.

The scenario is rerun with individual ACS errors enabled to identify performance drivers, with stability and combined slew and settle time results summarized in Table 9. Steady state pointing stability is driven primarily by RWA error models, including tachometer and drag/cogging effects. Case A-F1 settle time is mainly driven by the torque limited slew, while Case F2-G shows that reaction wheel drag and cogging errors increase the settling time. This is mostly driven by wheel friction introducing additional slew control error and subsequently larger transients while exiting the slew maneuver. For the RWA Only ACS Architecture, performance is not limited by FGS noise.

**Table 9** Performance Contribution of Each ACS Error Source, RWA Only ACS Architecture.

| Case ID | ACS Errors Included | Stability (mas) | | | Settle Time (min) |
|---------|---------------------|------|------|------|------|
| | | X | Y | Z | |
| A | No Errors | 4.14 | 0.08 | 0.13 | 38.2 |
| B | SRP Variation | 4.14 | 0.10 | 0.33 | 38.2 |
| C | FGS | 10.40 | 0.22 | 0.22 | 38.4 |
| D | SRU/IRU | 3.95 | 0.27 | 0.20 | 38.3 |
| E | Micro-thruster | N/A | N/A | N/A | N/A |
| F1 | RWA (Tachometer) | 11.74 | 10.20 | 9.77 | 39.0 |
| F2 | RWA (Drag, Cogging) | 12.14 | 10.13 | 9.45 | 52.1 |
| G | All Errors | 15.14 | 11.30 | 10.63 | 52.1 |

*4.1.4  ACS Architecture #2: Hybrid (RWA and Micro-Thruster) Performance Analysis*

An example OS-1 roll maneuver and coronagraph target observation is simulated to demonstrate the operational timeline and evaluate the performance of the Hybrid ACS architecture. Slew and settling time is assessed and pointing stability is evaluated over the 1.75-hour observation period. Figure 74 presents the results with all ACS error models enabled. The operational baseline assumes system momentum is continuously unloaded using micro-thrusters. The maneuver is executed using reaction wheels, with a small bias applied to avoid zero-speed crossings during post-slew settling. The wheels are then spun down to zero speed as actuation is transitioned to the micro-thrusters.

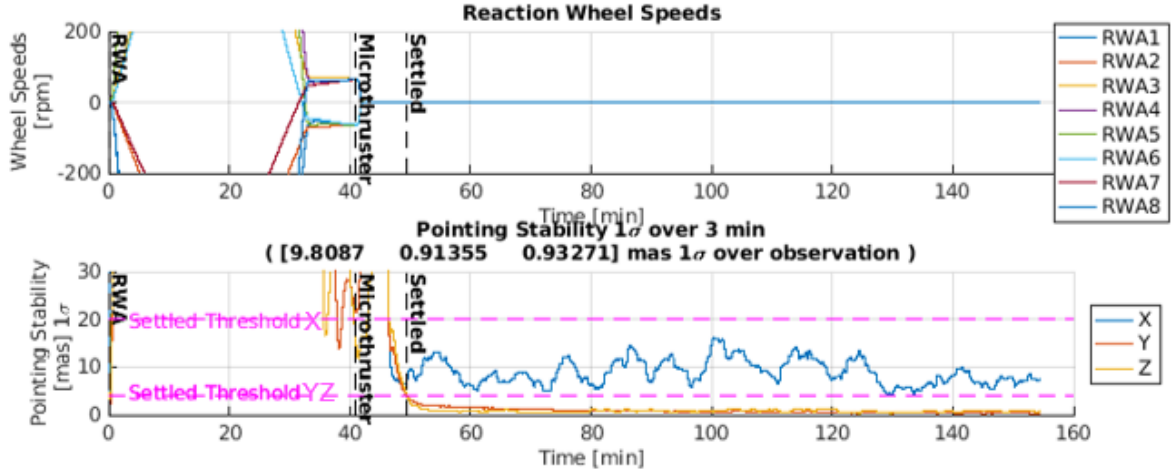

**Fig. 74** OS-1 Roll Maneuver and Target Observation, Hybrid ACS Architecture.

For the Hybrid ACS configuration, steady state pointing stability is <1 mas, 1σ in tip/tilt and <10 mas, 1σ in roll. The roll maneuver slew and settle time is 50 minutes for the baseline RWA sizing. Settling is defined when the pointing stability over 3 minutes reaches 4 mas, 1σ in tip/tilt and 20 mas, 1σ in roll. Steady state stability performance is maintained indefinitely using micro-thruster control with the reaction wheels at zero speed.

The scenario is rerun with individual ACS errors enabled to identify performance drivers, with results summarized in Table 10. Steady state pointing stability is sensor limited and driven by the FGS noise error feeding into ACS controller. Similarly to the Hybrid Case, settling time is driven by the torque limited slew with RWA drag/cogging affecting the transient behavior following the slew maneuver. Additionally, wheel friction causes the actuation handoff maneuver from reaction wheels to micro-thrusters to produce a body torque error while spinning down the wheels. This transient behavior introduces additional time to settle and causes a small increase in the computed pointing stability over the observation period.

**Table 10** Performance Contribution of Each ACS Error Source, Hybrid ACS Architecture.

| Case ID | ACS Errors Included | Stability (mas) | | | Settle Time (min) |
|---------|---------------------|------|------|------|-------------------|
| | | **X** | **Y** | **Z** | |
| A | No Errors | 1.04 | 0.35 | 0.11 | 39.1 |
| B | SRP Variation | 1.01 | 0.34 | 0.30 | 39.1 |
| C | FGS | 9.87 | 0.54 | 0.78 | 39.1 |
| D | SRU/IRU | 0.50 | 0.24 | 0.06 | 40.6 |
| E | Micro-thruster | 1.04 | 0.35 | 0.11 | 39.1 |
| F1 | RWA (Tachometer) | 1.31 | 0.29 | 0.13 | 39.1 |
| F2 | RWA (Drag, Cogging) | 1.22 | 1.48 | 0.44 | 47.5 |
| G | All Errors | 9.81 | 0.91 | 0.93 | 49.5 |

*4.1.5  ACS Analysis Conclusions*

EAC1 ACS analysis used time-domain simulations of OS-1 to assess ACS pointing stability, agility, and performance drivers for each ACS architecture option. High-fidelity models of observatory dynamics, SRP disturbances, reaction wheels, micro-thrusters, and sensors were exercised to establish ACS operational baselines and inform pointing error budgets.

As expected, the Hybrid ACS architecture shows an increased performance over the current RWA Only ACS architecture (with benchmarked dual passive isolation at the RWA + Payload) with regards to pointing stability, achieving sub-milliarcsecond tip/tilt stability and stable long-duration observations. The RWA Only ACS architecture performance is driven by reaction wheel error sources (tachometer error and drag/cogging) and continuous wheel speed migration under SRP torque. In effort to improve RWA Only architecture performance, preliminary jitter analyses have introduced active isolation at the payload, discussed in Sec. 4.2.1.4. Future studies will also consider active isolation at an RWA pallet, discussed in Sec. 4.2.2.2.

For the Hybrid ACS architecture, steady state stability performance is no longer actuator-limited but sensor-limited, with pointing stability driven primarily by the Fine Guidance Sensor noise. Settling time remains influenced by reaction wheel drag and cogging during post-slew transients but does not limit long-term performance. Ongoing work is focused on incorporating the high-fidelity FGS model into the ACS performance simulation and development associated estimation techniques. Refinements to the FGS noise characterization and dynamic behavior are expected to provide a more accurate assessment of achievable pointing stability and further optimize the Hybrid ACS design.

*4.2 Jitter Analysis*

EAC1 Jitter analysis was performed in both the time domain and the frequency domain. The closed-loop time domain simulation captures the ACS error models and control logic, as well as active stability correction (FSM and PMSA actuation) beyond ACS. The frequency domain analysis is essentially open loop (disturbance to optical error), capturing 'high frequency' jitter error that is above the active control bandwidths. The two analyses are complementary: time domain analysis provides nominal performance predictions that capture controller interactions for a specific observing scenario, while the steady-state frequency domain analyses are conservative and capture worst-case conditions for isolation system design. Furthermore, due to computational efficiency, the frequency domain analysis is well-suited for high level architecture trades and down-selection of cases to pass to the time domain simulations.

The EAC1 jitter analyses were performed using state-space models in modal coordinates. The eigen solution is extracted from the structural FEM and includes all flexible modes 0-100 Hz and residual vectors to capture flexibility at active control and disturbance input locations. The truncation of modes > 100 Hz was chosen primarily for computational efficiency; the preliminary FEM has many low mass panel modes that have negligible impact on jitter and are burdensome to carry. Furthermore, it is expected that the critical jitter modes are less than 100 Hz: global PMA, local PMSA on mount, and global SMA modes, which are included in the analysis at ~20 Hz, ~50 Hz, and ~11 Hz, respectively. The baseline model utilizes the following structural damping schedule:

- 0.10% blanket (representative of composite)
- 0.25% on spacecraft (aluminum)
- 0.25% at discrete locations on the large deployables (MBA, Solar Arrays, LDRS)
- 2.00% on spacecraft isolation system (0.5-1.0Hz)
- 5.00% on RWA isolator systems (3-6.5Hz)
- 0.25% on RWA bearings

The LOM is connected to the jitter state-space model by post-processing optical outputs by the LOM sensitivities (time domain) or by direct inclusion in the C-matrix (frequency domain).

### 4.2.1 Time Domain ACS + Jitter Simulation

Time domain simulations were performed to evaluate the ACS architectures defined below. Note that options 1 and 2 are consistent with ACS analysis, while option 3 is an alternative improvement to the RWA Only architecture and is a focus of ongoing efforts:

1) Option 1: ACS architecture #1, RWA only with benchmarked passive payload and RWA isolation systems

2) Option 2: ACS architecture #2, Hybrid RWA (slews) and μ-Thrusters (momentum control and observations)

3) Option 3: ACS architecture #1, RWA only with active payload and passive RWA isolation systems

The time domain jitter simulation is built upon the low frequency ACS model, as described in Sec. 4.1.1. To capture both ACS and jitter contributions to stability error, the high-fidelity jitter analysis includes the following additional features and sub-models:

- Flexible dynamics from the FEM to 100Hz (identical FEM for options #1 and #2, connected in separate pieces and connected via isolator for option #3)

- Spinning reaction wheels (mechanism modes) with added gyroscopic forces/moments, the FSM Assembly (mirror, reaction mass, and actuators), articulating primary mirror segment mounts

- LOM derived optical sensitivities for LOS for the CI and GIs, WFE for the CI, and beamwalk for the CI

- Control loops: RWA, micro-thrusters, FSM, 19 M1 segments and M2, and 6 active isolator struts (option #3)

- Sensors: Gyro, star tracker, fine guidance sensor (FGS), low order wavefront sensor (LOWFS), RWA tachometer, and metrology for M1 and M2

- RWA actuator Induced Vibration (IV)

- Thermal deformation of front-end optics from STOP analysis, used as time series playback

The working "desirements" for coronagraph jitter during an observation are LOS error < 0.1 mas RMS (YZ-axes), and RMS WFE < 1 pm RMS, as tabulated in Sec. 1.2. At this point, error budgets do not include suballocations for pointing stability about the boresight (X-axis) or beamwalk. Nonetheless, these metrics are calculated and reported in the time domain analysis to support error budget development.

### 4.2.1.1 Simulation Architecture

Figure 75 shows the jitter simulation architecture, emphasizing the new blocks added to the basic ACS simulation. The operational differences between the RWA only and the Hybrid RWA/Micro-thruster Architectures occur primarily in the ACS Control block, and are described in Sec. 4.1.2. In short, for RWA Only control, the RWA speeds are biased at the end of the slew, then drift throughout observation to counteract SRP torque. For Hybrid control, the RWAs are spun to zero at the end of the slew, and the micro-thrusters counteract SRP torque throughout the observation.

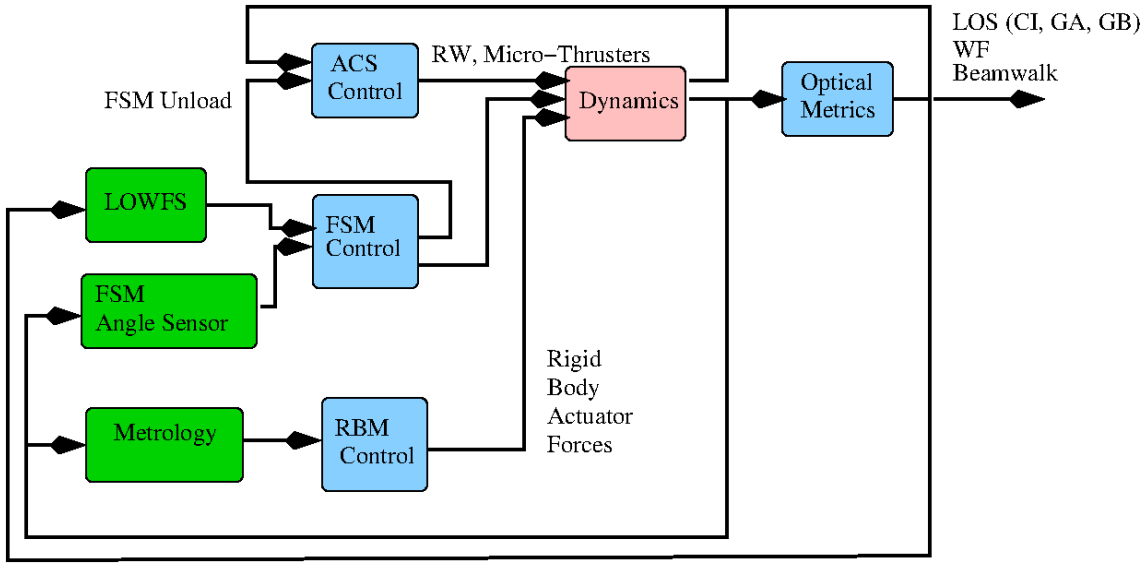

**Fig. 75** Hybrid RW/Micro-thruster jitter time domain model.

The FSM control takes the LOS, measured by the LOWFS and FSM angles, and sends torque commands (10 Hz bandwidth) to the FSM/RM dynamics as well as a very low bandwidth unload signal to the ACS to keep the FSM angles small.

The M1 segment and M2 rigid body control takes in metrology sensors (laser metrology, gap, piston, and shear sensors) and sends 120 force commands to the M1 segment and M2 actuators with a 0.05Hz bandwidth. Computed optical metrics include LOS for the CI and the GIs used in the FGS as well as WFE and beamwalk on the front end optics (to the CI entrance pupil).

### 4.2.1.2 Option 1 Performance: ACS Architecture #1, RWA only with Passive PL / Passive RWA Isolation

The following figures show option #1 performance over a portion of the imaging duration. Note that this is not a true steady state case, as the solar pressure torques cause the system momentum to build up and, consequently, cause the wheels to speed up; this causes excitation of a succession of modes. Also note the somewhat larger ACS errors compared to option #2 (Hybrid micro-thrusters), which lead to larger FSM corrections and subsequent beamwalk jitter; the LOS is still controlled to the desired level by the relatively high bandwidth FSM, with stability less than 0.1 mas. As mentioned in Sec. 4.1.3, the ACS errors are driven primarily by RWA errors including the tachometer. Ongoing analyses consider methods of mitigating this contribution, including architectural/control methods (less reliance on or more filtering of the wheel speeds), or higher precision tachometer measurements.

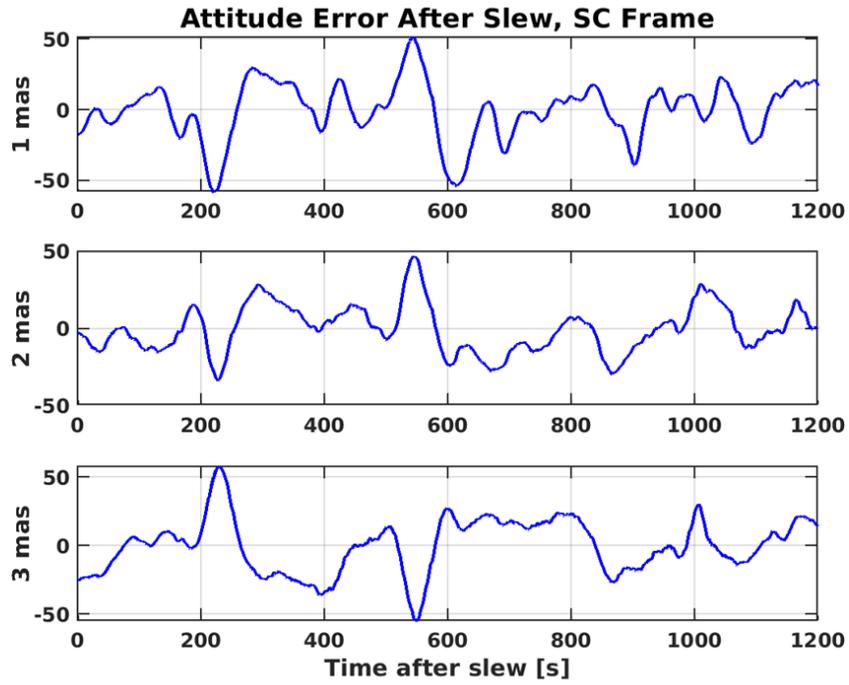

**Fig. 76** Option #1 (RWA only, passive payload / passive RWA isolation), ACS Performance.

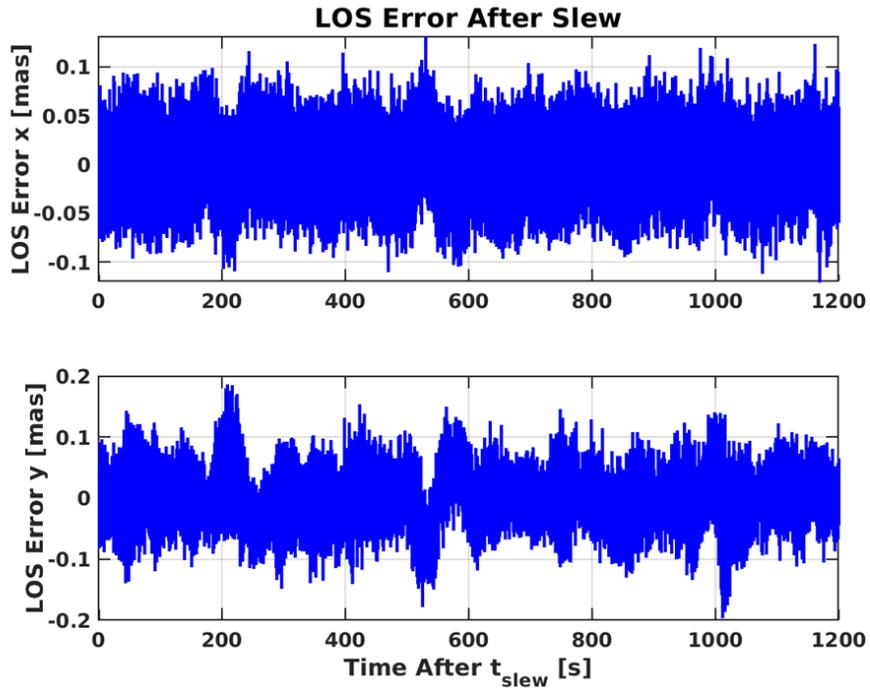

**Fig. 77** Option #1 (RWA only, passive payload / passive RWA isolation), LOS Performance (with FSM).

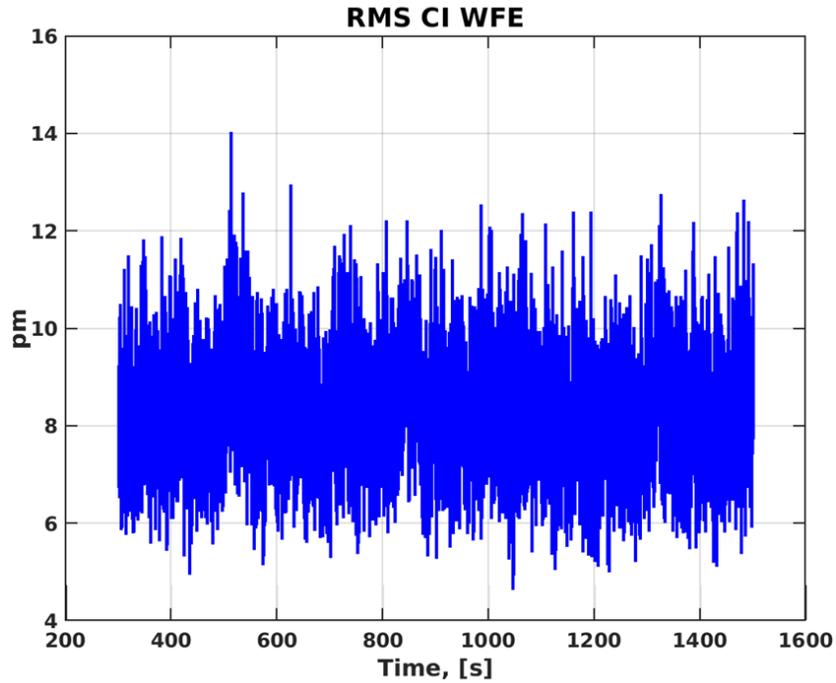

**Fig. 78** Option #1 (RWA only, passive payload / passive RWA isolation), RMS WFE Performance.

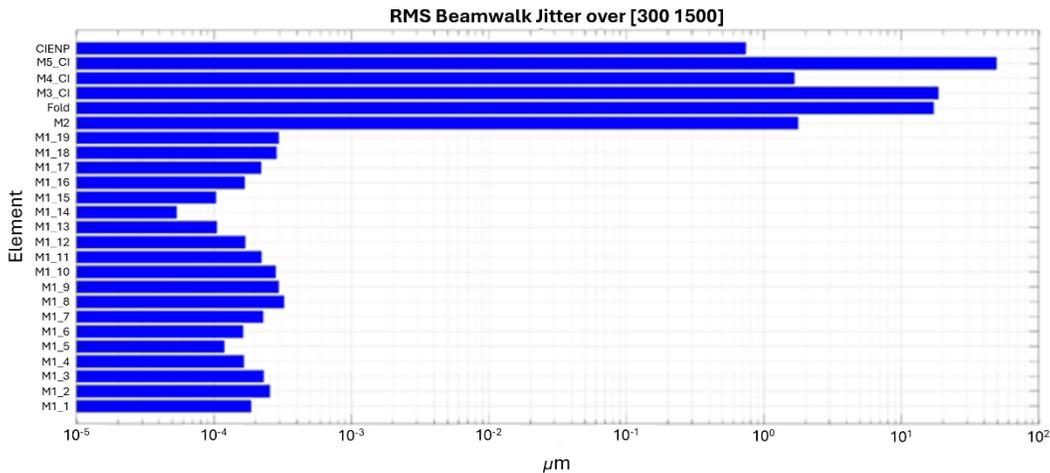

**Fig. 79** Option #1 (RWA only, passive payload / passive RWA isolation), Beamwalk Performance

### 4.2.1.3 Option 2 Performance: ACS Architecture #2, Hybrid RWA + Micro-Thruster

The following figures show the steady-state performance of option #2 over the imaging duration (6,300s), after the transition to micro-thrusters; this is the first roll case from the OS-1 scenario. Consistent with the ACS analysis in Sec. 4.1.4, the ACS performance is driven by the FGS noise. Here we see the LOS driven down to stability less than 0.1 mas by the FSM. The WFE is driven down to single digit pm by the segment controls. We include an initial assessment of beamwalk jitter as this is likely a significant performance consideration for the coronagraph. We note the beamwalk is largest at the M5 mirror, which is furthest from a pupil.

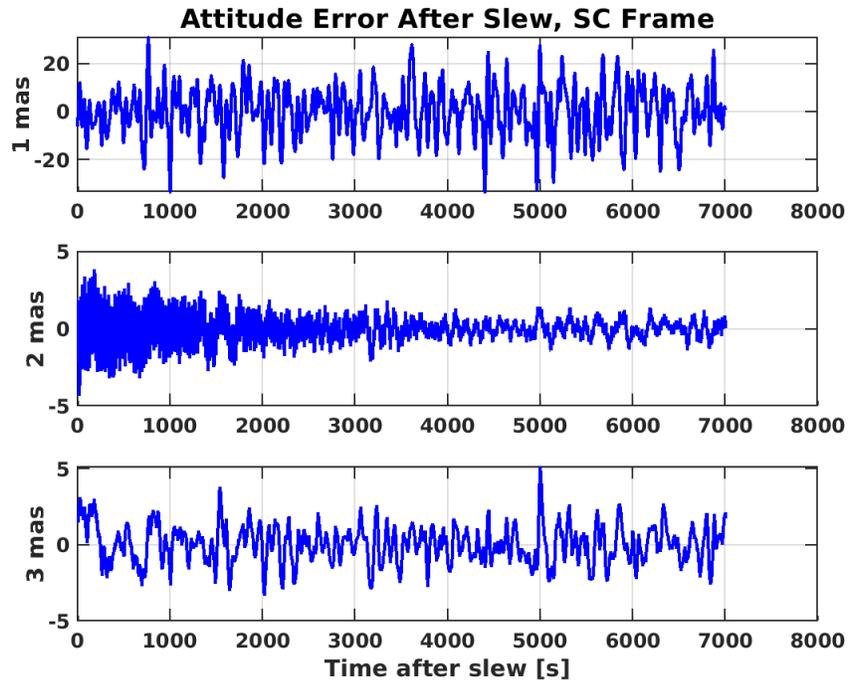

**Fig. 80** Option #2 (Hybrid RWAs + Micro-Thrusters), Steady-State ACS Performance.

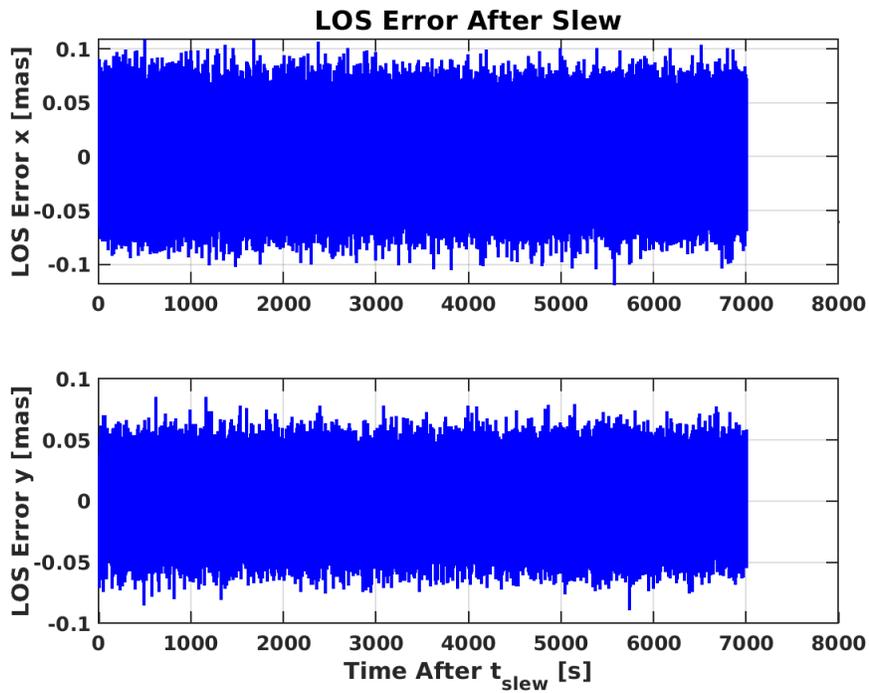

**Fig. 81** Option #2 (Hybrid RWAs + Micro-Thrusters), Steady-State LOS Performance.

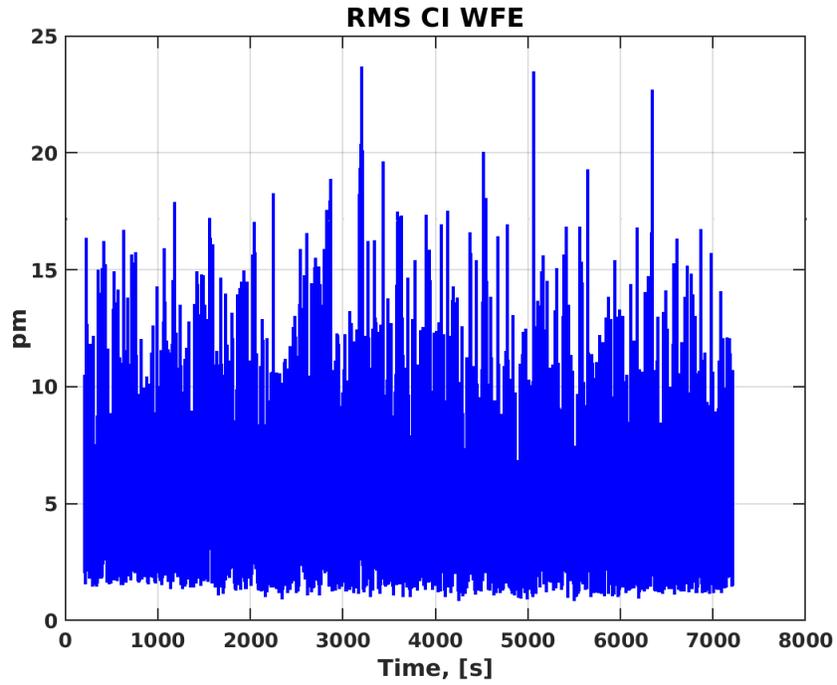

**Fig. 82** Option #2 (Hybrid RWAs + Micro-Thrusters), Steady-State RMS WFE Performance.

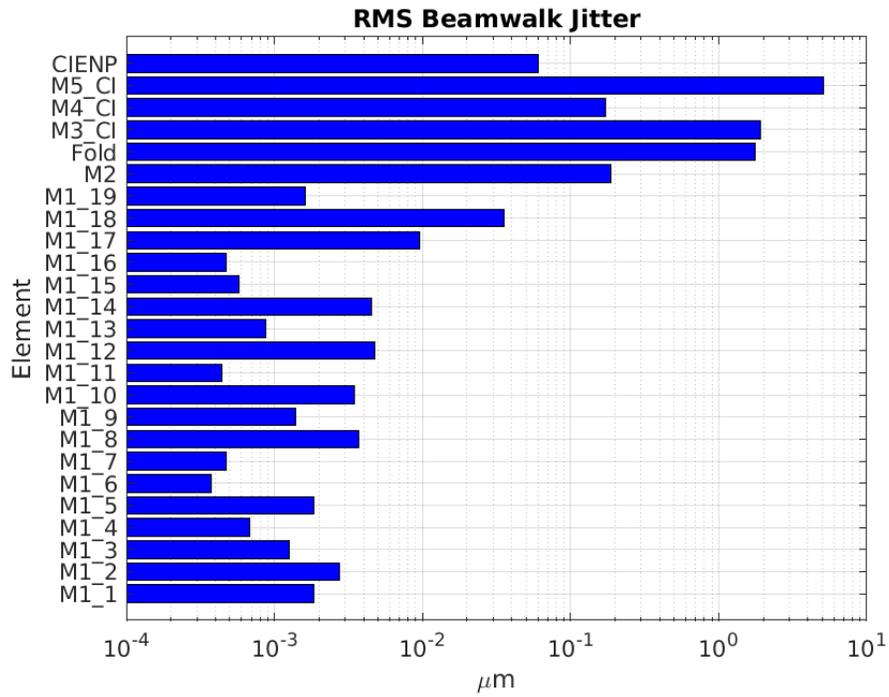

**Fig. 83** Option #2 (Hybrid RWAs + Micro-Thrusters), Steady-State Beamwalk Performance.

*4.2.1.4  Option 3 Performance: ACS Architecture #1, RWA Only with active PL / passive RWA isolation*

Preliminary simulations were performed to assess this alternative hybrid control option. In this option, attitude is maintained using RWAs only (no micro-thrusters), but the passive spacecraft/payload isolator is replaced with an Active Strut Isolation System (ASIS); the passive isolation system is maintained at the RWAs. The active payload struts are in a hexapod arrangement and have a very low passive stiffness (given about mHz system modes) in parallel with active forces. The 6 degrees of freedom (DOF) of the active struts are partitioned into 2 sets of 3, for the following three ACS control loops.

1)  The FGS inertial rotation knowledge drives a 3 DOF subset of the 6 strut forces to control the torque on the instrument (0.06 Hz bandwidth)

2)  The strut displacements are mapped to relative translation between instrument and bus, and used to drive the orthogonal 3 DOF subset of the strut forces (0.003 Hz bandwidth)

3)  The strut displacements mapped to relative rotation between the instrument and the bus are used to drive the reaction wheels on the bus, so that the bus follows the motion of the instrument (0.003 Hz bandwidth)

The following figures show the performance of option #3 during imaging. As with option #1, these results are not quite steady state as the RWAs are spinning up as they soak up the momentum from the solar torques. In this case, due to the softer payload isolation, we see much less excitation of the system modes. Here the FGS is again the primary driver of ACS errors.

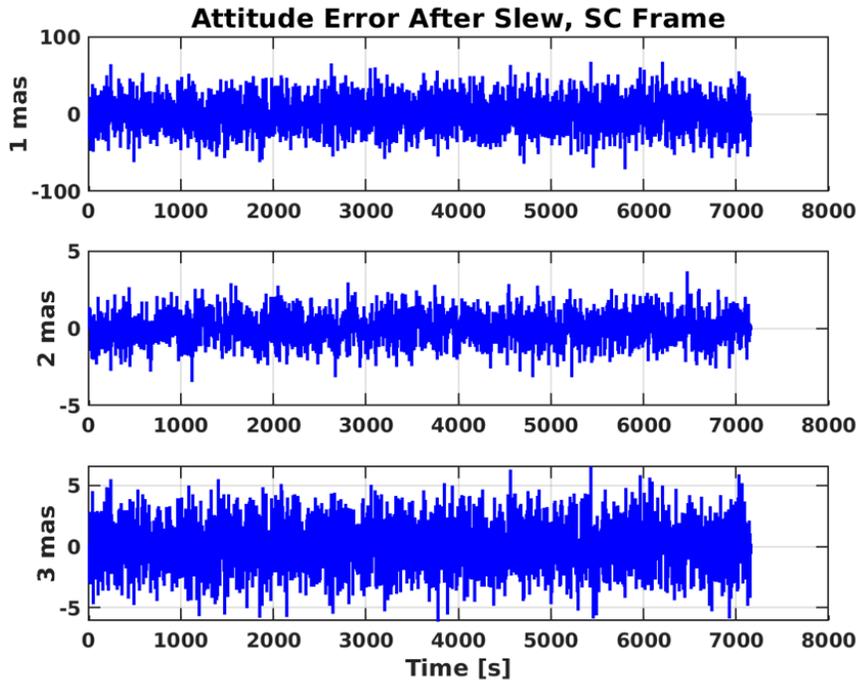

**Fig. 84** Option #3 (RWA Only, active payload / passive RWA isolation), ACS Performance.

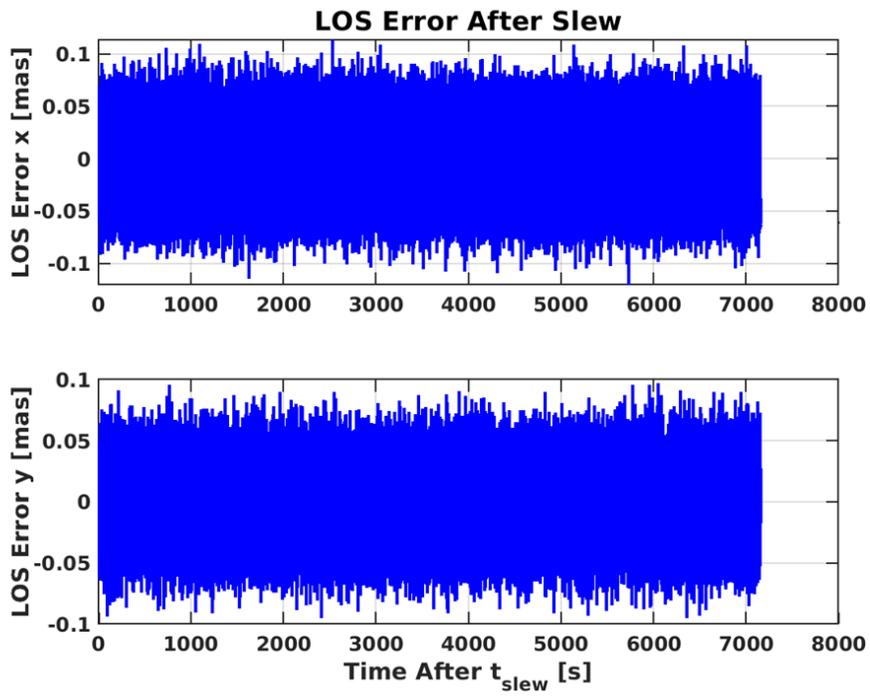

**Fig. 85** Option #3 (RWA Only, active payload / passive RWA isolation), LOS Performance.

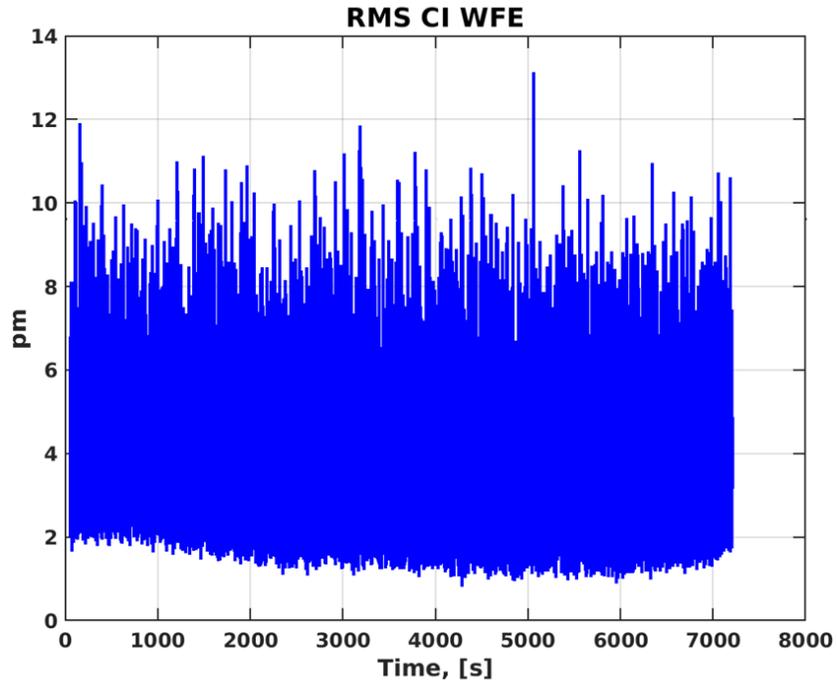

**Fig. 86** Option #3 (RWA Only, active payload / passive RWA isolation), RMS WFE Performance.

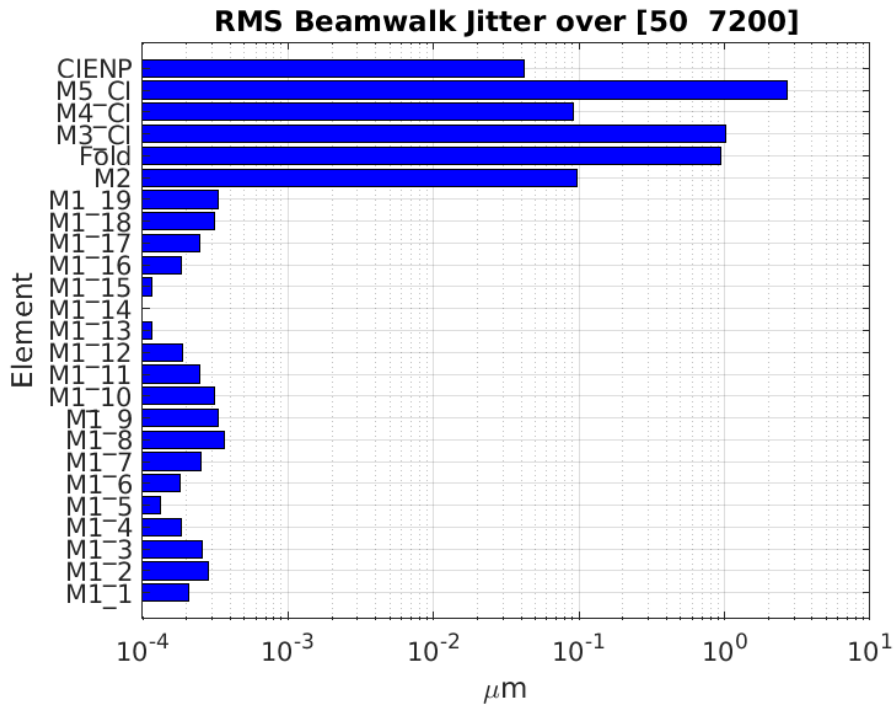

**Fig. 87** Option #3 (RWA Only, active payload / passive RWA isolation), Beamwalk Performance.

*4.2.1.5 Time Domain Jitter Summary*

Table 11 shows a summary of the performance for the evaluated jitter mitigation techniques. The baseline RWA only case (option 1) has larger numbers, marginally meeting the LOS goal, but exceeding the WFE goal by ~10x. The hybrid and active isolation cases are promising, meeting the LOS goal with healthy margin and exceeding the WFE goal by only ~5x. Future work will focus on improving WFE performance by tuning WFS&C parameters, and extension of the ACS and jitter analysis for the active payload isolation (option 3). Additional future studies will look at different variations of the RWA only ACS architecture, with RWAs mounted on a passively or actively isolated pallet. See the following discussion on frequency domain jitter trades.

**Table 11** Time Domain Jitter Summary

| Control Case | $T_w$ (s) | Attitude Stability (mas) | | | LOS Stability (mas) | | RMS WFE (pm) |
|---|---|---|---|---|---|---|---|
| | | x(bs) | y | z | | | |
| RW Passive Isolators | 6,300 | 61.9 | 92.5 | 98.7 | 0.074 | 0.11 | 11.2 |
| Micro-thrusters | 6,300 | 10.4 | 0.960 | 1.23 | 0.024 | 0.018 | 4.85 |
| RW, ASIS | 6,300 | 18.8 | 0.90 | 1.7 | 0.024 | 0.022 | 3.77 |

*4.2.2 Frequency Domain Jitter Trades*

As a first-pass strategy to attenuate RWA vibration to the mirrors, the baseline EAC1 design includes dual-stage passive isolation systems at each individual RWA-Spacecraft interface and at the SC-Payload interface. The baseline RWA breakpoint frequencies are 3.0-6.5 Hz (benchmarked Roman performance) and the baseline payload breakpoint frequencies are 0.5-1.0 Hz (benchmarked JWST performance). Preliminary time domain analyses suggest it will be challenging to meet jitter requirements in this option #1 configuration (ACS architecture #1: RWA Only with passive payload / passive RWA isolation) and these benchmarked isolation frequencies. Frequency domain architecture trades were performed to identify isolation system requirements for the RWA only ACS architecture to meet jitter requirements and to down-select corresponding models to pass to the high-fidelity time domain simulation.

In lieu of a sub-allocation for the RWA IV disturbance assessed in the frequency domain, the analysis has worked toward the top-level 'desirements' tabulated in Sec. 1.2: LOS error < 0.1 mas RMS and WFE < 1.0 pm RMS. Figure 88 shows a summary of the analysis pipeline. High pass filters were applied to results to represent active jitter rejection (LOS = 20 Hz [FSM], WFE = 0.0016 Hz [WFSC]). Beamwalk jitter was not evaluated in the frequency domain.

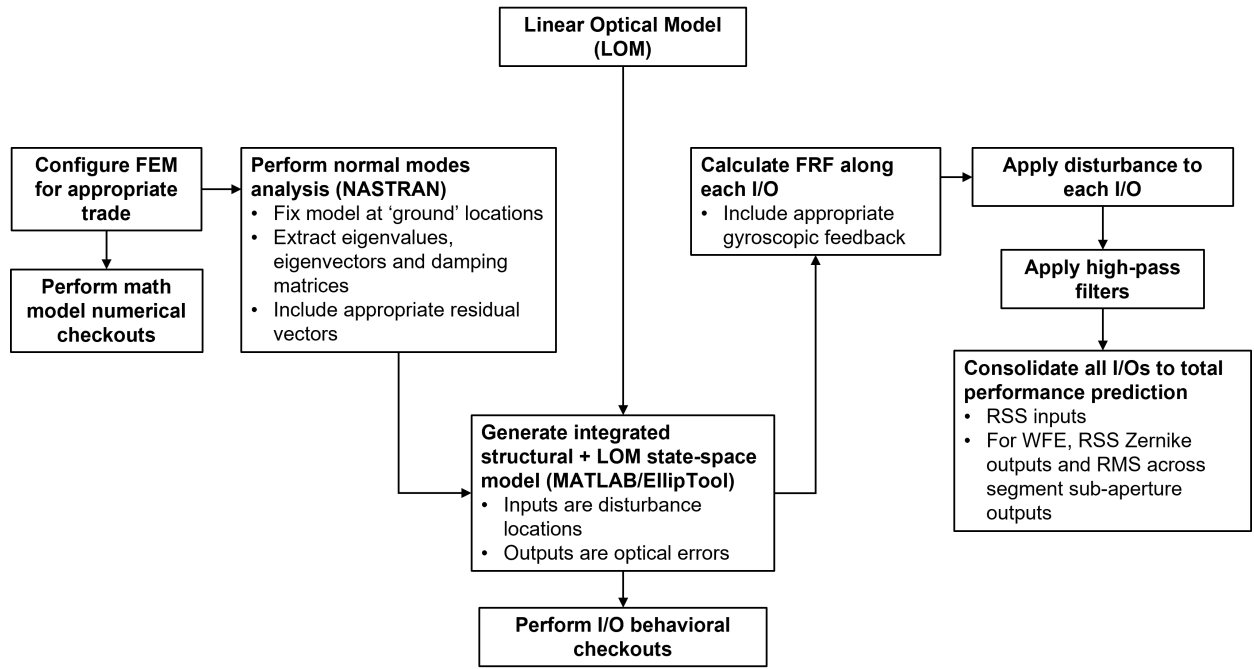

**Fig. 88** Frequency Domain Jitter Analysis Pipeline.

### 4.2.2.1 Dual Stage RWA/Payload Isolation Studies

For the first study, the isolator spring element stiffnesses (CBUSH) were simply reduced in the FEM to generate the matrix of configurations shown in Table 12. For brevity, the table is also annotated to show relative performance results for each configuration. Due to the simple passive spring representation, all cases provide 40 dB/decade attenuation vs. the >80 dB/decade that can be achieved with active control.[23] This was intentional, as the ideal outcome of the study was to find a passive/passive isolation solution for low cost/complexity.

Table 12 Dual Stage RWA/Payload Isolation Trade Matrix.

| | | | | | Case 1 | | Case 2 | | Case 3 | | Case 4 | | Baseline Architecture Case 5 | |
|---|---|---|---|---|---|---|---|---|---|---|---|---|---|---|
| | | | | | \<RWA Isolation Frequency Range (Hz)\> | | | | | | | | | |
| | | | | | Low | High | Low | High | Low | High | Low | High | **Low** | **High** |
| | | | | | 0.10 | 0.22 | 0.23 | 0.51 | 0.55 | 1.19 | 1.28 | 2.78 | **3.00** | **6.50** |
| | Payload Isolation Frequency Range (Hz) | Case A | Low | 0.10 | <u>1A</u> | | <u>2A</u> | | <u>3A</u> | | <u>4A</u> | | <u>5A</u> | |
| | | | High | 0.20 | | | | | | | | | | |
| | | Case B | Low | 0.17 | <u>1B</u> | | <u>2B</u> | | <u>3B</u> | | **4B** | | ***<u>5B</u>*** | |
| | | | High | 0.34 | | | | | | | | | | |
| | | Case C | Low | 0.29 | <u>1C</u> | | <u>2C</u> | | **3C** | | ***<u>4C</u>*** | | ***<u>5C</u>*** | |
| | | | High | 0.58 | | | | | | | | | | |
| Baseline Architecture | | **Case D** | **Low** | **0.50** | <u>1D</u> | | <u>2D</u> | | ***<u>3D</u>*** | | ***<u>4D</u>*** | | ***<u>5D</u>*** | |
| | | | **High** | **1.00** | | | | | | | | | | |

| Performance Key: | <u>RMS jitter < desirement (underline)</u> |
|---|---|
| | **RMS jitter < desirement, with minor poke-through at some (bold)** |
| | ***<u>RMS jitter > desirement (bold, underline, and italic)</u>*** |

To bound conservatism in the study, the RWA disturbance model used mean +1sigma imbalances/harmonics from as-built wheels manufactured for heritage missions and sized consistent with HWO ACS needs. For reference, the assumed values are < 0.5 advertised/specified, but are consistent with as-delivered values across multiple programs. At each wheelspeed, the fundamental imbalance and harmonics are applied at the CG of each wheel, at appropriate discrete frequencies, and includes gyroscopic feedback terms. It is assumed that all wheels are operating at the same speed (conservative) and that phasing is uncorrelated between wheels, harmonics, forces, and moments; total error is RSS across all input/output paths. The final analysis output is a plot of total RMS jitter (LOS or WFE) vs. Wheelspeed. The pass/fail criteria for any configuration is a "quiet wheelspeed range" (jitter below desirement) 0-1000 RPM, which is the range required by ACS to maintain pointing stability over an OS-1 science observation.

Analysis results show that WFE jitter is the differentiating performance between cases. All cases maintain RMS LOS error < 0.1 mas 0-1000 RPM. Case 2D was selected for further study, as it maintains RMS WFE < 1 pm 0-1000 RPM, while maintaining the benchmarked *passive* payload isolation system. It should be noted that there are several caveats to this performance: 1) The simple spring representation of the isolation systems does not include surge modes that will typically flatten the attenuation curve at 1.5-2x the breakpoint frequencies and 2) it will be very challenging to achieve *passive* 0.2-0.5 Hz RWA isolation. Still, the 2D case model was passed to time domain simulation to understand the performance when the additional error sources are included.

The case 2D time simulation went unstable as the RWAs approached ~10 rev/sec, due to bifurcation of the isolation modes at increasing wheelspeed. This behavior is well documented in Ref. 24 and summarized for a single EAC1 isolated RWA in Fig. 89: As the wheelspeed increases, the rotor on bearing rocking mode pair and RWA on isolator rocking mode pair each split into a posigrade and retrograde whirl mode. The frequency and damping of the retrograde isolator mode degrade rapidly, creating instability when the mode frequency crosses over the ACS bandwidth.

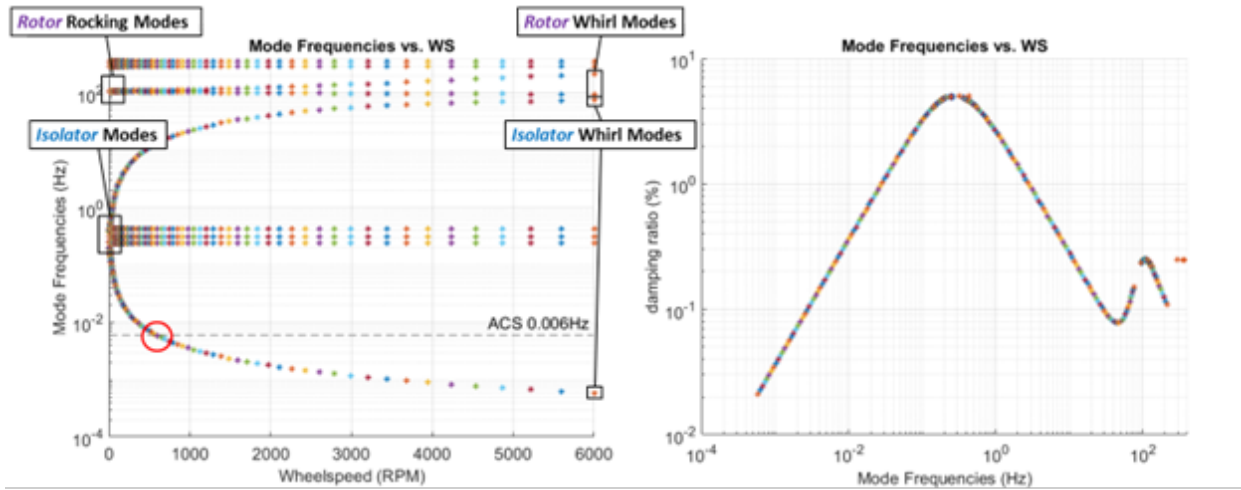

**Fig. 89** RWA Isolator Gyroscopic Behavior (Case 2D).

The instability could potentially be corrected with a lower ACS bandwidth; however, this degrades ACS disturbance rejection performance at low frequencies. Further, the 'soft springs' required for 0.2-0.5 Hz isolation bring additional challenges, such as large deflections and verification of 0g performance during ground testing. The study suggests the best solution to the 'RWA only' jitter problem is likely not to simply reduce the passive isolation frequencies at the individual RWAs. One alternative is to improve the payload isolation stage instead of the RWA isolation stage by introducing active control. A preliminary assessment of this control architecture is discussed in Sec. 4.2.1.4. A second alternative is to hardmount all of the RWAs on a single isolated pallet; this architecture is discussed in the following section.

### 4.2.2.2  RWA Pallet Study

There are several immediate benefits to isolating a single RWA pallet instead of individual RWAs. The total pallet system inertia is higher, which reduces sensitivity to gyroscopic effects and parasitic shunt path (harness) stiffness. It also reduces the integration and test of individual isolation systems to a single system. The typical pallet configuration challenge is packaging of the large assembly into the Spacecraft.

A conceptual EAC1 pallet design was generated by moving the RWAs to a central location and flipping 4 of 8 spin axes for symmetry; a simple composite delta frame was added between the wheels. To realize the benefits of higher inertia, the allowable assembly mass was assumed to be 2x the standalone RWAs (total mass = 300 kg = 150 kg RWAs + 100 kg pallet structure +50 kg pallet NSM). The entire pallet was placed on a hexapod, with strut lines of action chosen for tight but even spacing between 6DOF isolation modes. The hexapod enables sliding of the isolation mode frequencies with a single parameter (axial strut stiffness). Figure 90 and Fig. 91 show the pallet design and isolation modes, respectively.

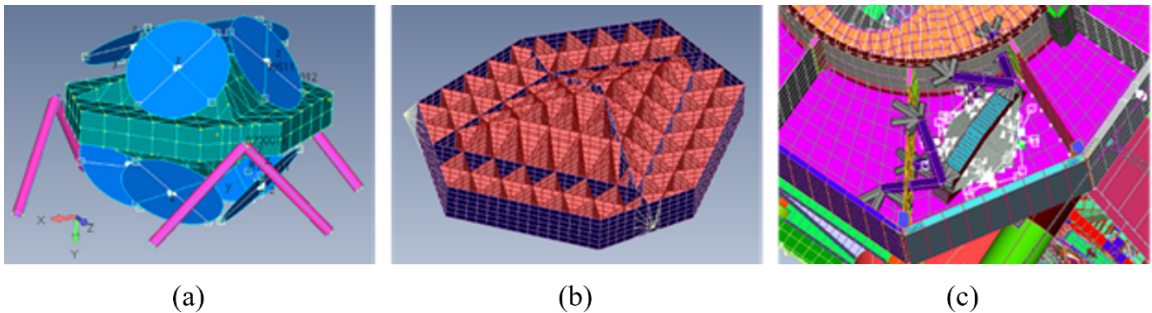

(a)             (b)            (c)

**Fig. 90** RWA Pallet: (a) Design Geometry, (b) Rib Structure, and (c) Integrated observatory.

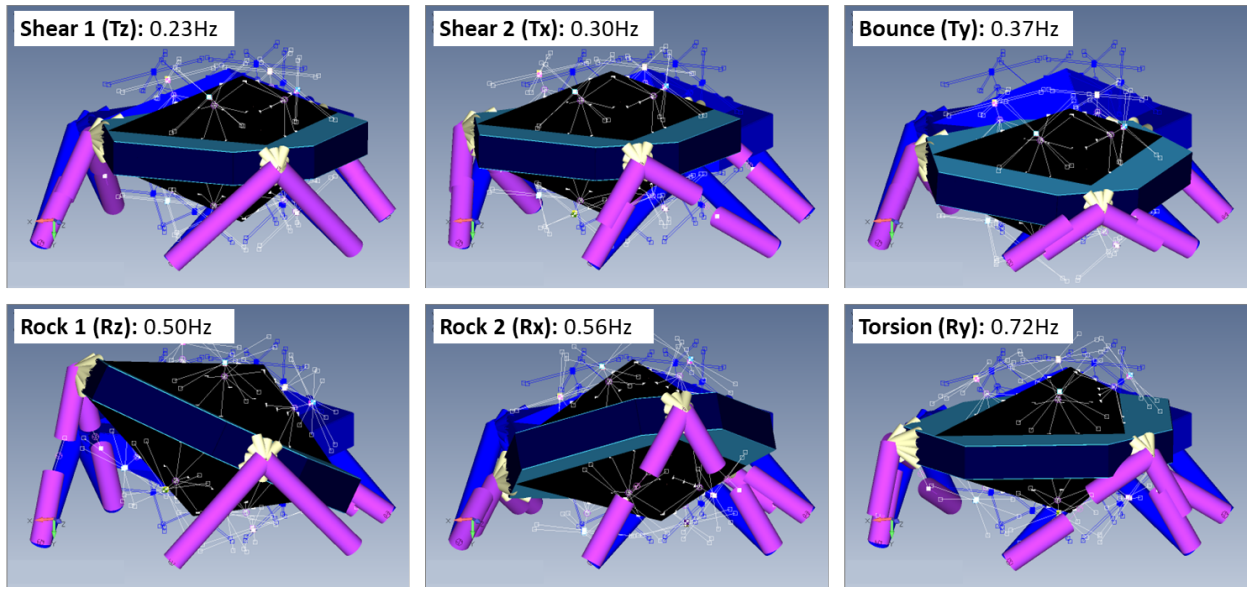

**Fig. 91** RWA Pallet Isolation Modes (Case 2D, see matrix below).

A behavioral study was performed for the RWA pallet to verify the reduced sensitivity to gyroscopics (and therefore lower risk of interaction with ACS). The gyroscopic feedback is dependent on the total momentum on the pallet (sum of all wheels) and therefore, a sphere of momentum vectors (and corresponding wheelspeed spreads) was considered. The behavior for all cases is somewhat analogous to the individual RWA: As wheel speed increases, the 3 rotational modes split into 2 whirl modes and 1 torsional mode about the momentum axis, and the translational isolation modes are unchanged. Most importantly, for all cases, the isolation modes do not cross-over the ACS bandwidth as the max wheelspeed approaches 6,000 RPM. Figure 92 shows the mode frequency and damping for the worst-case momentum vector (largest frequency drop).

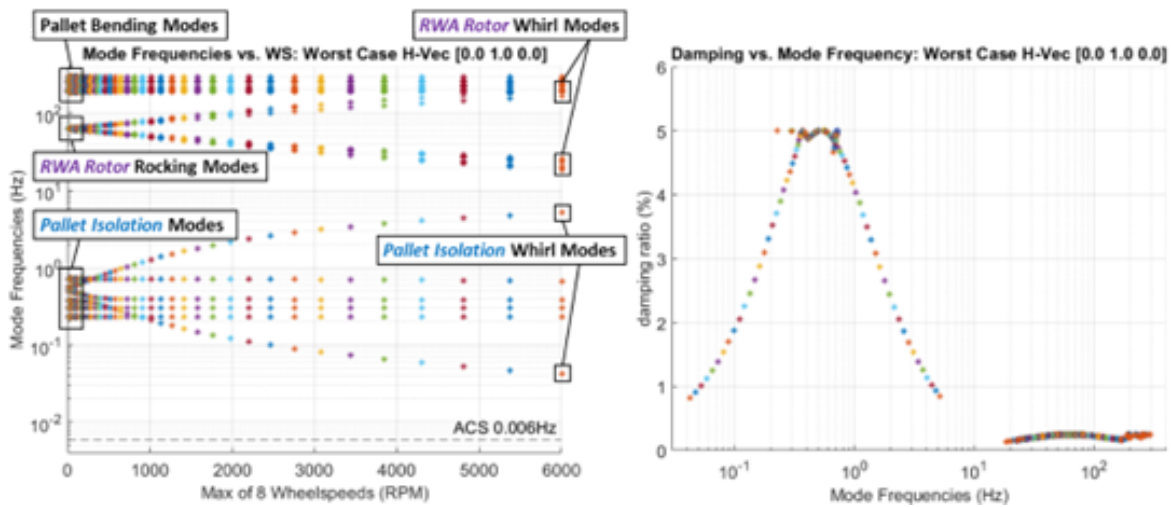

**Fig. 92** Pallet Isolator Gyroscopic Behavior (Case 2D, see matrix below).

Finally, the jitter analysis was repeated for the matrix of configurations shown in Table 13. All cases met the LOS error desirement 0-1000 RPM. WFE was again the more challenging metric, requiring a large drop in either the payload isolation frequency or pallet isolation frequency (or combination of both) to meet the WFE desirement 0-1000 RPM. Cases 2D and 3D were chosen as favorite candidates; Fig. 93 shows their WFE responses. Case 2D meets the desirement while maintaining the benchmarked payload isolation system, but the associated pallet isolation frequencies may not be achievable passively. Case 3D nearly meets the desirement with more realistic pallet isolation frequencies.

In general, the RWA or pallet isolation system has two goals. The first is to work in tandem with the payload isolation system to attenuate higher order harmonic disturbances before they can excite the payload optical modes; this is accomplished with relative ease, without pushing the breakpoint frequencies to extreme lows. The second comes from the isolation modes themselves; these can couple with the payload isolation modes and low frequency deployable modes, creating WFE from quasi-static deflection of the payload (analogous to 1g sag). The pallet isolation modes must be properly spaced from these modes and/or heavily damped. Future studies will focus on damping treatment and/or active isolation to mitigate these interactions.

**Table 13** RWA Pallet Isolation Trade Matrix

| | | | | Case 1 | | Case 2 | | Case 3 | | Case 4 | | Case 5 | |
| --- | --- | --- | --- | --- | --- | --- | --- | --- | --- | --- | --- | --- | --- |
| | | | | $k_{strut}$ = 113 | | $k_{strut}$ = 589 | | $k_{strut}$ = 3081 | | $k_{strut}$ = 16119 | | **$k_{strut}$ = 84330** | |
| | | | | Low | High | Low | High | Low | High | Low | High | **Low** | **High** |
| | | | | 0.10 | 0.31 | 0.23 | 0.72 | 0.52 | 1.64 | 1.20 | 3.75 | **2.74** | **8.57** |
| | Payload Isolation Frequency Range (Hz) | Case A | Low | 0.10 | | | | | | | | | |
| | | | High | 0.20 | | <u>1A</u> | | <u>2A</u> | | <u>3A</u> | | <u>4A</u> | | <u>5A</u> |
| | | Case B | Low | 0.17 | | | | | | | | | |
| | | | High | 0.34 | | <u>1B</u> | | <u>2B</u> | | <u>3B</u> | | **4B** | | <u>**_5B_**</u> |
| | | Case C | Low | 0.29 | | | | | | | | | |
| | | | High | 0.58 | | <u>1C</u> | | <u>2C</u> | | **3C** | | <u>**_4C_**</u> | | <u>**_5C_**</u> |
| **Baseline Architecture** | | **Case D** | **Low** | **0.50** | | | | | | | | | |
| | | | **High** | **1.00** | | <u>1D</u> | | <u>2D</u> | | <u>**_3D_**</u> | | <u>**_4D_**</u> | | <u>**_5D_**</u> |

Over the Case 5 columns, the header cell spans: **Baseline Architecture**

| Performance Key: | <u>RMS jitter < desirement (underline)</u> |
| --- | --- |
| | **RMS jitter < desirement, with minor poke-through at some (bold)** |
| | <u>**_RMS jitter > desirement (bold, underline, and italic)_**</u> |

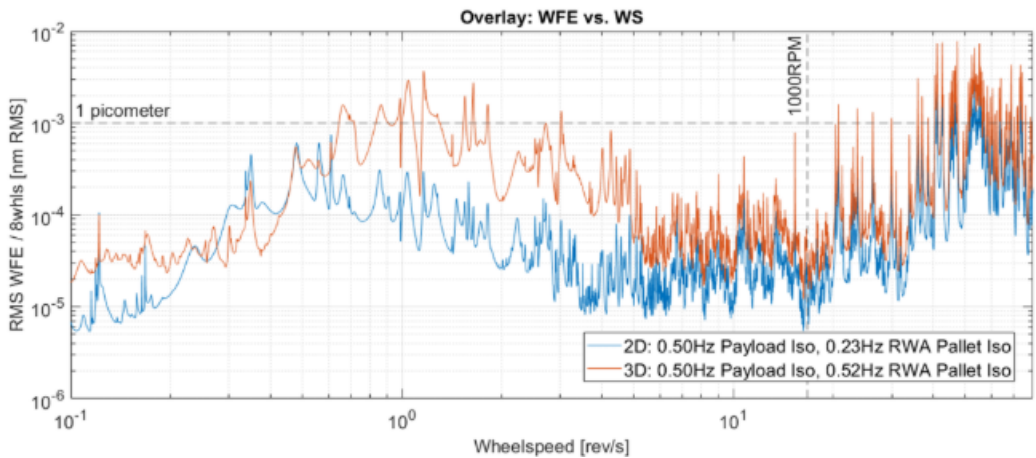

**Fig. 93** WFE Response for Cases 2D and 2D.

Models for both 2D and 3D cases have been passed to the time domain for a more complete assessment. There is hope the time simulation will generate lower predictions, as it reduces conservative frequency domain assumptions that all wheels are at equal speeds and all responses are steady state.

### 4.2.3 Jitter Analysis Conclusions

EAC1 jitter analysis identified four ACS + isolation system architectures with potential to meet stringent optical stability requirements:

1. Option 1: ACS architecture #1, RWAs only, with passive payload isolation and passive RWA isolation. For this architecture, a common RWA pallet is preferred to avoid interaction of gyroscopic modes and ACS logic. While analysis shows promising results, it requires low passive isolation frequencies, which may create implementation challenges. Ongoing time domain analysis with different passive isolator cases will help determine isolation requirements and feasibility of this architecture.

2. Option 2: ACS architecture #2, Hybrid RWA slews and Micro-thrusters. This architecture shows promise due to inherently low noise micro-thruster operation. This option requires further maturation and demonstration of micro-thruster technology.

3. Option 3: ACS architecture #1, RWAs only, with active payload isolation and passive RWA isolation. Preliminary frequency domain studies and time domain analysis suggest this architecture will adequately attenuate induced vibrations from the RWAs. An additional benefit of the active payload isolation is that it can attenuate multiple disturbance sources on the Spacecraft, not only the RWAs (high gain antenna, thermal 'pinging', etc.). Preliminary assessments of this architecture will be extended in future time domain simulations. This option requires development of an active system for a large HWO scale payload.

4. Option 4: ACS architecture #1, RWAs only, with passive payload isolation and active RWA pallet isolation. This architecture is an extension of option #1; if the required isolation breakpoint frequencies are not achievable passively, an active system could be introduced. While this would still require technology development, the active system would be smaller scale than option #3. This architecture will be an additional focus of future time domain simulations. As with any telescope, it may be challenging to package the RWA pallet into the Spacecraft.

All ACS + isolation system architectures are potential options, with the final baseline pending technology maturation and evaluation of other emerging technologies, such as ultra-low disturbance mechanisms, which will be included in future studies.

## 4.3 Structural-Thermal-Optical Performance Analysis

STOP (Structural-Thermal-Optical Performance) analysis utilizes discipline physics-based models to predict thermally induced optical performance errors. The process begins with thermal predictions derived from the thermal model and ends with processing through the optical models to recover optical metrics of interest. Recall, for HWO, the IM pipeline extends the STOP analysis process to include applicable WFS&C loops to evaluate stability between pre and post corrected performance as well as incorporates diffraction analysis to generate contrast stability and speckle time-series for post-processing evaluation.

### 4.3.1 Temperature Mapping Process and Distortion Analysis

Temperature mapping is performed to transfer temperature predictions onto the structural FEM nodes. For quasi-static analyses, temperatures are mapped at a single time point representing steady-state conditions. For transient analyses requiring time-varying distortion predictions, temperatures are mapped at multiple temporal points throughout the thermal transient to capture the dynamic response of the structure.

The temperature mapping process covers the majority of structural FEM nodes directly, though some nodes remain intentionally unmapped due to differences in thermal and structural model discretization. To provide complete temperature boundary conditions, unmapped nodes are "filled in" using a conduction-only thermal analysis. This is accomplished by converting the structural FEM to thermal usage and running nonlinear heat transfer, using the mapped temperatures as boundary conditions to solve for temperatures at all remaining nodes.

With complete temperature definitions across all structural FEM nodes and time points, thermoelastic distortion analyses are performed using static structural analysis for each thermal state. These analyses calculate the structural response to thermal loading by applying the temperature differential from a reference state (typically room temperature) to the actual thermal condition. The distortions are determined by the structural stiffness properties and, most critically, the coefficient of thermal expansion (CTE) of each material; CTE assumptions are therefore critical to accurate distortion predictions. Even within single components, such as glass optics, CTE exhibits spatial variation due to manufacturing processes and material inhomogeneities. Current CTE variation assumptions are based on previous work on similar optical systems, but as the HWO program matures, component test data will inform the true CTE variations and enable more accurate distortion modeling.

The resulting nodal displacements represent the thermally-induced distortions of the observatory structure and serve two purposes in optical analysis: 1) displacement outputs provide the primary interface to the LOM for rapid optical performance assessment and 2) distortions can be used to create perturbed versions of the CODE V ray trace optical model through SigFit, which serves as the intermediary software between physical distortions and CODE V, enabling detailed ray-trace analysis of thermally-distorted optical systems.

The thermoelastic distortions capture both rigid body motions and figure changes of optical elements, enabling comprehensive assessment of thermal impacts on observatory performance through both linear optical modeling and detailed ray-trace analysis.

### 4.3.1.1 Case and Attitude Definitions

Thermal analyses fall into one of two categories:

1) Quasi-steady state cases sampling the Field of Regard (FOR)

2) Transient solutions utilizing the coronagraph Observing Scenarios (OS)

FOR analysis samples optical performance across observatory attitudes using quasi-steady state solutions at specific attitudes; this approach enables rapid architecture comparisons by omitting thermal mass effects, particularly useful when thermal stability requirements aren't established. The FOR reflects observatory attitudes based on ±45° pitch and ±22.5° roll and analyses typically include cases at seven points to "sample" the FOR, as shown by Fig. 94. For reference, Fig. 95 shows the observatory thermal model oriented in two attitudes in the FOR; note the change with respect to solar vector impingement on the observatory at different attitudes – this effect impacts quasi-steady state behavior as well as transient behavior as the observatory slews from one attitude to another. The STOP team processes these quasi-steady state cases to identify slews that generate the largest temperature deltas and, consequently, the largest optical metric deltas; Fig. 96 shows an example of this evaluation for the figure portion of WFE. While this simplified approach is useful for rapid evaluations, transient analyses are needed to provide additional insight into the more complex, time-dependent behavior that governs observatory performance.

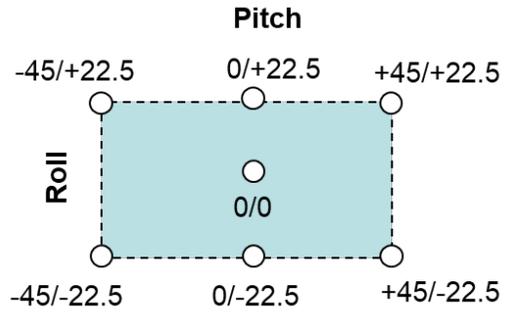

**Fig. 94** Thermal Field of Regard.

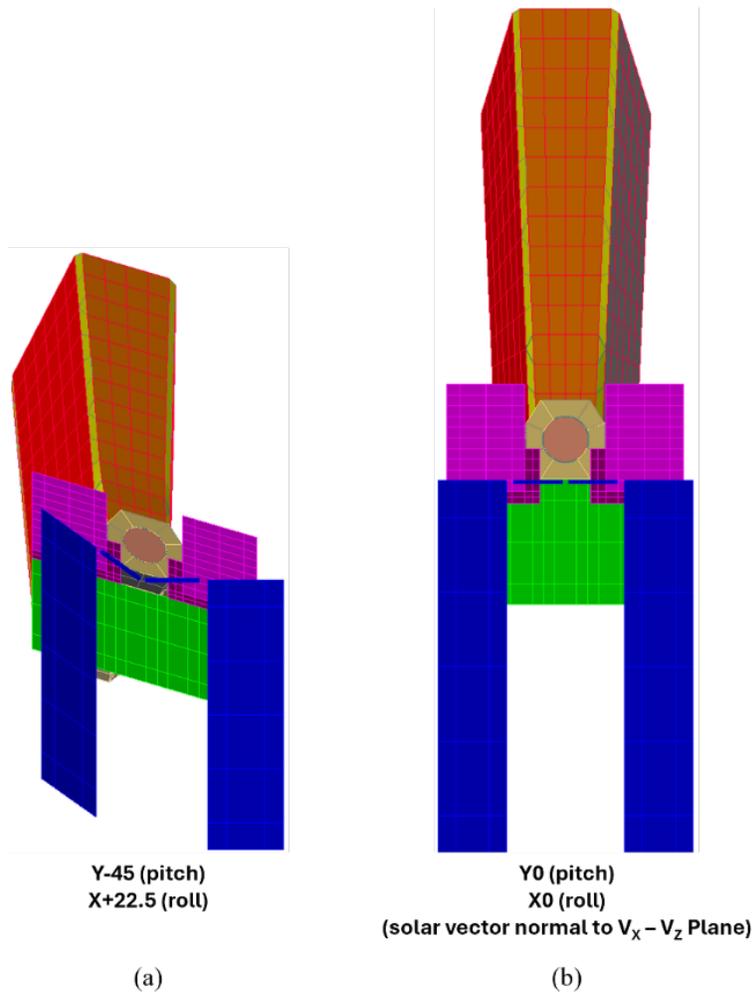

**Fig. 95** Field of Regard Example Attitudes: (a) Y-45 X+22.5 and (b) Y0 X0.

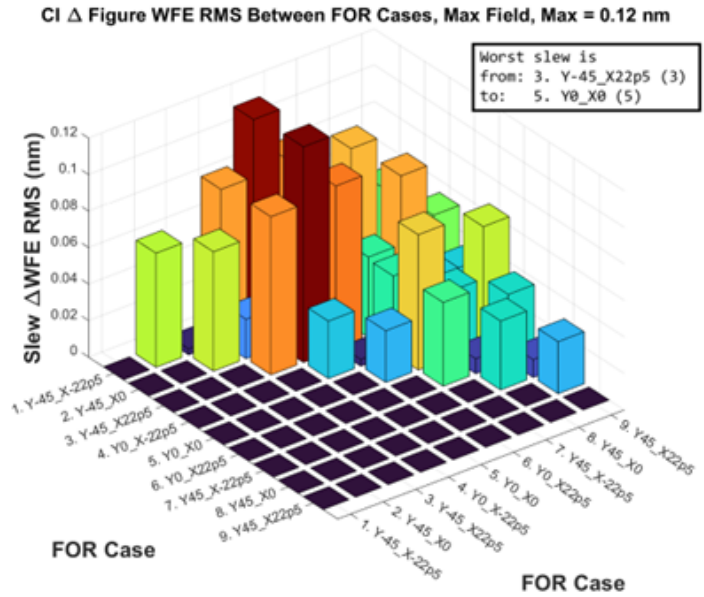

**Fig. 96** Example Analysis across the Field of Regard.

Figure 97 shows the timeline for a transient coronagraph OS-1. It is based on Roman analysis, as described in Sec. 1.1, and includes a settling period and 50-hour Wavefront Sensing and Control calibration with time-dependent power profiles for electronics, mechanisms, and deformable optics.

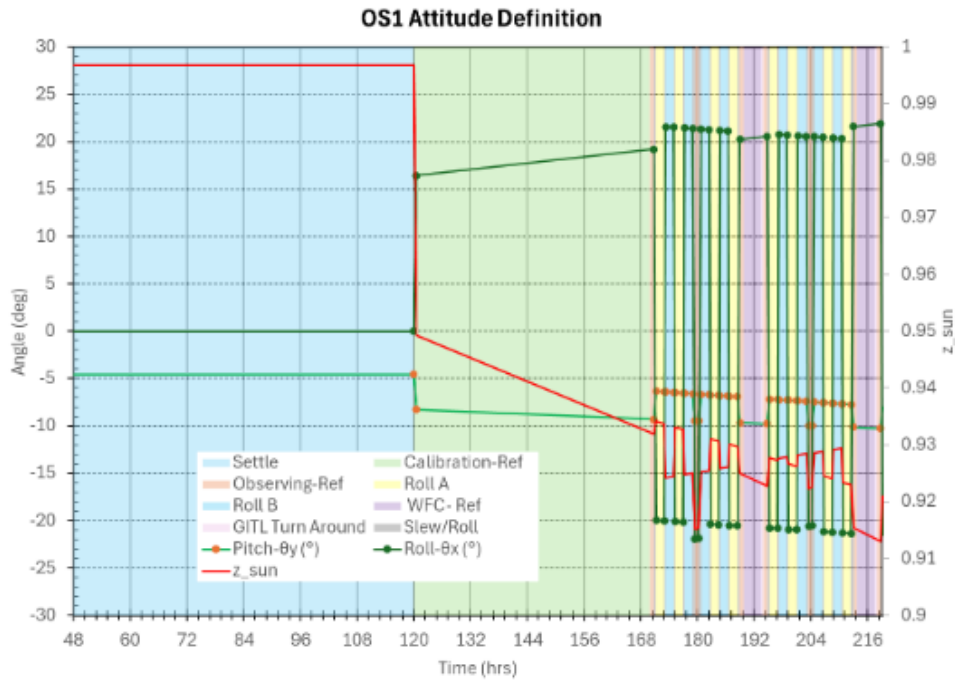

**Fig. 97** Coronagraph Observing Scenario (OS) Modeled on Roman CGI.

After calibration, the OS-1 "dance" begins, similar to Roman's OS: pitch change from reference to target star, followed by repeated roll maneuvers between extremes with Wavefront Sensing touch-ups and Ground-in-the-loop turnarounds (see Fig. 98). Given its material contribution to stability performance, this reflects the portion of OS-1 that the IM team is closely studying.

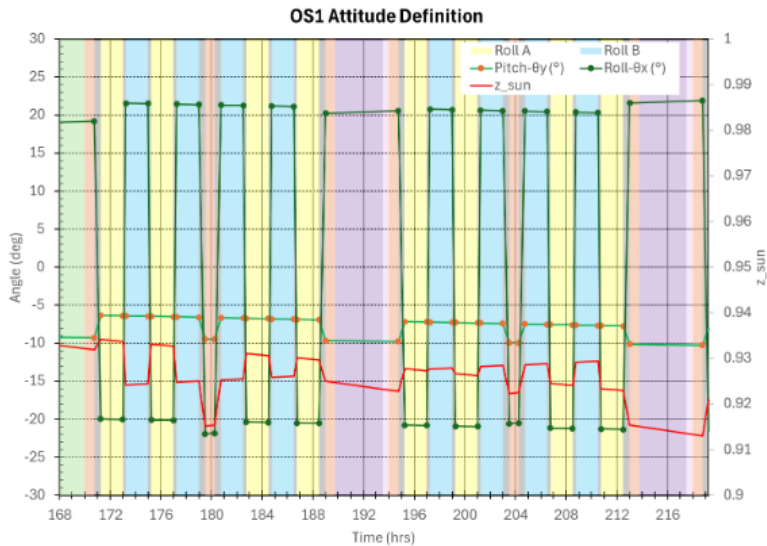

**Fig. 98** "The Dance" Portion of a Coronagraph OS.

All analyses (both FOR and OS) begin by stabilizing temperatures at an initial solution with radiation couplings calculated by Thermal Desktop. Environmental heating rates are calculated based on the specified parameters:

- FOR cases use a fixed orbit defined by observatory pitch and roll
- Coronagraph OS cases use a time-dependent solar-vector list

After thermal predictions are generated, temperatures are mapped to the structural model and sent through the STOP pipeline. In FOR analysis, as it is a steady state comparison, only the final timestep is mapped for each case. In OS analysis, data is mapped every 300 seconds from the end of the calibration period through the end of the run.

### 4.3.1.2 Postprocessing Results using Sink Temperatures

In anticipation of growing model complexity and associated increases in computing time, the IM team has been exploring thermal proxies for optical performance in the EAC1 model. The most useful proxy is tracking of radiative sink temperatures in the baffle cavity that quantify the radiative environment in the boresight direction of the PM segments.

Throughout the evolution of the EAC1 thermal model, analysis results demonstrate a strong correlation between thermal stability of these sinks and WFE due to Figure change. Thermal stability is then defined as the delta of the average boresight sink temperature between points of interest. In FOR analysis, this is the difference between the two attitudes with the largest difference in WFE change. In OS-1, this is the delta from the end of the calibration period to other times within "the dance."

### 4.3.1.3 Discussion of Results

One key takeaway from EAC1 is that the attitudes exhibiting the largest difference in WFE correspond consistently to the extremes of solar baffle loading: maximum solar energy input (pitch 0 / roll 0, or Y0 X0) and minimum solar energy input. Note, Y0 X0 reflects the attitude with the baffle's highest view factor to the sun, causing it to absorb the most energy, leading to the warmest temperatures. Attitudes with the most pitch and roll, particularly when the baffle is pitched furthest away from the sun (Y-45 X22.5), reflect attitudes with the baffle's smallest view factor to the sun, absorbing the least energy, leading to the coldest baffle conditions. Minimizing or compensating for the change in energy flowing into the baffle, as that directly affects the baffle cavity's radiative environment, is critical to achieving overall stability performance.

To demonstrate the effect, Table 14 juxtaposes cases 3800 and 3800-C. Case 3800 represents a fully passive baffle, without thermal control, while case 3800-C holds the inner layer of the baffle at a boundary temperature. This mimics "perfect" thermal control and eliminates any variability in front of the PM. The delta in sync temperatures between the worst FOR attitudes drops from 780 mK to 1.2 mK and the associated WFE due to segment surface distortion drops from 118 pm to 0.43 pm—more than 99% of the optical instability is attributable to temperature changes on the inner layer of the baffle.

**Table 14** Sink stability and Optical Performance in FOR Analysis.

| Case Description | | Boresight Avg. Sink Metrics | | | Figure Error (pm) in FOR |
|---|---|---|---|---|---|
| Case # | Boundary Control of Baffle | Max (K) Y0 X0 | Min (K) Y-45 X22.5 | diff (mK) | |
| 3800 | None | 201.474 | 200.694 | 779.8 | 118 |
| 3800-C | Inner Layer of Baffle Held as Boundary | 293.267 | 293.266 | 1.19 | 0.43 |

In OS-1 analysis, the IM team tracks stability from calibration end (t=170hr). Figure 99 shows the average boresight sink temperature changes for this case and, compared to Fig. 97 and Fig. 98, qualitatively correlate with attitude profiles and optical performance.

The OS-1 initial pitch change from calibration to target star drives sink instability and Mean Figure Error (defined below). Both optical results and sink stability are especially sensitive to pitch changes that alter the baffle's solar view factor; while roll changes solar load direction, the magnitude remains constant between positive and negative angles. Thus, STOP analysis indicates that designing pitch-resilient systems is key to stability performance.

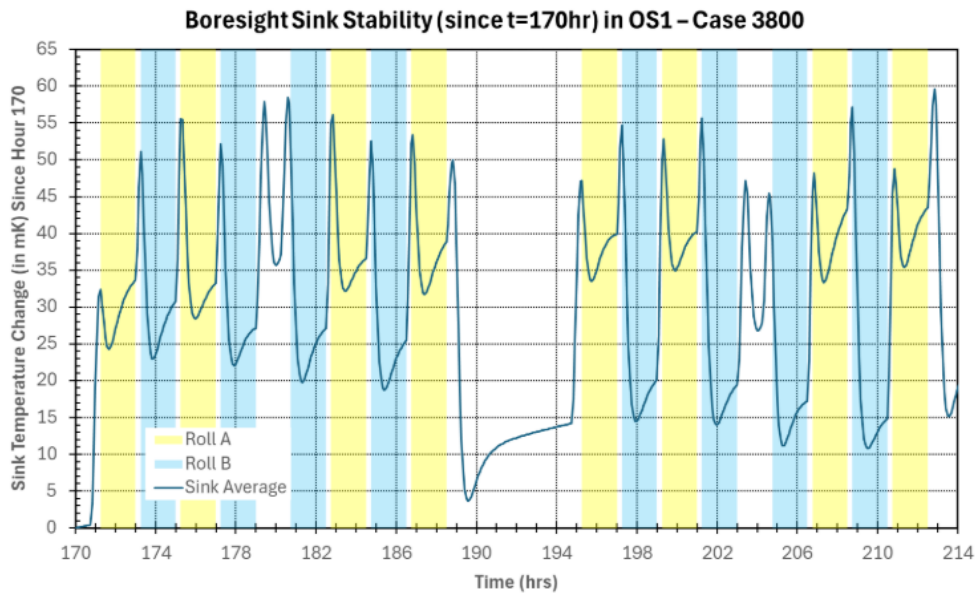

**Fig. 99** Sink Stability in Coronagraph OS Analysis.

The following reflect several, additional findings from STOP analysis and subsequent on-going work:

- The thermal team is currently exploring results from OS-1 analyses, specifically how to account for the time constant of the mirrors and proxies for Variance and Delta optical metrics
- The current OS-1 profile defines only roll endpoints, excluding intermediate attitudes (including zero-crossing); this limits capture of peak solar loading and represents an area for improvement in future analysis iterations
- The model currently reflects telescope heaters as boundary nodes for computational efficiency, excluding the effects of power variations and resulting thermal gradients across heater zones; planned studies will incorporate PID control modeling for enhanced surface fidelity

### 4.3.2 Effect of Rigid Body Control System

Active pose correction of individual primary mirror segments, described in Sec. 4.3.3 and in Ref. 19, is an enabling technology for the HWO mission. Figure 100 and Fig. 101 show WFE response broken down into Alignment and Figure components, where Alignment represents rigid body position of the segments while Figure represents their flexible internal surface deformation. The rigid body control system is able to correct Alignment from the nano-meter range into the single-digit pico-meter range, but is limited by noise and measurement errors.

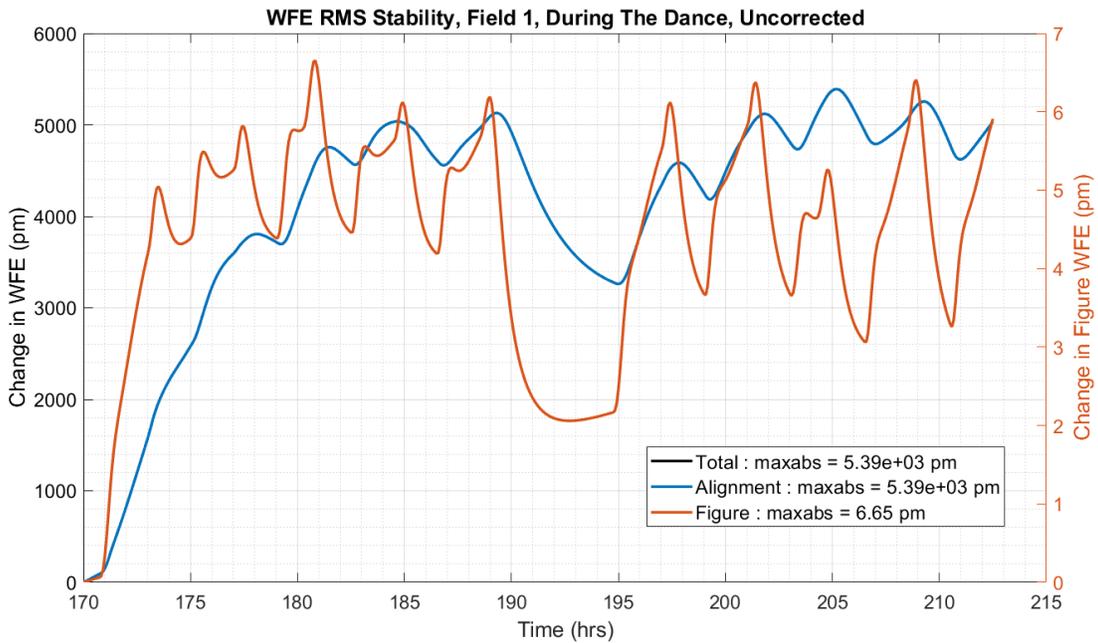

**Fig. 100** Uncorrected WFE during the Dance, Figure and Alignment Contributions; WFE is driven by Alignment (alignment and total lines are on top of each other).

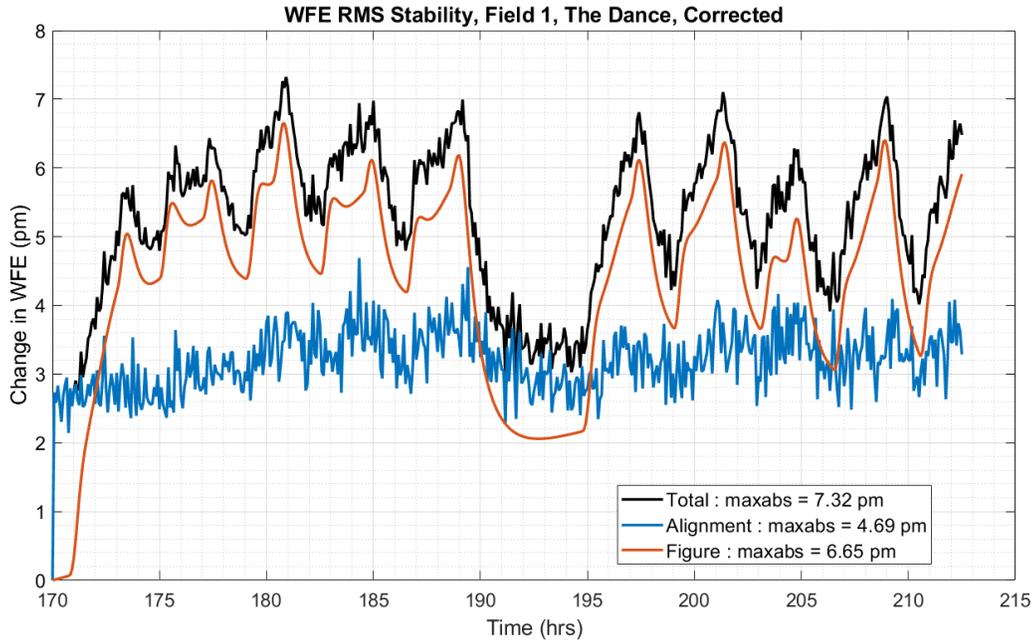

**Fig. 101** WFE during the Dance after Primary Mirror Segment Rigid Body Correction.

Given the assumed CTE distribution in the single instantiation of the telescope analyzed in EAC1 and the thermal architecture behind the mirror segments, the Figure portion of WFE looks like a scaled version of segment-level power shown in Fig. 102. This signature cannot be corrected at this level of analysis but can be addressed by either out-of-field WFS&C or DM control loops inside the coronagraph instrument itself. The DM controlled residual errors are shown later in the paper.

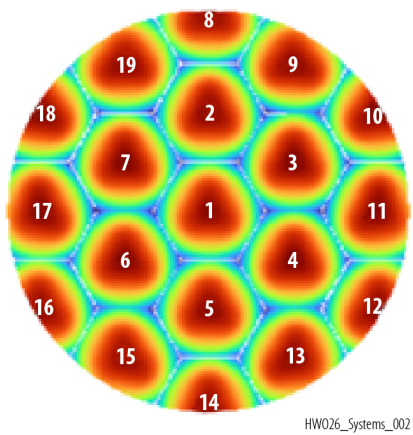

**Fig. 102** Typical Figure WFE Signature in EAC1.

Coronagraphic technique baselined for HWO uses ADI and is focused on discerning the planet from the noise in the images by rolling the observatory around its boresight axis while pointing at the target star. The scenario is thus divided into Roll A and Roll B attitudes, with stability defined as Mean from the end of wavefront sensing and control (t=170hrs) and Variance across the entire observation, as well as Delta Mean and Delta Variance between Rolls A and B. The quantities are shown graphically for the Figure portion of the response in Fig. 103 and Fig. 104 and for the total response in Fig. 105 and Fig. 106.

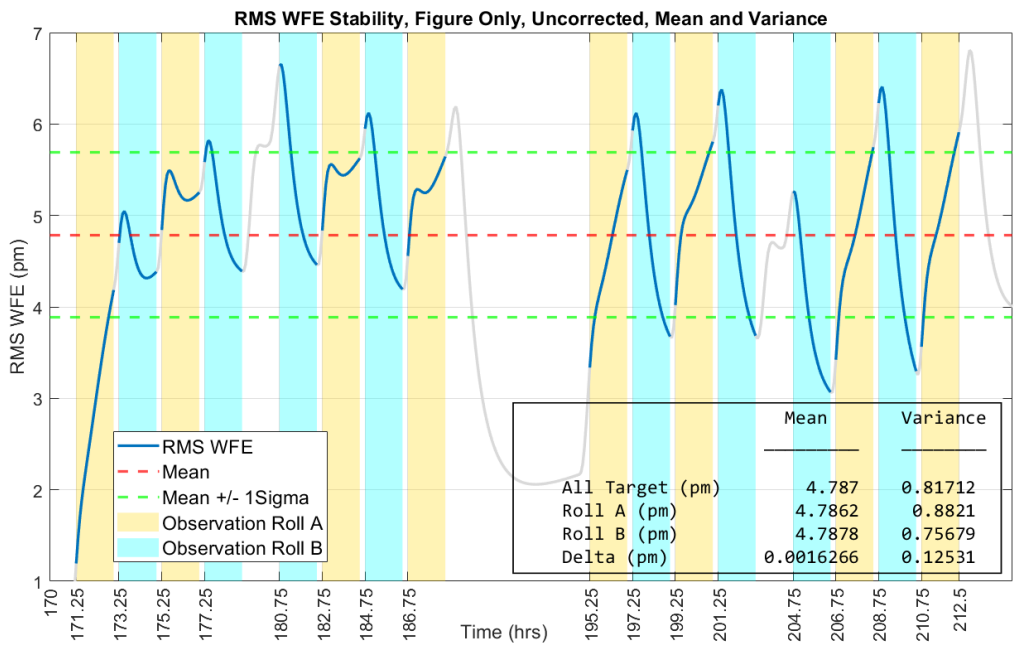

**Fig. 103** Visual Representation of Mean and Variance CI Stability Metrics for Figure WFE.

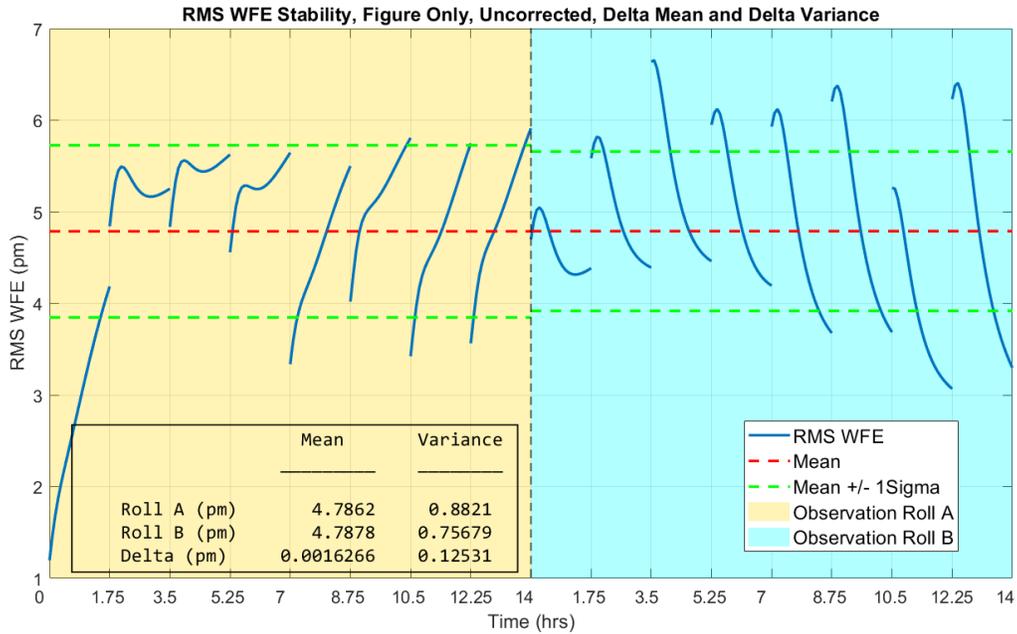

**Fig. 104** Visual Representation for Delta Mean and Delta Variance Metrics for Figure WFE.

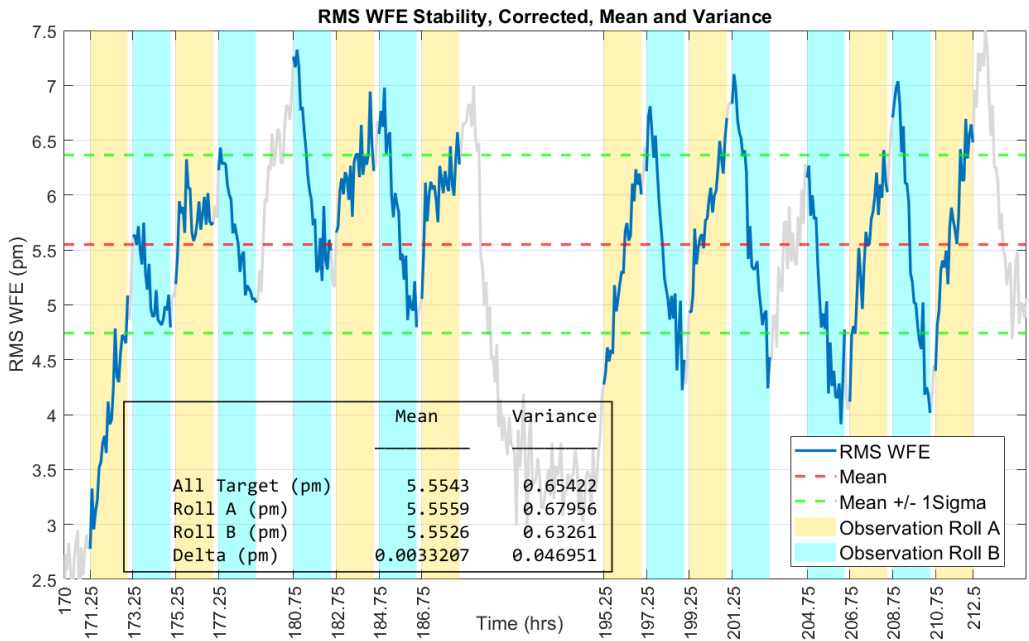

**Fig. 105** Total WFE after Rigid Body Correction, Mean and Variance.

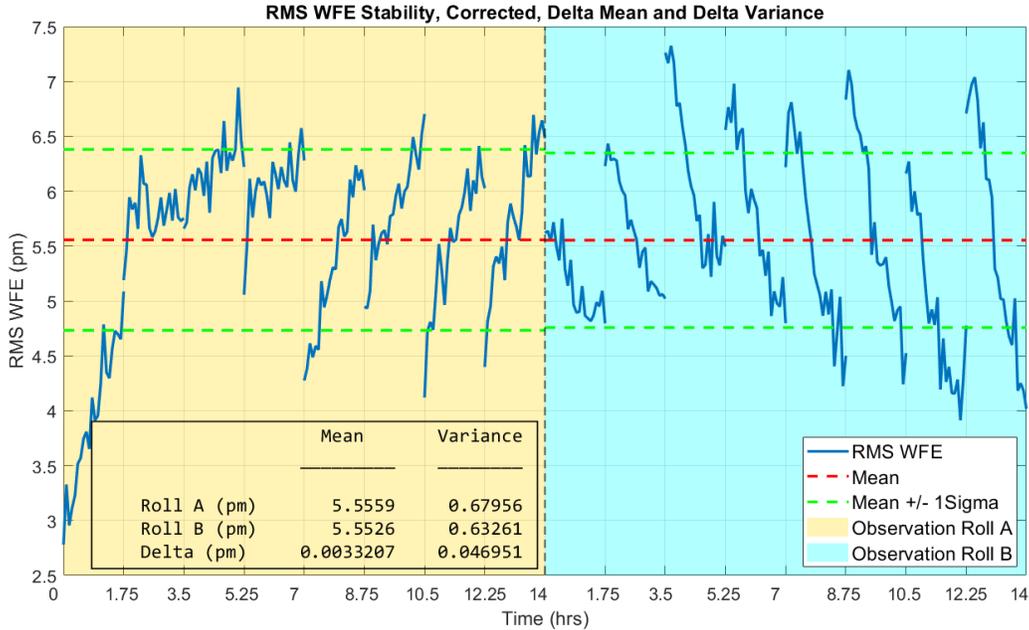

**Fig. 106** Total WFE after Rigid Body Correction, Delta Mean and Delta Variance.

Initial results show promising single-digit picometer wavefront error stability during the OS-1 "dance," indicating strong potential for maintaining the optical quality required for exoplanet detection.

### 4.3.3 Wavefront Sensing and Control Analysis

In the IM pipeline as described in Sec. 3.1, the Optical Path Difference (OPD) at the entrance pupil of the coronagraph from the STOP analysis for a given OS, including the rigid-body control of the PM segments and SM, is provided as the input to the coronagraph WFS&C model of DM control. This section shows an example implementation of the LOWFS uses a vector Zernike Wavefront Sensor (vZWFS) to estimate variations in the complex electric field, and provide inner-loop DM control used to maintain contrast stability and reject thermally imparted changes in the incoming wavefront that telescope rigid-body control could not correct.

#### 4.3.3.1 Vector Zernike Wavefront Sensor Model Overview

The vZWFS-based control uses out-of-band starlight reflected off a dichroic beamsplitter in the apodizer pupil, relayed to a phase-shifting mask made of liquid crystal or metasurfaces, and then the two phase-shifted polarization states are separated via a Wollaston prism and imaged onto a single focal plane, for reconstruction of the complex electric field. Throughout the coronagraph observations, the vZWFS-sensed complex electric field is compared to the target electric field determined during initial Electric Field Conjugation (a simplification for now, in the absence of outer-loop high-order wavefront control from the Direct Imaging Focal Plane). The DM control law is built from the DM actuator Jacobian matrices about the target electric field, with weak actuators removed and a tunable regularization constant selected through trial wavefront error convergence. At this time, the DMs are assumed to each have 64x64 actuators, and the Inner Working Angle (IWA) and Outer Working Angle (OWA) of the vZWFS-based control are 2 $\lambda$/D and 12 $\lambda$/D, respectively, though future analyses may vary the number of actuators or the OWA.

Observing scenarios are simulated by starting from a pre-computed initial DM solution that conjugates the electric field at the end of the calibration period (time = 170 hours) and pre-computed DM actuator Jacobian matrices about the target wavefront map. For each subsequent timestep, a compact diffraction model is used to propagate the wavefront phase errors at the coronagraph entrance pupil to predict the complex electric field in the apodizer pupil, at a single monochromatic wavelength. Images are generated at the vZWFS focal plane for a given target stellar magnitude and integration time, and photon noise is added, prior to reconstruction of the electric field. Then the DM control law is applied to command the DMs to a new solution for the next vZWFS integration window. The process is then repeated across all timesteps in the dataset.

Additional details on the vZWFS sensor and DM control law are described in Ref. 19.

### 4.3.3.2 Closed-Loop DM Control Performance

The vZWFS-based DM controller model was applied to the post-rigid-body-control STOP results presented in Sec. 4.3.2, assuming a one minute integration time and DM update rate, for representative target stars of apparent magnitude 0.0, 2.5, and 5.0 on the Vega-scale. The resulting Normalized Intensity (NI) between 2 $\lambda$/D and 12 $\lambda$/D, as well as the change in NI relative to time = 172 hours, were calculated for the closed-loop cases and compared to the open-loop inputs without DM control in Fig. 107.

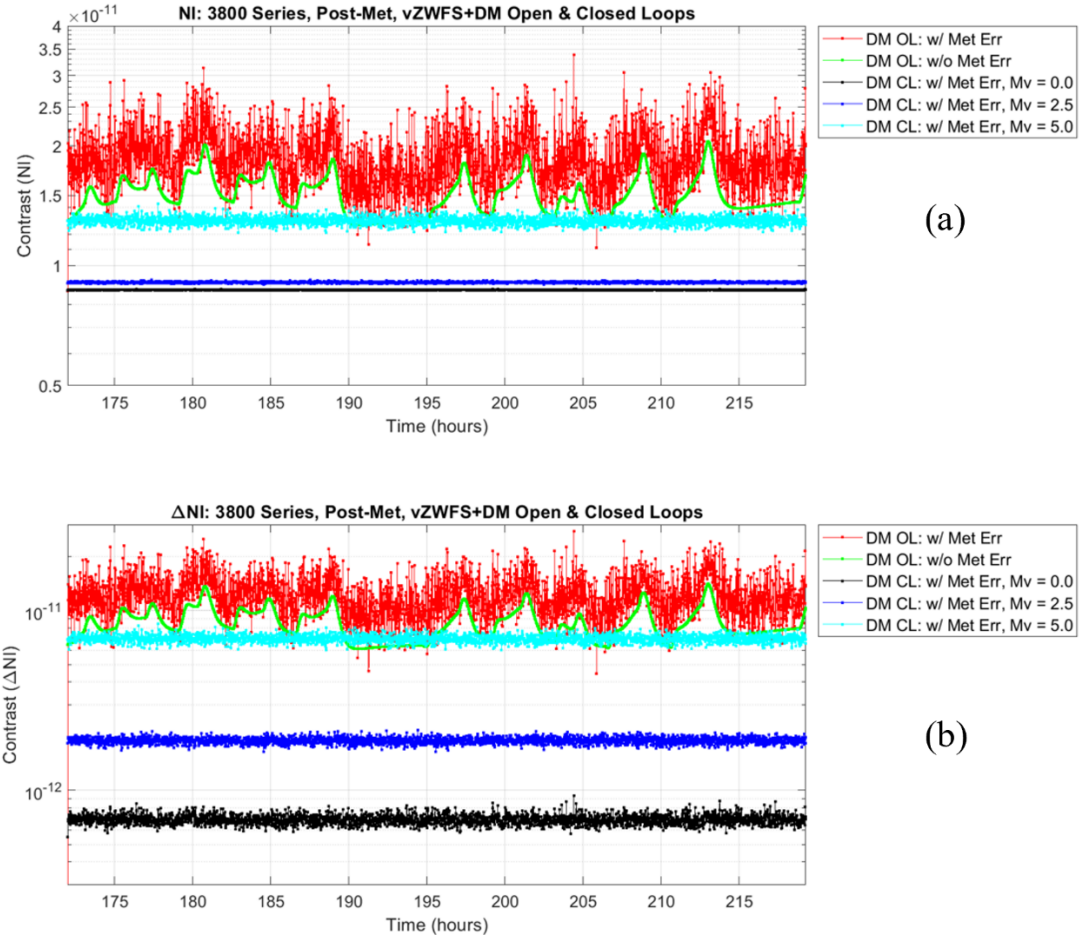

**Fig. 107** Normalized Intensity Performance of Closed-Loop vZWFS-based DM Control for OS-1: (a) Total NI and (b) Change in NI

The vZWFS-based control is able to sense and reject the residual PM segment Figure errors arising from thermal deformations during the observatory roll maneuvers, with residual NI driven primarily by the noise floor of the sensing, given the available starlight from the target and the integration time. For bright-enough targets, the thermally induced NI could be maintained below the $10^{-11}$ level. In order to extend performance to dimmer target stars, the WFS&C team will explore lengthening the integration time and optimize the trade-off between the noise floor's impact on the FRN budget's Variance and Delta Variance terms and the low temporal frequency residual thermal Figure errors' impact on the FRN budget's Delta Mean terms. The impact of chromatic effects within the vZWFS bandwidth on sensor performance will also be studied.

### 4.3.4  Diffraction Model and Analysis

Diffraction modeling and analysis is the final step in the IM STOP pipeline. This section describes the diffraction and coronagraph models and presents analysis results. These results are generated using the RBC wavefront errors from Sec. 4.3.2 as inputs, producing speckle time-series as outputs.

#### 4.3.4.1  Diffraction and Coronagraph Models

The full, nonlinear diffraction model leverages lessons learned from the Roman CGI. It includes all optics starting from the segmented PM (M1) through the entrance pupil of one channel of the coronagraph instrument (see Fig. 108).

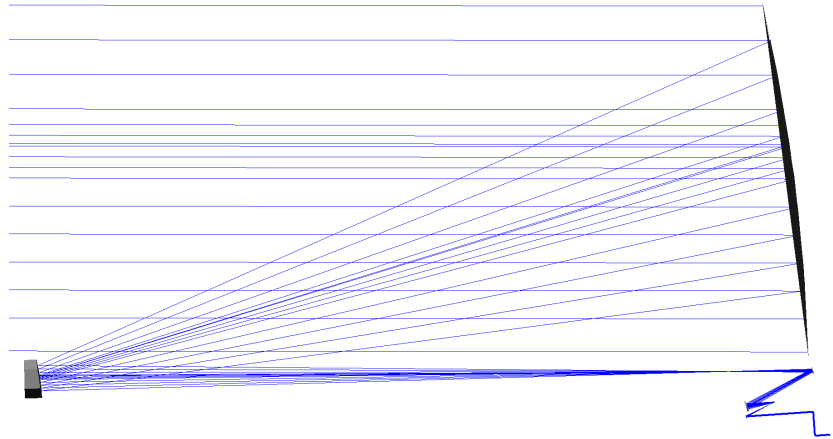

**Fig. 108** Front End Optical System Feeding the CI – This Portion of the Observatory is Included in the EAC1 Diffraction Model.

The optical system is unfolded into a linear sequence of elements while maintaining their effective focal lengths and the propagation distances between optics. Representative static surface aberrations that emulate potential manufacturing errors are applied to each optical element. As each time step of the thermal and structural models are processed, dynamic rigid body misalignment of these optics as well as surface deformations of the PM segments and SM are applied. For each instantiation, the segmented PM surface map is an amalgamation of surface deformations and residual rigid body motions in six degrees of freedom for all 19 segments. An example of a surface map of the full PM is shown in Fig. 109.

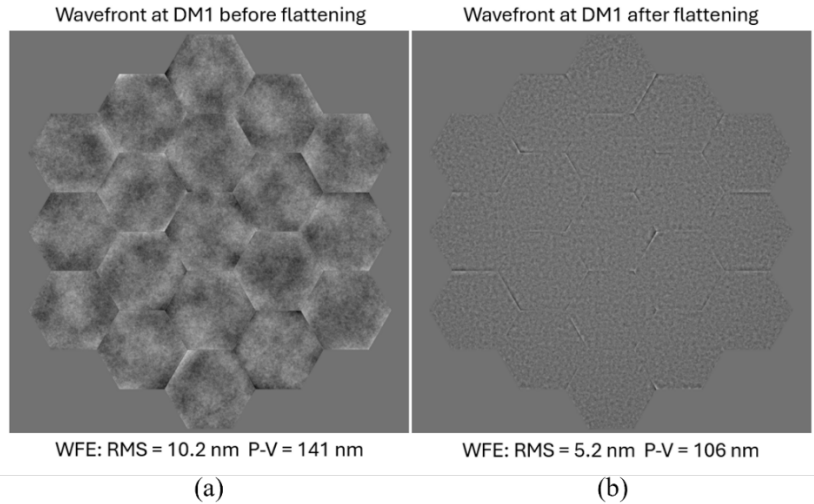

**Fig. 109** Representative Wavefront Map at the Coronagraph Entrance Pupil: (a) Before and (b) after Flattening with the Deformable Mirrors.

Fresnel electric-field propagation from optic to optic is modeled using the PROPER library and produces an electric-field map at the coronagraph entrance pupil for each time step. This electric-field is combined with the respective Code V OPD map to produce a total electric-field entering the Coronagraph Instrument.

While the telescope front-end is fully modeled, the HWO EAC1 coronagraph instrument itself is modeled in a compact form, excluding all optics except the two 96x96 DMs, ideal pupil planes for the apodizer and Lyot stop, and ideal focal planes for the coronagraph mask and detector. Although no surface errors within the coronagraph itself are modeled, polarization-dependent aberrations derived from ray tracing the full observatory and instrument are added at the first DM, labeled DM1 and located at a pupil plane. The apodizing mask, located at a pupil plane before the coronagraph mask, is an idealized realization designed to work with the HWO entrance pupil as defined by the segmented primary mirror. The charge 6 vortex focal plane mask is also represented by an idealized model. The Lyot mask is located at the pupil plane immediately following the focal plane mask. The EAC1 version of these key coronagraph masks do not include manufacturing defects or other such deficiencies at this time.

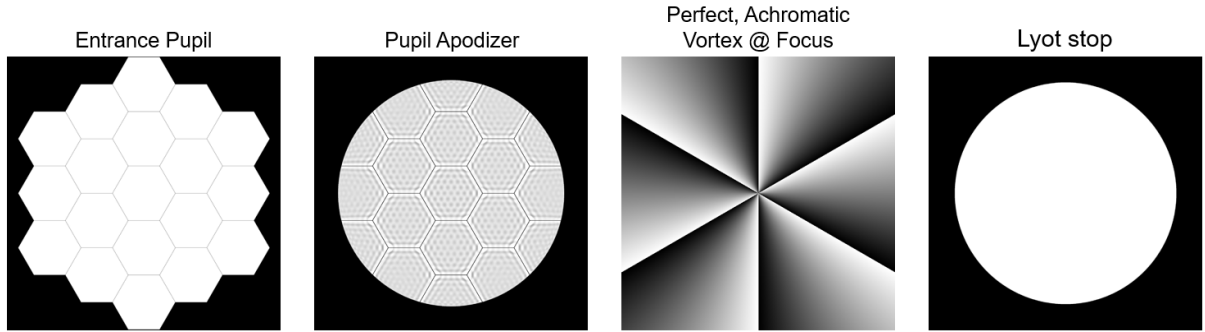

**Fig. 110** The HWO EAC1 Entrance Pupil with a Set of Masks for an Apodized Scalar Vortex Coronagraph (designs provided by Susan Redmond).

In this compact version of the Coronagraph Instrument, Fresnel propagation is only applied between the two DMs. Otherwise, direct propagation is modeled by applying Fourier transforms to and from pupil planes (apodizer and Lyot stop) and focal planes (vortex mask and detector).

### 4.3.4.2 Diffraction and Coronagraph Model Analysis

The initial coronagraph raw contrast is assessed at hour 170 of the time series by applying EFC to minimize the normalized intensity within the "dark hole" for a 20% bandpass centered at 550 nm based on the state of the optical model at that point in time. It is assumed all PM segments have been optimally positioned and the observatory has been properly pointed at the reference star, centering it on the vortex mask.

Thereafter, the time series analysis is focused on the stability of the telescope and coronagraph instrument and how it affects the contrast within the "dark hole" relative to the state at 170 hours. While results include RBCS control, as described in Sec. 4.3.2, they do not yet include wavefront sensing and control (WFSC), as presented in Sec. 4.3.3, to actively offset any changes to the optical wavefront. The diffraction analysis using WFSC is ongoing and expected to demonstrate even better contrast stability.

The output of the coronagraph model in the IM pipeline is a set of normalized intensity images containing speckle patterns called a speckle time series. These are processed and converted into a variety of analysis charts that represent the performance of the HWO coronagraph instrument. These charts include speckle patterns over the region of interest around the target star using the ADI technique (Fig. 111) and normalized intensity averaged over various annular regions of interest (Fig. 112).

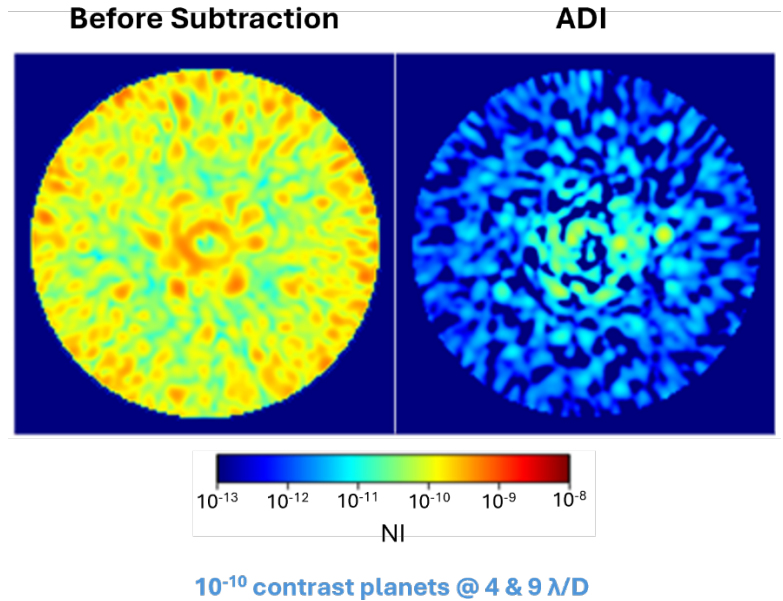

**Fig. 111** A Speckle Time Series Normalized Intensity (NI) Image along with Angular Differential Imaging (ADI).

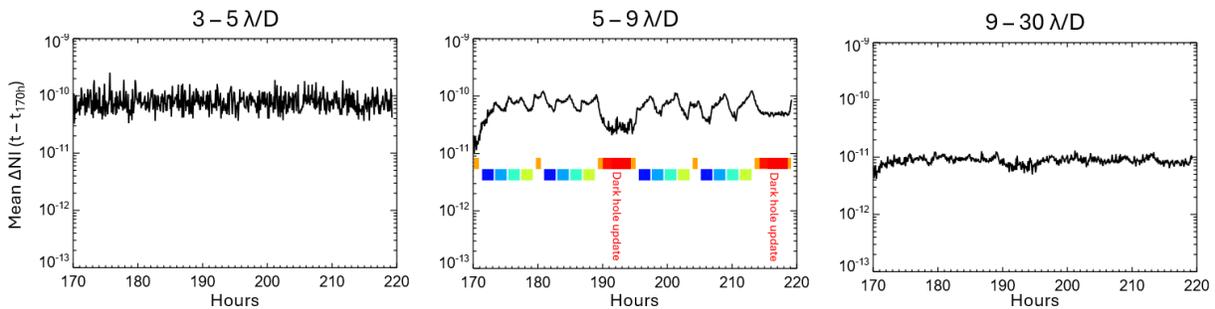

**Fig. 112** Relative Normalized Intensity vs. Time for Three Annular Regions around the Target Star: 3 - 5 l/D, 5 - 9 l/D, and 9 - 30 l/D Referenced to Timestep 170 h; Note the High Frequency Noise from the RBCS is most Evident at 3 - 5 l/D and 9 - 30 l/D.

High frequency noise from the Rigid Body Control System model is observed primarily within the $3 - 5\ \lambda/D$ and $9 - 30\ \lambda/D$ annular regions. The noise within the $3 - 5\ \lambda/D$ region is due to the segmentation of the primary mirror into approximately five equal partitions across its diameter while noise within the $9 - 30\ \lambda/D$ region is most likely due to the higher spatial frequencies attributed to the edges of the mirror segments. Future EACs may consider modifying the segmentation of the primary mirror to reduce noise within the inner-most annular region where Earth-like exoplanets are more likely to be found.

Although instrument performance as shown in Fig. 111 and Fig. 112 looks reasonable for an HWO coronagraph, it must be reiterated that it is still very early in the developmental stage and the EAC1 models include aspirational noise limits that require further technology development to achieve while excluding many other factors and deficiencies that will further degrade coronagraphic capabilities and require additional control models to overcome. As HWO models are further developed, the speckle time series data sets will be stored and made available for more in depth post-processing studies. In the meantime, Ref. 16 provides an in-depth discussion of end-to-end modeling of the Roman CGI coronagraph that is a preview of what will come.

*4.3.5  STOP Summary*

The team has developed a comprehensive integrated modeling capability for STOP analysis that extends beyond typical thermal distortion evaluation to include WFS&C and diffraction analysis. As can be seen from the preliminary results presented here, the ability to understand and identify performance drivers adds a powerful tool to the SET's toolbox. Critically, this tool allows designers to trade off thermal control capability against telescope rigid-body control and wavefront sensing and DM control, determining the temporal and spatial frequency overlaps necessary to achieve the requisite optical stability. STOP analyses results to-date, both FOR and OS-1, suggest the current EAC designs are close to meeting figure variation requirements with active optical control. In response, the SET has laid out a STOP analysis plan for the near-to-intermediate future that focuses on a series of trades to further understand the performance drivers and close on design options that bridge the requirements gap; the following describes a few of these trades,

- PM segment thermal control: To what extent can active thermal control architecture for the PM segments improve performance?
- Perfect baffle vs. passive baffle: can a design with a completely passive baffle, with no active thermal control, allow closure to requirements? To what extent can baffle design constraints relax if the observatory architecture also implements a sunshade to mitigate thermal variation?
- Baffle architecture: Does a non-scarfed baffle, much easier to implement mechanically but problematic thermally given views into the cavity, impose too high a penalty?

The trade studies, only briefly described here, reflect only the "tip of the iceberg" with respect to future analysis plans. Going forward, the SET will continue to leverage the integrated modeling STOP process to characterize design performance and shape design decisions through trades.

## 5 Summary and Future Work

The SET, comprised of team members across NASA centers and JPL, successfully completed design architectures for EACs 1-3 and developed end-to-end modeling for EAC1. This team was established to support the HWO TMPO in performing design explorations and formulating technology requirements. Using model-driven practices based on JWST heritage and enhanced by Roman experience, the team developed several tools to perform parametric design studies and trades. In parallel, the models were shared with industry partners to support their technology maturation efforts and independent architecture studies.

While many trades are on-going, the team has identified several key findings from architecture trades that will inform future EAC design objectives and assumptions.

Science Driven Trades – Aperture Size

a) The aperture size and associated science benefits were revisited at the end of EACs 1-3 efforts. Larger apertures provide three key advantages: greater collecting area, smaller coronagraph inner working angle (IWA), and smaller PSF scale. These improvements enable shorter exposure times, access to more habitable zones/target stars, and reduced exozodi contamination.[25] Smaller IWA also means planets are observed at larger λ/D in the coronagraph dark zone, where planet transmission and detection confidence are.[26] Consequently, exposure times can scale as D^-5 and exoEarth yield as D^2.[27] Near-Infrared benefits are even greater: measuring $CO_2$ and $CH_4$ on potentially Earth-like planets requires near-infrared observations where coronagraphic IWA is the primary constraint. Here, aperture size is critical and NIR yields can exceed D^2 scaling.[28] Additionally, larger apertures provide critical risk reductions by mitigating exozodi uncertainty, addressing coronagraphic noise floor challenges, and enabling relaxed instrument requirements that reduce technology investment.[28] These factors drive EAC 4 and 5 designs to accommodate a larger aperture to explore this trade space.

Payload Engineering Trades – OTE Prescription

a) A critical architectural trade for EACs 1-3 is on-axis vs. off-axis telescope geometry, with coronagraph performance as the deciding factor. The off-axis architecture eliminates on-axis pupil obscurations from the secondary mirror and support structure, which diffract light and create speckles that interfere with exoplanet detection. While mitigation techniques exist (pupil masks, deformable mirrors), they reduce throughput, bandwidth, and aberration tolerance - all critical to science yield. The off-axis design provides a more straightforward path to achieving coronagraphic performance for exo-Earth detection.

b) However, off-axis designs require twice the angle-of-incidence (AOI) of on-axis designs for the same focal length, producing polarization aberrations that degrade contrast, especially at short wavelengths and small separations. Both LUVOIR and HabEx[29] limited AOI to ≤12° to maintain contrast. The off-axis geometry also doubles the focal distance, potentially increasing volume and reducing structural rigidity. Furthermore, Lack of axial symmetry increases segment types for a segmented primary, raising manufacturing costs and schedule.

c) Note, EACs 1 and 2 designs demonstrated that the SM tower can be packaged while EAC3 struggled with creating a stiff secondary mirror tower. Ultimately, the off-axis architecture offers advantages that improve overall coronagraph performance and represents the most stressing case for packaging. Therefore, EACs 4 and 5 will adopt the off-axis architecture.

Observatory/Spacecraft Trades – Sunshade

a) Design efforts have revealed the telescope baffle/barrel as a key engineering challenge. This large deployable structure needs to fit inside the rocket fairing, deploy after launch, be thermally stable, accommodate thermal control system hardware (for an active thermal system), provide micro-meteoroid protection, be light-tight, and accommodate stray light vanes, to name a few of the challenging requirements. Addition of a large JWST-like flat sunshade removes a few of the challenging requirements—thermal stability, light-tightness, and stray light requirements could be significantly relaxed and ease the design challenges. While a large sunshade was not part of the trade consideration for EACs 1-3, it will be studied as part of the EACs 4 and 5.

With the engineering team and integrated modeling pipeline in place, and EACs 1-3 established, the team has identified the following near-term future work:

- Completing dynamics, jitter, and STOP trades using EAC1 IM pipeline
- Developing observing strategies and post-processing algorithms
- Generating EAC 4 and 5 designs to accommodate a larger primary mirror and improve stability margin[1]
- Forming closer industry collaborations on mission design exploration and sharing lessons learned

- Providing mass/power/volume and mechanical and thermal interface assumptions for the upcoming instrument study calls

The SET will continue to advance the EAC designs with lessons learned from EACs 1-3 and modeling/analysis results from EAC1. The goal is to develop two new EAC designs for 4 and 5, expand the trade study breadth, and perform IM on two additional designs to inform the follow-on EAC that will lead to a successful Mission Confirmation Review.

*Disclosures*

The authors declare that there are no financial interests, commercial affiliations, or other potential conflicts of interest that could have influenced the objectivity of this research or the writing of this paper.

*Acknowledgments*

The author list recognizes those who directly contributed to the manuscript for this paper. The authors would like to acknowledge the following teams and individuals who made significant contributions to EACs 1-3 design and modeling efforts. The authors would also like to give special acknowledgment to Mike Menzel, who served as the Mission Systems Engineer and led the SET through the completion of EAC2.

- System: David Hughes,[a] Mike Menzel,[*] and Julie Van Campen[a]
- ACS: Oscar Alvarez-Salazar,[b] David Arndt,[b] Tupper Hyde,[*] and Eric Stoneking[a]
- Mechanical: Colby Azersky,[a] Marcel Bluth,[i] Chas Carlson,[c] Stephen Cheney,[h] Brian Childs,[c] James Cooper,[c] Ryan Green,[*] Chris Hopkins,[h] Drew Jones,[*] Daniel Kolenz,[b] Ben Mignosa,[a] Alberto Rosanova,[a] Mike Schein,[*] and Jim Whipple[a]
- Optical: Brandon Dube,[*] Guangjun Gao,[a] and Qian Gong,[*] and Mike Rodgers[b]
- Structural: Kevin Carpenter,[b] Steve Cheney,[h] Katie Cheng,[i] Mike Eisenhower,[k] Chris Hopkins,[h] Will Krieger,[b] Jaganathan Ranganathan,[h] and Alex Sprunt[b]
- Thermal: Greg Allen,[b] Ed Canavan,[a] Emily Maheras,[f] Chris May,[c] Sang Park,[k] and Hume Peabody[a]
- Starlight Suppression and Diffraction Analysis: Martina Atanassova,[i] Der-you Kao,[i] John Krist,[b] Greg Michels,[j] Navtej Saini,[b] and Scott Will[*]
- Wavefront Sensing and Control Modeling: Alden Jurling,[a] John Lou,[b] David Redding,[b] Hari Subedi,[a] Kaitlyn Summey,[a] Jon Tesch,[b] Mitchell Troy,[b] and Hanying Zhou[b]

[a]NASA Goddard Space Flight Center, 8800 Greenbelt Rd., Greenbelt, USA

[b]Jet Propulsion Laboratory, California Institute of Technology, 4800 Oak Grove Dr., La Cañada Flintridge, USA

[c]Aerodyne Industries, LLC, 8910 Astronaut Blvd., Suite 208, Cape Canaveral, USA

[d]Quartus Engineering, 2300 Dulles Station Blvd., Suite 650, Herndon, USA

[e]NASA Ames Research Center, Moffett Field, USA

[f]Vertex Aerospace, LLC, P.O. Box 192, Grasonville, USA

[g]Tellus1 Scientific, LLC, 8401 Whitesburg Dr. SE, Unit 4662, Huntsville, AL, USA

[h]NASA Marshall Space Flight Center, Martin Rd. SW, USA

[i]KBR, 7701 Greenbelt Rd., Suite 400, Greenbelt, MD, USA

[j]Sigmadyne, 1200C Scottsville Rd., Suite 350, Rochester, NY, USA

[k]Smithsonian Astrophysical Observatory, Cambridge, MA, USA

[*]These individuals are not currently with the project at the time of this paper's publication

The Jet Propulsion Laboratory, California Institute of Technology, contributed to elements of the HWO systems design and analyses under a contract with the National Aeronautics and Space Administration (80NM0018D0004). @2026 All rights reserved. California Institute of Technology. Government sponsorship acknowledged.

ChatGSFC was used for limited language editing.

**Alice Liu** serves as the HWO Mission Systems Engineer and the Integrated Modeling and Error Budget Lead for the Nancy Grace Roman Space Telescope. She has worked at NASA Goddard Space Flight Center for over 23 years, supporting various flight missions while holding multiple leadership roles. She earned her PhD from the Massachusetts Institute of Technology in Dynamics and Control, focusing on optimizing staged control systems for large space interferometers, with support from a Michelson Fellowship. She began her career as the Jitter Lead for the Solar Dynamics Observatory and the Attitude Control System Lead for the Terrestrial Planet Finder-Coronagraph study. Throughout her career, she has accumulated significant experience in payload pointing, stability, and jitter mitigation, regularly providing consultation to proposal teams, design labs, and multiple flight projects to address increasingly stringent stability requirements.

**Marie Levine** received a Ph.D. from the California Institute of Technology and has been working at the Jet Propulsion Laboratory for 35 years as an expert in precision optical systems and integrated modeling. She is currently the HWO Integrated Modeling lead within the NASA Mission Systems Engineering team. She held a similar position for nearly 14 years on the James Webb Space Telescope. Prior to that Marie was the Technology Manager of the Exoplanet Exploration Program, the Systems Manager for the Terrestrial Planet Finder Coronagraph, and the PI on two flight experiments IPEX-I and II for the on-orbit characterization of microdynamics in precision structures. Marie is also a member of the NASA Standing Review Board for the Roman Space Telescope. She has received many awards including the NASA Exceptional Engineering Medal, the NASA Exceptional Service Medal and was designated Woman of the Year by US Representative Adam Schiff where here accomplishments were read in front of Congress.

**Charley Noecker** is an optical engineer at NASA Jet Propulsion Laboratory/Caltech. He received his BA degree in physics from Swarthmore College, Pennsylvania, in 1981 and his PhD in physics from the University of Michigan in 1988. He has been involved in exoplanet studies and space optics since early 1990 at the Smithsonian Astrophysical Observatory in Cambridge, Massachusetts, and then at Ball Aerospace in Boulder, Colorado and NASA JPL in Pasadena California. He has authored or coauthored more than 100 journal papers and a U.S. patent. His interests continue to be in advanced optical systems and exoplanet research. He is a member of SPIE.

**Jon Lawrence** has worked at NASA Goddard Space Flight Center for 34 years; 25 of which he was the Mechanical Systems Lead and Launch Vehicle Liaison for the James Webb Space Telescope. Subsequently, he was the Mechanical Systems Lead on the Capture Containment Return System (Mars Sample Return) and now acts as Mechanical Systems and Launcher Lead for the Habitable Worlds Observatory. Prior to JWST, he was NASA's Deployables Lead on the Tropical Rainfall Measuring Mission. Before joining NASA, he worked for 7 years at General Dynamics Space Systems in the Advanced Composites Design group on multiple space projects. He received his B.S. degree in Mechanical Engineering from Virginia Tech.

Biographies and photographs for additional authors are not included for brevity.

## Caption List